# Satoshi Nakamoto and the Origins of Bitcoin – The Profile of a 1-in-a-Billion Genius

*Jens Ducrée, School of Physical Sciences, Dublin City University, Ireland, email: jens.ducree@dcu.ie*

## Abstract

The mystery about the ingenious creator of Bitcoin concealing behind the pseudonym Satoshi Nakamoto has been fascinating the global public for more than a decade. Suddenly jumping out of the dark in 2008, this persona hurled the decentralized electronic cash system "Bitcoin", which has reached a peak market capitalization in the region of 1 trillion USD. In a purposely agnostic, and meticulous "leaving no stone unturned" approach, this study presents new hard facts, which evidently slipped through Satoshi Nakamoto's elaborate privacy shield, and derives meaningful pointers that are primarily inferred from Bitcoin's whitepaper, its blockchain parameters, and data that were widely up to his discretion.

This ample stack of established and novel evidence is systematically categorized, analyzed, and then connected to its related, real-world ambient, like relevant locations and happenings in the past, and at the time. Evidence compounds towards a substantial role of the Benelux cryptography ecosystem, with strong transatlantic links, in the creation of Bitcoin. A consistent biography, a psychogram, and gripping story of an ingenious, multi-talented, autodidactic, reticent, and capricious polymath transpire, which are absolutely unique from a history of science and technology perspective. A cohort of previously fielded and best matches emerging from the investigations are probed against an unprecedently restrictive, multi-stage exclusion filter, which can, with maximum certainty, rule out most "Satoshi Nakamoto" candidates, while some of them remain to be confirmed.

### Short Abstract

*With this article, you will be able to decide who is <u>not, or highly unlikely</u> to be Satoshi Nakamoto, be equipped with an ample stack of systematically categorized evidence and efficient methodologies to find suitable candidates, and can possibly unveil the real identity of the creator of Bitcoin - if you want.*

# Table of Contents

# Introduction

## Unprecedented Mystery

The now notorious pseudonym Satoshi Nakamoto [1] stands for the remarkable inventor of one of the most ground-breaking innovations of the ongoing 21st century: the cryptocurrency [2] Bitcoin [3] (ticker symbol [4]: BTC) with its underlying blockchain [5] concept; this disruptive technology has enabled the "Web3" [6] by equipping the internet with an innate value layer. Yet, even up after more than a decade after its launch in 2008/2009, only little is known about this individual or group, who spent extraordinarily sophisticated efforts on hiding his/her/their true identity, biography, and whereabouts.

Under his alias, Satoshi Nakamoto was only interacting with the community for about two years between the second half of 2008 and end of 2010, followed by emails to peers in the following spring, before completely vanishing end of April 2011 without clear farewell or guidance message. Afterwards, Satoshi Nakamoto appears to have only surfaced with brief statements, one in an attempt to stop a media-driven manhunt for a Japanese-American sharing his "name" [7], to dismiss a mainstream media hype [8], and to intervene in a self-destructive civil war raging within the Bitcoin community [9]; however, serious doubt has been cast on whether these postings truly originated from the elusive architect of Bitcoin [10]; and if not, why would the new owner have only used them so sparsely.

Astonishingly, the mastermind of a banking revolution has (so far) never touched (any significant share) of his rather abundantly filled coffers of digital assets, which are estimated to accrue of the order of 1 million Bitcoin (ticker: BTC), and which he is believed to have been hoarded in 2009 as the dominant miner on the nascent blockchain; assuming a trading pair of 20,000 US$ / BTC (at the time of writing), these holdings roughly correspond to a whopping 20 billion US$, ranking him amongst the richest people on the planet. Some pundits forecast a (volatile) growth towards 100,000 US$ / BTC, 500,000 US$ / BTC, and even 1 million US$ / BTC, which would lift him, possibly before the end of the current decade, to the top of the Forbes list of billionaires [11].

This is, to the author's best knowledge, the first time that the inventor of a "unicorn"-like [12] technology, that has reached, along quite a roller-coaster ride, a maximum total value on the range of 1 trillion US$, has inapprehensibly departed, remained completely anonymous, and stayed abstinent from his massive fortune, fame and impact. This behavior utterly contrasts business celebrities like Bill Gates [13], Steve Jobs [14], Elon Musk [15], Mark Zuckerberg [16], Jeff Bezos [17], or Jack Dorsey [18] in the related IT/Web2.0 [19] space, or prominent blockchain pioneers and promoters like Vitalik Buterin [20, 21], Charles Hoskinson [22, 23], Gavin Wood [24, 25], Sam Bankman-Fried [26-28], Roger Ver [29], or Changpeng Zhao ("CZ") [30, 31].

Quite fascinatingly, and despite the fact that Bitcoin was published and launched, from a historic point of view, rather recently in 2008/2009, and still "lives" on the internet, the methodology for unravelling its origins tends to resemble more the reconstruction of a long-lost ancient culture, or a cold-case criminal investigation [32]; this is surprising in the age of information where data is ostensibly eternalized, and usually leaves an indelible digital trace. The problem is that the search quasi exclusively relies on sterile electronic, rather than physical artefacts carrying telling "dirt". In this work, traces comparable to DNA, fingerprints, or clothes fibers left in crime scenes are retrieved; and most remarkably, this decisive evidence has been openly available in Bitcoin whitepaper's and blockchain parameters that has been in the public domain from the very beginning of the story.

## Pool of Candidates

Numerous initiatives aimed to unravel the identity of Satoshi Nakamoto [33-68]. There is a short list of credible candidates [69-72], possibly also [73, 74], who are, fairly consensually, considered by the crypto community to have been capable of assembling Bitcoin in 2008; all of them, whether still alive or prior to their demise, vehemently deny. A wider circle of "suspects" has, more or less seriously, been "nominated" [42, 46, 47, 58, 72-93]; however, most experts would severely challenge these their involvement in the creation of Bitcoin. Likewise, these contenders have overwhelmingly, and categorically dismissed their involvement.

Yet, from a group of otherwise quickly debunked self-nominations [94, 95], one Australian-born candidate shot up in December 2015; he stubbornly insists having been the main figure in orchestrating Bitcoin [39, 57, 96-100], and even filed libel cases targeting some of those openly challenging his inventorship [101-108]. Over recent years, he also filed a large number of blockchain-relevant patents [109-111] through his employer [112].

A slew of further names has been dropped over the years [15, 58, 83, 85, 113-135], without sustained traction, or even consensus. At the end, and despite countless search efforts, we still do not know the person(s) behind the alias Satoshi Nakamoto; as a matter of fact, the crypto-community overwhelmingly agrees that this mythical genesis story bears great benefit for Bitcoin and its offspring, i.e., second- and third-generation blockchains [21, 23, 25, 136-141]. In certain, susceptible communities, a (quite concerning) development resembling a quasi-monotheistic [142], sect-like "Church of Satoshi" cult, with a narrative alluding to a celestial origin of Bitcoin is observed.

## About the Author

While having published in the field of blockchain in recent years [143-151], coming from a background [152, 153] in physics [154-158] and microsystems engineering [159-164], the author of this article was neither pertaining to the cryptographer nor the so-called cypherpunk scene [165-168] during the birth of Bitcoin, nor is he affiliated with, or intends to specifically promote any of the potential "Satoshi Nakamoto" candidates. However, along his endeavor, the author also had the opportunity to converse with select insiders, who tend to be very cagey and reclusive, asking for anonymity in the context of this work.

Furthermore, the author's multi-decade experience with scientific publishing and the research community [169, 170], as well as his close interaction with the corporate sector and public organizations, turned out to be a decisive enabler for detecting the stack of new quality evidence presented here. This auspicious constellation offers the benefit of a rather neutral, as much as possible evidence-oriented approach to eventually draft an unbiased and soundly-reasoned Satoshi Nakamoto profile.

## Motivation and Methodology

### Solidity of Evidence

A major part of this work results from turning every stone, some of them revealing exciting stories, others do not appear to be leading to anything substantial; still, while resulting in a long article, all avenues are documented, underpinned by a plethora of citations, also to empower subsequent research, e.g., by data mining approaches. The length of this paper, and its high level or detail, is crucial for a proper scientific methodology; this is because any, at first glance, even miniscule evidence that irreconcilably contradicts a given "Satoshi Nakamoto" hypothesis immediately annihilates such a binary identity claim in its entirety. The particular selection and dismissal of Satoshi Nakamoto candidates that will be examined in this work reflects this exclusion strategy.

All evidence, such as metadata from postings with their timestamps [171], time zones [172], Internet Protocol (IP) [173] addresses [174], and their linked geographical locations, leave a certain chance of having been manipulated by the inventor of Bitcoin, especially when originating from the digital and online arena. Still, some fresh "analog" evidence has been carved out, which has obviously slipped past Satoshi Nakamoto's highly refined privacy screen, and which has, quite startingly, been widely overlooked in previous investigations; these new snippets mostly concern the conspicuous collection of citations in the Bitcoin whitepaper [175], and the circumstances of the Bitcoin genesis block [176, 177], which are carefully examined and put into context in this investigation.

### Traces in Numbers?

In addition, further contributions to the story presented here are derived from the name, and, as per the common crypto-motto "Vires in Numeris" (English: "strength in numbers"), numbers and dates set, with a significant extent of discretion, by Satoshi Nakamoto, and their connection to the creation of the Bitcoin blockchain and its contemporary and historical embedding. It may certainly be argued whether these are, on the one hand, overinterpretations of unintended coincidences, or, on the other hand, deliberately placed breadcrumbs, e.g., for humoristic purposes, for encouraging a (mostly virtual) "Indiana Jones" [178] or "Da Vinci Code" [179] like treasure hunt similar to cryptograms familiar from notorious pirate stories [180], or crime [181]. The relevance of such, at first ostensibly weak indirect evidence [182], tends to consolidate when occurring in multiple independent contexts.

### Benefits of Foundational Mystery

The search for Satoshi Nakamoto also resembles the deciphering of cultural puzzles or understanding the objectives pursued, and techniques utilized by historic civilizations for constructing monuments, e.g., prehistoric Stonehenge [183], The Egyptian Pyramids [184, 185], Nabataean Petra [186, 187], and others [188-198], or the provenance of, or methods for forming, the relics like the Shroud of Turin [199], the Seamless robe of Jesus [200], or the Shrine of the Three Kings [201]. At times, the author is even further jinxed by his fascination for the origin of Bitcoin, irresistibly prompting him to draw parallels to scouring for the historic root of old stories, such as the (fictional) island of Atlantis [202], the medieval saga of Rheingold [203], King Arthur [204], or the famed search for the Holy Grail [179, 205-207]. People may spot further allegories [208] in Satoshi Nakamoto's writings, and even venture to find parallels to the still perplexing, medieval-age prophecies by Nostradamus [209], and delve into their manifold options for their interpretation. But this, of course, stretching it extremely / too far.

However, while the Bitcoin story unfolded in the present digital era, literally in front of our eyes, and Bitcoin technology is fully understood [210, 211], progressively advanced and increasingly utilized in the real world, the story of its beginnings, and the very peculiar mindset of its ingenious creator(s) may remain cloaked in eternal obscurity.

It is highly doubtful whether this was part of his ploy, but, at the end of the day, and in addition to removing a single point of failure by pivotal decentralization, the foundational mystery has also succeeded as a priceless marketing strategy for Bitcoin itself, and its crypto-descendants; it also lifted Satoshi Nakamoto to the top celebrity amongst core internet technologists, making him even better recognized in the general public than, for instance, Tim Berners-Lee ("TimBL") [212], the key figure behind the World Wide Web (WWW) [213].

### Unique Methodology

Alternative to most other investigations on the creator of Bitcoin, this study avoids starting out with a specific candidate, and then exclusively presenting evidence supporting their hypothesis, while putting a blind eye on inconsistencies. Instead, an outline of a Satoshi Nakamoto persona, including prerequisites and elimination criteria underpinned by best evidence, is agnostically developed, before screening this template against known and newly identified public figures.

Note that due to the striking rarity of material in the form of vigorously peer-reviewed papers that are the best-practice in scientific publishing, the author decided to include (a subset of) the abundance of webpages and online feeds that have been produced over time, entirely independently from this work; these sources can, unfortunately, seldomly be properly validated, and may even be susceptible to manipulation, e.g., by Satoshi Nakamoto himself, by an imposter, his promoters, opponents, or investors. Furthermore, insiders or confidants are apparently not inclined to come out of the woods to tell the story, or not even to expose Faketoshis, because they may be happy to leave the public in the dark, and protect their own, and potentially also Satoshi Nakamoto's privacy.

Hence, in absence of "smoking guns" or "caught in the acts" type of direct, eyewitness evidence, this article is poking in deep and muddy open waters for a treasure not meant to be found, so, unavoidably, numerous modifiers such as "likely", "probably", "possibly", "presumably", "apparently", "supposedly", "seemingly", "ostensibly", "hypothetically", "theoretically", "maybe", and "perhaps" are spread throughout the text.

Even though some circumstantial evidence presented here are unlikely to have been intended by Satoshi Nakamoto at the time, they are an essential part of the "every stone turned" approach vigorously pursued in this investigation. Also, historic correlations to certain forgotten techniques and historic events may often seem to be farfetched at first sight, but the methodology for developing lines of argument, to rank alternative interpretations according to their likeliness in order to arrive at the most plausible hypothesis definitely applies to the quest for the origins of Bitcoin.

## Objectives

This study is mainly motivated from a history of science and technology perspective, following the overall objective [214] "Those who cannot remember the past are condemned to repeat it." Furthermore, the idea that this game-changing, potential "person of the century" may (hopefully) still be among us, like a "guy next door", is definitely enthralling. A celebrity with a similar quest for anonymity is known in the scene of contemporary street art uses the pseudonym "Banksy" [215].

## Real-World Impact

From a practical point of view, Satoshi Nakamoto's biography, personality, motivations, and concerns derived here may substantially contribute to a sound risk analysis on the future of Bitcoin and cryptocurrencies, e.g., regarding the flooding of the market with early mined BTC [216], copyright [217] and other intellectual property [218] matters. However, more than a decade after Satoshi Nakamoto's (online) disappearance, the fields of blockchain and cryptocurrency have considerably moved on, and thus became widely autonomous, and technically detached from its enigmatic founding father.

## Number of References

As the subject of Bitcoin and its origins is of highly interdisciplinary nature, the reader is referred to a rather unusually comprehensive, at first glance, excessive collection of explanatory webpages, primarily to (the English-language version of) Wikipedia [219]. It is fully understood that the resulting, vast number of citations that are decorating the text might slightly impair the reading experience; yet the author prefers to properly define topics. Furthermore, unraveling the enormous conundrum of the origin of Bitcoin is, and will be decisively boosted, by systematic collection of even the slightest cues, and aided by advanced data mining [220] the comprehensive repertoire of circumstantial evidence, e.g., in conjunction with (more advanced) semantic web [221] / knowledge graphs [222] technologies, machine learning [223], and artificial intelligence (AI) [224] services [225-229].

## Date and Time Format

Note that unless specified otherwise, the (European) date format DD/MM/YYYY [230] is used throughout the text and references; note that specified dates may vary by $\pm 1$ day, depending on the local time zone [231], and whether standard [232], or daylight savings time (DST) [233] are applied during summer in the respective global hemisphere [234, 235]. Consider that whether and when an entire country, or its individual federal states and regions, switch between standard and "summer time" (DST) might vary, and their routine may have changed over the years [236, 237].

While the digital world utilizes coordinated universal time (UTC) [238], e.g., to avoid complications around DST [233], considerations of events in local time zones [231], with Greenwich Mean Time (GMT) [239] as default, are of outstanding importance to this study; at the end of the day, Satoshi Nakamoto is a human being whose chronobiology [240], i.e., biological rhythm ("clock"), makes activity during daylight hours most natural. Note that the time zone specified on postings on curated newsgroups are typically imprinted by the corresponding clock setting of their (central) server, i.e., typically to the time zone of their own, and not the sender's geolocation. Moreover, select time zones run split, e.g., 30- [241-246] or 45-minute [247-249], rather than integer-hour offsets with respect to UTC [238].

## Structure or Article

### Evolution of Bitcoin

This work starts with a survey of the evolution of Bitcoin over the period when Satoshi Nakamoto was actively involved in Bitcoin between 2007/2008 and 2010/2011. Furthermore, possible scientific and ideological backgrounds of Bitcoin in the cryptography research [250] and cypherpunk [165] communities are introduced. Afterwards, the financial, legal and personal that might have impacted Satoshi Nakamoto's path to, and away from Bitcoin is outlined.

### Bitcoin Whitepaper

As a pivotal element of this study, the Bitcoin whitepaper is carefully screened; as a result, novel, hard evidence emerges, in particular from its, at second look, somewhat bizarre list of citations and metadata. The various cues discovered might have slipped through Satoshi Nakamoto's otherwise very efficient, cunningly designed shield of anonymity.

### Bitcoin Dates

Next, the calendric dates of the release of the Bitcoin whitepaper and the start of the Bitcoin blockchain with block 0 and 1 are examined for potential correlations and their respective interpretations in the context of the origins of Bitcoin.

### Personality and Bio

Then, key fragments of Satoshi Nakamoto's character, professional and private vita are plotted from a diverse pool of information. While incorporating numerous existing findings, this effort is substantially supported by methodical analysis of the free parameters the creator of Bitcoin had available for his pseudonym, his self-defined date and place of birth, as well as his clever and wide-ranging strategy for privacy and anonymity.

### Blockchain Parameters

In the following, the Bitcoin blockchain is scrutinized for the parameters Satoshi Nakamoto could deliberately choose across a rather wide range of potential values. Variation of the eventually hardcoded settings is tested to check whether the reasoning of the number is purely technically reasoned, or may contain hidden messages. The theoretically less interested reader may fast-forward through the rather mathematical content to the last subsection delivering the take-home messages.

### Telling Numbers?

The subsequent section, for the first time, makes a best effort to interpretate the previously gained cues within the framework of math, alphanumeric systems, geometry, symbolism, history, and the sciences. Especially the numbers 21, and also 42, appear to act as pointers, e.g., to pertinent locations and events. However, if not entirely incidental, these wittily "encrypted" pieces are undoubtedly not meant to disclose Satoshi Nakamoto's formal identity; they rather seem to reflect his propensity for playing droll "brain games" and riddles with his audiences; such inclination is not uncommon for ingenious minds, and reveal facets of his attitude when conceiving, giving birth, nurturing, and weaning off his baby Bitcoin.

### Final Sections

Afterwards, the complex and multilayer stock of known and newly uncovered circumstantial evidence is thoroughly categorized. The most substantial parts of this assortment are wrapped around a scaffold to assemble a likely biography and psychogram of the individual concealing behind the moniker Satoshi Nakamoto. Next, a multi-stage sieving method composed of “Satoshiness” criteria is devised, and ultimately applied to best, potentially surprising matches immediately arising from this work, and various already circulating nominations.

### Appendix

An extended range of possibly relevant happenings in the context of Satoshi Nakamoto and Bitcoin regarding history, politics, economy, finance, money, society, art, science and sports during key years for Bitcoin is compiled in the Appendix. Its final section collects fun facts about the potential meanings of numbers highlighted in this investigation.

# The Birth and Infancy of Bitcoin

## Hitting the Surface

In July / August 2008, and thus just a few weeks prior to its launch, a “Satoshi Nakamoto” lunged, seemingly out of nowhere, onto the cryptographer scene when ostensibly finishing up his now famous Bitcoin whitepaper [175]; he used email addresses from highly privacy-oriented service providers [251-256], and found a way to incognito register the domain bitcoin.org [60, 257, 258] on 18 August 2008. Yet, interacting under a pseudonym was, and continues to be quite common in online fora, especially amongst cryptographers.

## Whitepaper

Around these days in July / August 2008, Satoshi Nakamoto contacted key stakeholders [60, 259-267] about prior art and their proper citation in the field of digital currencies [268-272], which he was, if not part of a crafty plot, not (fully) aware of at that time [272, 273]. He stated that “I actually did this kind of backwards. I had to write all the code before I could convince myself that I could solve every problem, then I wrote the paper” [274], that he contends to have been working on Bitcoin for about 1.5 years, i.e., since 2007 [275]. He stated “Much more of the work was designing than coding.” [276] and ”Writing a description for this thing for general audiences is bloody hard. There's nothing to relate it to” [277, 278].

On Friday, 31 October 2008 [279] at 14:10:00 EDT [233, 280], i.e., on a weekend-spirited early evening hour in European [281] time zones [231, 239], the Bitcoin whitepaper [175] was publicly announced [282] on the “cryptography - The Cryptography and Cryptography Policy Mailing List” [283] maintained by metzdowd.com [284]. The posting featured a text-based abstract, with double-spacing after the period [285], and a download link [175] to the PDF [286] file deposited at bitcoin.org [257]. The two spaces after full stop routine also seems to have been implemented throughout the Bitcoin whitepaper, even though the word processor would automatically adjust such gaps.

## Name Bitcoin

It has been conjectured that due to their very concept and fabric the name Bitcoin might have been inspired by its (arguable) precursors “Bit Gold” [71, 262], or “BitTorrent” [287-289]. If this was the case, it is very remarkable that both are not cited in the Bitcoin whitepaper.

## Launch of Bitcoin

The initial reception of the Bitcoin whitepaper [175] on this cryptography list [283] in November 2008 was overwhelmingly skeptical; only a handful of early adopters [70, 290-295] reacted during the first couple of weeks; even fewer programmers joined Satoshi Nakamoto in developing Bitcoin [70, 118, 294, 296-300] in this nascent period, and often bowed out quite soon. The Bitcoin source code files were shared by Satoshi Nakamoto on 16/11/2008 [66, 301].

On Saturday, 03/01/2009 at 6:15 pm GMT [302], the Bitcoin blockchain was kicked off with the Genesis block 0 [177, 302], which included the headline "The Times 03/Jan/2009 Chancellor on brink of second bailout for banks" [176] of the British newspaper "The Times" [303, 304]. Block 1 [305] was mined on, depending on the time zone [231], 08 or 09/01/2009 at 2:54 AM GMT [239], i.e., after a, so far, not conclusively explained, six-day hiatus.

Bitcoin v0.1 was released on Thursday, 08/01/2009 [306-308] at 14:27:40 EST [280] (i.e., 19:27 GMT [239]), and on Wednesday, 11/02/2009 [309] (without time of the day) at P2P Foundation [310] for download on SourceForge [311] and bitcoin.org [257], respectively; nowadays, the code resides at GitHub [312, 313]. Satoshi Nakamoto mentioned on the same day at bitcoin-list along the release of Bitcoin version 0.1.2 "… I wasn't able to test it in the wild until now" [314]. Note the telling "I" in this statement; if Bitcoin has been invented by a team, it would certainly have made great sense to resort to it as a multi-eyed testbed.

On the 11/01/2009 at 03:33 am, cypherpunk and early contributor Hal Finney [70, 315, 316] deposited his now famous tweet "Running Bitcoin" [317], making him the, most likely, the second miner after Satoshi Nakamoto. On 12/01/2009 at 03:30 (am?) [318], Finney received the first transaction of 10 BTC (i.e., input 50 BTC, output: 10 BTC to Hal, change of 40 BTC to a new Bitcoin address generated by Satoshi Nakamoto in the UTXO model) from Satoshi Nakamoto, mined in block 170 [319]. On 21/01/2009 at 05:29 pm, Finney added "Looking at ways to add more anonymity to bitcoin" [320], and on 27/01/2009 at 08:14 pm, he uttered, very presciently [321-323], "Thinking about how to reduce $CO_2$ emissions from a widespread Bitcoin implementation" [324].

## Bitcoin.org

The domain was originally owned by Bitcoin's first two developers, Satoshi Nakamoto and Martti Malmi [296]. After leaving the project, Satoshi Nakamoto transferred ownership of bitcoin.org to additional people, some of them separate from the Bitcoin developers, to spread responsibility and prevent any one person or group from easily gaining control over the Bitcoin project [325].

From 2011 to 2013, the site was primarily used for releasing new versions of the software now called Bitcoin Core. In 2013, the site was redesigned, adding numerous pages, listing additional Bitcoin software, and creating the translation system [325].

To the author's best knowledge, bitcoin.org was moved to "Louhi Net Oy" [326] in Finland [327] on 18 May 2011 [328], then registered to "WhoisGuard" [329] in Panama [330] on 2016-12-26 [331-333], and on 2021-09-23 [334] to "namecheap" [335] to with a country tag from Iceland [336].

## Bitcoin Community

Satoshi Nakamoto continued to actively participate in discussion and development, e.g., on the cryptography mailing list [283], BitcoinTalk.org [337], or the P2P Foundation [338, 339]. About a year after its birth, on 22 November 2009, Satoshi Nakamoto started the BitcoinTalk.org forum [337, 340]. The community helped with improvements and bug fixes of the code, e.g., the notorious "overflow bug" [341] that created more than 184 billion (!) BTC in block 74,638 of 10/08/2010 (before being corrected only 5 hours later by a soft fork). Around mid-2010, he successively handed over the responsibility of the Bitcoin project.

## Departure

On 12/12/2010 [342-346], i.e., after roughly two years, Satoshi Nakamoto unexpectedly and unglamorously disappeared from all online fora with a rather inconspicuous posting "There's more work to do on DoS, but I'm doing a quick build of what I have so far in case it's needed, before venturing into more complex ideas" [343]. By then, Satoshi Nakamoto had transferred the control over the Bitcoin project, e.g., bitcoin.org [257, 258, 328], to his trusted peers, which indicates a longer, and well-planned, gradual exodus, rather than a spontaneous emotional decision, or personal accident. His last forum posting does also not (directly) hint to a clinical cut, Satoshi Nakamoto rather just wanted to fade away, probably to avoid turmoil or begging to stay by the budding Bitcoin community. Even though highly speculative, this plot somewhat insinuates ascension narratives [347] as a central pillar supporting the origin myths [348] of various religions [349]. In that sense (of humor?), will there be, one day, a Second Coming [350] of the founder of Bitcoin?

Satoshi Nakamoto still continued sending emails [351, 352] for a short while to early contributors and trustees like Gavin Andresen [75], Mike Hearn [351, 353-355] and Martti Malmi [296], who took charge of bitcoin.org [257, 258, 356]. Here are excerpts from of the final messages in April / May 2011.

- Mike Hearn to Satoshi Nakamoto on 20/04/2011 [351]: "Are you planning on rejoining the community at some point (e.g., for code reviews), or is your plan to permanently step back from the limelight?", Satoshi Nakamoto: "I've moved on to other things. It's in good hands with Gavin and everyone."
- Satoshi Nakamoto to Gavin Andersen in 26/04/2011 [357]: "I wish you wouldn't keep talking about me as a mysterious shadowy figure, the press just turns that into a pirate currency angle. Maybe instead make it about the open-source project and give more credit to your dev contributors; it helps motivate them."
- Satoshi Nakamoto to Martti Malmi in early May 2011 [358]: "I've moved on to other things and probably won't be around in the future."

From his follow-on email communication with select peers, it appears that Satoshi Nakamoto wanted to mature his still nascent baby "Bitcoin" within its small developer community, and was strongly opposed, and apparently even scared of enhanced public exposure; this attitude manifested when Bitcoin was eyed by Wikileaks [346, 358-362] as a potential vehicle for bypassing its financial blockage in early 2011: "It would have been nice to get this attention in any other context. WikiLeaks has kicked the hornet's nest, and the swarm is headed towards us." He was also concerned when the CIA [363] invited an informative meeting on Bitcoin, where Gavin Andresen [75] presented in mid-2011.

On Sunday, 26 April 2011 [357, 363, 364], Gavin Andresen informed fellow coders: "Satoshi did suggest this morning that I (we) should try to de-emphasize the whole 'mysterious founder' thing when talking publicly about bitcoin." After this day, Satoshi Nakamoto stopped replying, even to personal emails of his inner circle. Bitcoin enthusiasts pondered somberly why Satoshi Nakamoto had departed.

Yet, by then, his epic invention had become adolescent, within a self-sovereign group of active supporters. Part of Bitcoin's long-term sustained success is often attributed to Satoshi Nakamoto's conservative strategy of deploying rather well-established and thoroughly field-tested methods developed in the 1970s-1990s [365], instead of resorting to highly advanced, but still experimental cryptographic techniques.

A potential factor of Satoshi Nakamoto's, at least from an outside perspective, incomprehensible exit might have also been that mailing lists are tough to cope with for the faint hearted; for instance, rudeness, lack of respect, and strong opinions without leaving room for compromise have led later to the notorious Bitcoin "civil war", which drove out some early key stakeholders [75, 353], who both directly conversed (through text messages) with Satoshi Nakamoto.

Gavin Andresen [75] described Satoshi Nakamoto as “a brilliant coder, but it was quirky” [366]. Others believed that Nakamoto might have been a group: Dan Kaminsky [367], a highly accomplished security researcher, who competently scrutinized the Bitcoin code, stated that Nakamoto could either be a "team of people", or must definitely have been a "genius"; a former Bitcoin core developer, Laszlo Hanyecz [368], also known for the first real-world purchase by Bitcoin [369, 370], uttered the feeling “Bitcoin seems awfully well designed for one person to crank out” [364]. One of the first Bitcoin adopters, Hal Finney [70, 315], who emailed Satoshi Nakamoto and helped him early on with the code, is quoted: “I thought I was dealing with a young man of Japanese ancestry who was very smart and sincere. I've had the good fortune to know many brilliant people over the course of my life, so I recognize the signs” [52, 315].

There were some later messages [7-9, 371, 372] sent by Satoshi Nakamoto via email or postings on, but their genuineness remains seriously disputed, as his associated online accounts might have been compromised, or simply repossessed after certain idle periods [373]. Yet, it would be worth contemplating why potential hackers did not make even more use of their power to speak in the name of Satoshi Nakamoto for the sake of their own interests, e.g., to unite or seriously damage the Bitcoin project, or to manipulate cryptocurrency markets.

The talismanic father of Bitcoin is assumed to have been a dominant miner in the first year 2009 according to the so-called “Patoshi patterns” [216, 374, 375]. The creator of Bitcoin chose not to include further messages in the blocks he succeeded to mine, such as the headline of “The Times” [303, 304] he engraved in the genesis block 0 [177, 302], or in transactions issued from Bitcoin addresses attributed to him [374]; this alternative texting mechanism is enabled by his blockchain’s intrinsic scripting language, while preserving his pseudonymity. Most remarkably, Satoshi Nakamoto seems to have never moved any (significant number of) coins, at least not from his about 1 million BTC he is firmly believed to have mined in 2009 [376] (but possibly from other addresses that he generated afterwards).

## Ideology

### Cypherpunk Movement

From his very motivation to engage into the monster effort of creating Bitcoin, as well as from his interactions with the community, it can be inferred that Satoshi Nakamoto was distinctively contemptuous of the legacy financial system, as expressed with the release of the code [362], and by the citation in the genesis block. It is unknown whether he fully subscribed to the more extreme ideologies [377], which were a signature of the cypherpunk [165-168] movement: libertarianism [378], crypto anarchism [93, 166, 379], and Austrian economics [380].

His libertarian attitude shines through rather rarely, e.g., in postings like [381] “Yes, [we will not find a solution to political problems in cryptography,] but we can win a major battle in the arms race and gain a new territory of freedom for several years. Governments are good at cutting off the heads of a centrally controlled network like Napster, but pure P2P networks like Gnutella and Tor seem to be holding their own.” on 07/11/2008 [382], and [383] “It's very attractive to the libertarian viewpoint if we can explain it properly. I'm better with code than with words though.” on 14/11/2008. His lead motive for developing Bitcoin appears to be his lack of trust in banks [310, 339].

However, an overly evangelistic approach is widely missing in his whitepaper and forum postings; they mainly feature a very rational and respectful, perhaps somewhat ideals-driven, and altruistic “techie”. Satoshi Nakamoto mostly engaged in factual expert discussion, rather than primarily acting as a commercially minded entrepreneur, or hot-headed preacher for the cypherpunk ideology. Whether intuitively or tactically, Satoshi Nakamoto did not boast about his own, early miners’ future wealth when extrapolating the potential value curve of BTC; a merely financial, capitalist-spirited investor or businessman might not have been able to resist emphasizing this "selfish" aspect.

### Anguilla Meeting

About 200 stakeholders of promoting of the new digital currency system e-Gold [384, 385] wanted to develop a strategy that could challenge central banks. They met in February 2000 [386, 387] at Anguilla [388], a British Overseas Territory in the Caribbean [389]. E-Gold folded in 2007 after its founders were indicted [390] by the United States Justice Department [391]. In accordance with the recent analysis of his motivation, skeptics argue that Satoshi Nakamoto lacked bias in implementing his new technology [53].

### Mysterious "X"

A number of libertarian-minded statements were posted on 9 & 10 December 2002 by an entity "X" under the topic "Virtual peer to peer banking" on the Usenet [392] groups alt.internet.p2p and uk.finance [393, 394]. Their later (re-)discovery in the context of the origins of Bitcoin sparked fiery speculation on Reddit [395] and Bitcointalk [396] whether these postings originated from the person later filing under the pseudonym "Satoshi" Nakamoto" [34].

Indeed, the text bits [397-399] "idea of a future with virtual peer to peer banking", "fixed total amount of money"," virtual coin", "deflation would replace inflation", "community can bypass the old powers (countries and governments)", "new p2p system", "current monetary systems were mainly backed with gold", "virtual peer to peer system could be other scarce resources, relatively easy to exchange via internet", "computer processing power, bandwidth and data storage", "virtual currency", "replace the system operator by a secure peer to peer system", "replace the underlying currency with something else, or slowly uncouple the underlying currency" exhibit many attitudes, objectives and technical features that entered Bitcoin a few years later in 2008/2009. Whether "X" was the individual later masquerading as "Satoshi Nakamoto" could (so far) not be verified. The posts of "X" were traced back [34] to IP addresses [174] in the Netherlands [400, 401], i.e., part of Benelux [401], and thus put the spotlight on some regional cryptographers [77, 113, 402, 403].

### Privacy

Highly gifted mathematical people, who are distinctly concerned about preserving their privacy, are often instinctively drawn into the fascinating field of cryptography [250]. Other than the general public who are mostly careless of how comprehensively data on their internet-connected devices and online activities are systematically tracked, analyzed, and monetized, cryptographers tend to be highly competent in protecting their data, and to efficiently conceal their (online) identity [404-410]. Measures of cover up, such as the registration of web domains and email addresses by fake names and addresses, like in the case of Satoshi Nakamoto, as well as air-gaped data storage, are believed to be part of their basic technical skillset.

### Distrust in Banking System

In his announcement of Bitcoin on the P2P Foundation [338], Satoshi Nakamoto clearly expressed his pronounced distrust in the traditional, central-bank issued money [339]:

"The root problem with conventional currency is all the trust that's required to make it work. The central bank must be trusted not to debase the currency, but the background of fiat currencies [411] is full of breaches of that trust. Banks must be trusted to hold our money and transfer it electronically, but they lend it out in waves of credit bubbles with barely a fraction in reserve. We have to trust them with our privacy, trust them not to let identity thieves drain our accounts. Their massive overhead costs make micropayments impossible."

It is important to consider that profound skepticism for the loose-handed monetary policies of central banks, the lax credit policies of lenders that fueled the immense subprime credit mortgage crisis, and the casino mentality of investment banking was not exclusive to libertarians and cypherpunks, but quite prevalent across societies at the time.

## Cryptography

### International Community

Major stakeholders in the cryptographer community are from the USA [50, 73, 80, 81, 122, 412-426], Canada [427, 428], Europe [281, 429] with the Benelux [113, 430-435], UK [436-440], Switzerland [423, 441, 442], Germany [122, 443-445], Denmark [446, 447]), and Israel [418, 436, 448-451], as well as, for instance, India [452], China [453] and Japan [454] in Asia [455, 456], and Australia [457, 458].

Significant crypto-expertise may also be found outside the academically dominated habitat, e.g., in nation states and their secret service and (anti-)cybercrime [459] organizations [460, 461], computer hardware and software giants (e.g., IBM, Intel, Microsoft, Apple, Google, Meta, Oracle, SAP, Hitachi, NTT), or special-interest groups like cypherpunks [165] and hackers [462].

### E-Cash

While the topic was not prevalent in the cryptographer community, electronic cash systems were sporadically published [417, 436, 463-484] in the cryptography community after its introduction [485] by David Chaum [73, 486] in the early 1980s. Interestingly, in the global mix of affiliations, a striking accumulation of authorships from Japan can be observed for the pre-Bitcoin period, most eminently from the team of Tatsuaki Okamoto [487-489] at NTT [490] in the neighborhood of the 1990s.

### CRYPTO Conference Series

It has been said that the cohort of top cryptographers equipped with the competence to launch Bitcoin at the time was rather small, estimated to a few hundred people worldwide [425, 491]. They conceivably belong to the network gathering at the annual “CRYPTO” flagship meeting that takes place at Santa Barbara [492], California [493], and its continental and more specialized offspring [429, 456] organized by the International Association for Cryptologic Research (IACR) [494]. Note, however, that papers related to digital cash / cryptocurrencies were occasionally presented [463, 467, 469, 470, 478], and probably discussed informally, but not a major session title on the official agenda of such conventions, particularly not those predating Bitcoin.

Still, it is worth mentioning that IACR was founded, and CRYPTO conference series was inaugurated by electronic cash pioneer David Chaum [73, 486, 495] in 1981 [486, 496, 497]. The legendary cypherpunk Hal Finney [70] presented on Reusable Proof-of-Work (RPOW) [316] (his attempt towards a cryptocurrency [315]) at a 2004 “Rump” session” [498]. Finney mention in 2005 RPOW as “a sort of play-money form of digital cash, an implementation of Nick Szabo's [71] concept of bit gold [262]" [499].

Possibly relevant in the context of “Satoshi Nakamoto”, teams from academic and corporate Japan are have been presenting at the annual CRYPTO meetings [500-504], and occasionally collaborating with its Western stakeholders; as not uncommon in Japan [454], a good portion of their names contain fractions / syllables like “-(a)toshi” and “-(k)(a)moto”, and even the first name “Satoshi” appears. While, for several reasons, the individual behind Satoshi Nakamoto is unlikely to have Japanese nationality, his pseudonym might still have been inspired by such names to plausibly distract attention from his true place of residence.

### CRYPTO & Bitcoin

A vast majority of the (senior) authors cited in its whitepaper [124, 419, 425, 426], of those who contributed essential cryptographic methods [70, 73] for enabling Bitcoin, or who play a certain role in the unfolding story of this article [80, 442, 446, 450, 505, 506], were (pre-Bitcoin) presenters, program committee or IACR [494] fellows [507] / board members / directors [508] / presidents, or accomplished veterans of the CRYPTO conference spectrum [74, 80, 81, 417, 420, 436, 441, 442, 448, 450, 495, 509-519].

## Wider Context

Satoshi Nakamoto's personal state of mind during the inception, creation, release, and maturation of Bitcoin along the years 2007-2010 might have been affected by other preceding and concurrent happenings in the domains of finance, law, and the personal ambient, that were, occasionally, less prominent in the mainstream media. For drafting his overall psychogram, also select, post-2010 events ought to be considered for Satoshi Nakamoto's fairly abrupt, and still perplexing departure from the project. Note that the events mentioned in the following are primarily viewed from an angle coined by the Western / US-American / British / European culture.

### Financial Crisis of 2007-2008

The harbingers of the global financial crisis [520] are ascribed to events that took place in August 2007 in the world of banking [521], which may well have triggered Satoshi Nakamoto's work on Bitcoin. This severe crisis further escalated during the year 2008 [522], peaking, in the public perception, with the filing of Chapter 11 bankruptcy protection [523] by Lehmann Brothers [524], the fourth-largest investment bank in the United States [525], on 15 September 2008 [526].

### Legal Cases

#### *Bernstein v. United States*

The notorious Bernstein v. United States series of court cases (1995-2003) [527] revolved about publication, and resulting export of cryptography technology from the United States, and the right guaranteeing freedom of speech as protected by the First Amendment [528] of the United States Constitution [529]. The plaintiff, Daniel J. Bernstein [122], has already been a top-notch cryptographer in the 2000s, i.e. the years preceding Bitcoin.

The court initially ruled that software source code was speech and that the governmental regulations preventing its publication were unconstitutional; then the case was reopened, and, on 15 October 2003 [530], i.e., after almost nine years, the judge eventually dismissed the case; Bernstein, who then represented himself, was asked to return when the government posed a "concrete threat".

#### *Napster & Megaupload*

Satoshi Nakamoto also referred to Napster [531, 532] that was founded as an independent peer-to-peer file sharing service in 1999. The legal challenges spearheaded by the music industry led to its shutdown in 2001. Similar court cases were emerging at the time [533]. Legally related centralization issues of Napster were referred to by Satoshi Nakamoto [381] only a week after the release of his Bitcoin whitepaper [175].

#### *Currency & Securities*

Satoshi Nakamoto might also have been aware that the issuance of a cryptocurrency might bring about certain legal matters, for instance, with the (US) Securities and Exchange Commission (SEC) [534], and with Article I, Section 8, Clause 5 of the US Constitution on the "Coinage Power of Congress" [535], or their international equivalents.

#### *Wikileaks*

In 2006, the whistleblower website WikiLeaks [359] was started, presumedly by Julian Assange [360, 536]. In 2007, the platform published the standard US army protocol at Guantanamo Bay [537]. There were further releases of classified papers in the pre-Bitcoin period 2006-2008 [538] and beyond, such as the disclosure of the 2010 "Iraq War Logs" [539] and "Afghan War Diary" [540] leading to imprison-ment of the whistleblower Chelsea Manning [541], who was assigned as intelligence analyst to an Army unit in Iraq in 2009.

The financial blockade of WikiLeaks in 2010/2011 triggered a discussion, including whether the Bitcoin community should promote BTC as alternative for funding in early 2011 (which eventually did not materialize). The ensuing public exposure and scrutiny of the infant Bitcoin project seriously concerned by Satoshi Nakamoto [362], and might even have prompted his gradual withdrawal, culminating with his final emails in April 2011 [357].

#### *Silk Road*

In February 2011, and pertaining to the dark web [542], the online black market and the first modern darknet market website "Silk Road" [543] was started; operated as a Tor [408] hidden service, platform attracted approximately 100,000 users. The FBI [544] shuttered the website in October 2014, and arrested its alleged founder Ross Ulbricht [119], alias "Dread Pirate Roberts". Satoshi Nakamoto occasionally pointed out that he did not want himself and Bitcoin to be associated with moral or economic "shadiness". These happenings might well have deepened Satoshi Nakamoto's choice for silence and anonymity over the years to come until now. Notably, there were reports about an alleged ties between Satoshi Nakamoto and the founder of Silk Road [545].

#### *Keccak-SHA-3 Controversy*

In November 2007, the US National Institute of Standards and Technology (NIST) [422] launched a competition to develop a new hash function called SHA-3 to complement the older SHA-1 and SHA-2 [546]. On October 2, 2012, NIST announced KECCAK [547] as the winning algorithm to be standardized as the new SHA-3 [548]. In the follow-up, a "weakening controversy" [549] emerged in the cryptographer community, even including claims that a backdoor was introduced [550-552] in light of the documents leaked by the former NSA [460] employee and contractor Edward Snowden [121, 553] in 2013. As already revealed in the discussion regarding the invited presentation of Bitcoin to the CIA [554] by Gavin Andresen [75] in summer 2011, Satoshi Nakamoto tended to be very cagey about the exposure of Bitcoin to institutional investigation, which might have influenced his resolute exit from the Bitcoin project.

### Bobby Fischer

On 17 January 2008, the extravagant, world-champion (1972-1975) chess genius Bobby Fischer passed away [555] in Reykjavik (interestingly, but most likely coincidentally, bitcoin.org [257, 356] is currently registered to this capital of Iceland [356]). With a passion for games challenging the prodigy intellect in his childhood and teenage years, it may be speculated that Bobby Fischer [555] has been the much-revered idol of an underage superbrain "Satoshi Nakamoto", possibly spending his spare time on rehearsing Fischer's chess strategies from the 1950s to 1970s.

The young Satoshi would then have intensively studied "The Game of the Century" [556] of 13-year-old Bobby Fischer in 1956, and his first major titles in 1957 (which corresponds to the year of the last citation [557] in the Bitcoin whitepaper [175]), and admired his "21-move brilliancy" [558] from 1963.

Satoshi Nakamoto would certainly also have been aware of the abnormalities and tragedies within Bobby Fischer's active days and post career, including his refusals to defend his lead in tournaments, his semi-retirements, forfeiture of title, sudden obscurity, detention, life as an émigré, eventually in Iceland. So, if this story holds, how could Satoshi not have been moved by the demise of his childhood hero Bobby Fischer in early 2008?

# Bitcoin Whitepaper

## Main Text

In his paper "Bitcoin: A Peer-to-Peer Electronic Cash System" that was released [559] to the "Cryptography" mailing list [283] at 14:10:00 EDT [233, 280] on 31 October 2008 [279], i.e., the evening of Halloween [560] in GMT [239] (or 01/11/2008 [561] in Japanese time zone [562]), Satoshi Nakamoto outlines all key technical elements of his invention in a flawless style of writing, and a logically well-arranged train of thought regarding the content and order of his sectioning; this observation displays the distinct signature of a genius who can compellingly explain the essential software components, and their astute interplay leading to his masterpiece. Nevertheless, Satoshi Nakamoto wrote on 14/11/2008 [383] "I'm better with code than with words though" – the author would say his talented straddled both terrains.

### Structure

The document [175] is structured in a common way of scientific publishing, starting with a title, author name (Satoshi Nakamoto), email address (satoshin@gmx.com), and affiliation (www.bitcoin.org [257]), followed by an abstract, introduction, sections 2-11 on the technical ingredients, and winding up with conclusions and references; it appears Satoshi Nakamoto had some experience in, or exposure to common academic publishing practices, or he at least wanted to convey this impression.

### Writing Style

Throughout the about 80,000 words making up Satoshi Nakamoto's text releases in the whitepaper and online postings, there is only a handful of minor typos. He rather, inconsistently, intersperses British [563] / Commonwealth English [564-570] spelling and expressions, like the (mildly) expletive "bloody", into his mainstay American English [571-574]. In the ASCII [575] format postings, he inserts a double-space after periods at the end of sentences. Very few typos are found in all his writings, so his texts were probably channeled through a spell checker which might have had a default setting, to British English, amended private dictionary containing American English terms, or vice versa.

### Terminology

It also strikes that Satoshi Nakamoto did not use certain terms; he used the established terms "electronic cash" [576], or short "e-cash", which include traditional payments systems like credit cards with a magnetic strip for swipe-based readout, but not "cryptocurrency" [577]; Satoshi Nakamoto also frequently referred to "blocks" and "chains" in the Bitcoin whitepaper [175], but did not introduce the phrase "blockchain". It seems both expressions were coined and disseminated in the community post 2008.

### Typesetting

According to its PDF [286] version of the Bitcoin whitepaper [175, 257, 282] available to the author (dated 24 March 2009}, i.e., post its original release on 31 October 2008), the Advanced Metadata' ('Properties ↦ Additional metadata ↦ Advanced ↦ PDF Properties' shows 'pdf:Producer: OpenOffice.org 2.4'), the Bitcoin whitepaper was authored with the 'Writer' [578] of the open-source office suite OpenOffice.org [579] (version 2.4), its rather primitive graphics probably generated by an external programme, and issued as a PDF [286] file (version 1.4). Further information has been tried to extract from the metadata [580], without conclusive outcome, e.g., on the original file path.

In 2001, the OpenOffice suite open-sourced as a competitor to other "What you see is what you get" ("WYSIWYG" [581]) kits, like Microsoft Office [582] featuring Microsoft Word [583] clearly prevailing the market. Since 2006, OpenOffice chose PDF as the standard printing outlet.

The metadata of the PDF file further lists (Properties ↦ Fonts) the set of built-in TrueType [584] fonts (as 'Embedded Subset'): ArialMT, CenturySchoolbook-Bold, CourierNewPSMT, OpenSymbol,

TimesNewRomanPS-BoldMT, TimesNewRomanPS-ItalicMT, TimesNewRomanPS; Arial was used in some figures, showing that these figures were not drawn with an external program; the main text used the Times New Roman [585] (amusingly in the context of the headline in the genesis block, the font commissioned by the British newspaper The Times [304] in 1931), which is also the default font of TEX [586] / LATEX [587].

Hence, speculations circulated that the Bitcoin whitepaper was written in TEX / LATEX, the package for document preparation that is popular in academic publishing, particularly in engineering, math, natural sciences, computing, and thus also cryptography; however, they could not be substantiated [588]. It would have taken a cumbersome, laborious and time-intensive effort of an advanced expert for managing to mimic common OpenOffice outputs by TEX / LATEX.

Given that instructions for conference proceedings, e.g., from the CRYPTO series [589], warmly recommend document preparation by LATEX [590]; hence, a cryptographer using OpenOffice in this community would have been very rare. The most popular WYSIWYG alternative to LATEX would have been Microsoft Word [583], which had a dominant market share of about 95% in the later 2000s, especially on Microsoft Windows (XP) [591] operating system that Satoshi Nakamoto ran around 2008. Consequently, given that the usage of OpenOffice does not reflect a carefully arranged smokescreen, it represents a rather unique identifier for Satoshi Nakamoto.

In any case, it might be worthwhile screening documents of "Satoshi Nakamoto" candidates for similar, very unique setups of word processing software, fonts, and other metadata.

### Creation Date

The 'Advanced Metadata' (Properties ↦ Additional metadata ↦ Advanced ↦ XMP Core Properties [592]) also reveal a creation time of 2009-03-24T11:33:15-06:00 (UTC [238]) [593]. In the United States [594], after 2007, daylight saving time (DST) [233] starts on the second Sunday in March, which was March 14 in 2009. Consequently, the majority of the federal states (or regions) in North America implementing DST would have already had forwarded their clock by one hour on 24 March 2009, and UTC-06:00 then corresponds to Mountain Time zones [595] (named after the Rocky Mountains [596]).

In the European Union [597] (back then still including pre-Brexit [598] UK [599]), the switch to DST [233] takes place in unison on the last Sunday of the same month since 1981 [600], i.e., on 28 March in 2009, so this 'old-world' continent was still on standard time at that very date of 2009-03-24 when the PDF was produced.

Note that the main page of the PDF's 'Document properties' specifies 'Created: 24/03/2009 18:33:15'. On the one hand, it uses European date format DD/MM/YYYY; on the other hand, a 7-hour offset to UTC-06:00 is observed, corresponding to UTC+01:00 [601], which suggested Central European Time (CET) [602] when the PDF was generated. However, these metadata may have been dynamically adjusted according to the localization settings of the author's own computer.

Just to note that the author could neither verify the authenticity of the PDF version created on 24 March 2009 examined in this section, nor was he able to retrieve Satoshi Nakamoto's original PDF released on 31 October 2008. The identity of these two files, especially their metadata, thus remains unconfirmed.

### Mindset

Satoshi Nakamoto also distinctively abstained from interweaving technically brilliant parts of his Bitcoin whitepaper with (personal) ideology, such as the cypherpunk manifesto [165, 166], clear anti-banking, or anti-government attitudes. Strong statements, or hate speech, are strikingly absent, also in his later emails and postings, which complies with sound practice and etiquette of scientific publishing. Note that the moderator Perry E. Metzger [603] of the cryptography mailing list [283], where the Satoshi Nakamoto published and discussed the Bitcoin whitepaper, intervened as early as 07 November 2008 to abstain from pitching monetary politics on his forum [604]. Note also that the Bitcoin whitepaper, which never went through peer-review, represents the only publication-like document that can unambiguously be attributed to the person known as Satoshi Nakamoto.

### Linguistics

The writing style of the Bitcoin whitepaper has been analyzed from a linguistic perspective [605, 606]. Amongst the candidates proposed at the time, it fits well with Nick Szabo [71], the inventor of Bit Gold [71, 262, 607-609]. Yet, the candidate sharply refutes, and also arguments that he was not a proficient coder and never implemented a cryptocurrency have been raised against this otherwise compelling hypothesis. His denial is further supported by secondary sources and reasoning [272], e.g., his forum postings on Bit Gold [262] when Bitcoin was already out in in late December 2008 [610].

### Versions

Noticeably, Satoshi Nakamoto seems to have sent out an earlier version of the Bitcoin whitepaper to peers [272], which contained a few edits with respect to the later ("final") version published via the cryptography mailing list [175], such as another email address satoshi@vistomail.com, and, most conspicuously, a different title "Electronic Cash Without a Trusted Third Party" [267]. Hence, instead of "Bitcoin", Satoshi Nakamoto might have initially used the abbreviation "ecash" (as it appears in the filename [272]), which may indicate that he was unaware of David Chaum's [73, 485, 486, 611] synonymous micropayment system [271, 611] that existed in the second half of the 1990s.

## Bibliography

Somewhat surprisingly, the bibliography [269, 270, 557, 612-616] turned out to be rather brief, selective, incomplete and outdated from a traditional scientific publishing point-of-view. Its composition looks quite unusual, with only one reference to a publication in a peer-reviewed journal [613], four papers from proceedings of US and European conferences in 1980 [616] and the 1990s [612, 614, 615], two web links [269, 270], and an archaic first edition of a standard textbook [557].

Formally, the bibliography numbered (followed by a period, instead of the more common brackets "[…]") according to the order of occurrence in the main text; alternatively, Satoshi Nakamoto might have opted for common alphabetical sorting according to the first author's (last) name.

To avoid confusion with the numbering of the "regular" citations in this article, the eight citations listed in the Bitcoin whitepaper will be designated in the following by round parentheses as (1)…(8). The six references (1) [270], (3-5) [613-615], (7) [616], (8) [557] originate from authors affiliated with the USA [617], (2) [612] with Belgium [618], and (6) [269] with the UK [599].

### Electronic Cash Systems (1,6)

The state of-the art is represented by technological precursors [269, 270]; anecdotally, these weblinks have been added, *post scriptum*, when Satoshi Nakamoto had already completed the draft of the Bitcoin whitepaper and contacted Adam Back [69] on hashcash [269], and then Wei Dai [72, 267] on b-money [270] for proper referencing in August 2008.

It is observed that b-money (1) [72, 270] from 1998 just shows as a text-based webpage elaborating a concept without any references to prior work in the field; it has never been implemented. Citation (6) [269] on hashcash [69, 269, 619] dated 2002 features 19 references, amongst them two to his prior publications on hashcash, and another one to b-money (1) [72, 270].

Intriguingly, one citation in (6) [269] refers to a personal communication in March 2002 between two cypherpunks [69, 70] who have been frequently mentioned as potential creators of Bitcoin; another entry references SYN cookies [620] by Daniel Bernstein [122] (who will play a role later in this article). Other than in the Bitcoin whitepaper [175], research on digital timestamping (2-5), or other work from the founding fathers of blockchain [425, 426], is completely absent in (6).

Notably, the bibliography of the Bitcoin whitepaper misses out on other important pioneering projects leading to electronic cash [73, 271, 485, 621-623], and its presumably closest ancestor Bit Gold [262, 607, 608], which a member or avid follower of the cypherpunk [165, 166] arena would probably have had on his radar at the time of conceiving Bitcoin. It appears that Satoshi Nakamoto only learnt about Nick Szabo's [71] Bit Gold [262] by Hal Finney's [70] post on the cryptographer mailing list on 07 November 2008 [624], i.e., after the release of his Bitcoin whitepaper.

The citation pattern thus suggests that Satoshi Nakamoto may not have only become successively become aware of prior work on electronic cash systems om academia, and related activities the cypherpunk [165, 166] community by the time he immersed into developing Bitcoin, so he does not seem to have been an integral, seasoned member of this rather radical interest group.

### Benelux Symposium (2)

Especially the second citation [124, 125, 612] in Bitcoin whitepaper would have surely raised eyebrows along the review process submission of a classical journal. On the one hand, it presents comparatively preliminary work with respect to the other references on the 1997 project "TIMESEC" [625-627] sponsored by the federal government of Belgium [618]. Furthermore, according to its "Proceedings of the 20th symposium in information theory in the Benelux" [628]issued by the "Werkgemeenschap voor Informatie- en Communicatietheorie (WIC)" [628-630] (as part of the IEEE Benelux chapter on information theory [631]), this symposium series primarily attracts local researchers from the Benelux [401] region, i.e., the states of Belgium [618], The Netherlands [400], and Luxembourg [632]; such meetings tend to serve networking purposes, with most attendees overwhelmingly affiliated with local academia.

Submissions are normally not vigorously peer-reviewed, and important findings are expected to be bundled and elaborated to full-fledged journal publications in the follow-up. Unusually, the table of contents [628] lists 29 contributions, all attributed to the same senior author, Peter Vanroose [633], who is also one of the editors of the proceedings (while showing up differently in the individual papers). Note that, per its very name, the focus of the symposium was not specifically on cryptography, and even less on electronic cash or cryptocurrencies.

Remarkably, citation (2) [612] of the 1999 symposium lists references (3-5) of Bitcoin whitepaper [175] (with some miniscule edits). Owing to its very local nature of publication, reference (2) in the Bitcoin whitepaper was thus rather poorly noticed, and hardly traceable and retrievable by international researchers who were not involved in this comparatively tiny regional meeting. The editor [633] of the symposium proceedings stated [634] that the book of proceedings "has been registered at the Koninklijke Bibliotheek Den Haag [635, 636] (The Netherlands) as ISBN 90-71048-14-4.

As was usual at the time (1999), publication was issued as a printed book, only. The creation date in the metadata of the online PDF [286] version states 04 June 2020. All attendees [628] of the symposium received a copy of the proceedings, which were also sent to several libraries (e.g., University libraries, and the mentioned Koninklijke Bibliotheek [635, 636])."

There are 46 contacts in the list of participants, primarily from local academic groups at KU Leuven (15) [430], two at UC Louvain [637], and others [433, 638] in Belgium [618], eight at TU Eindhoven [431] and others [639-642] in the Netherlands [400]. Despite their contribution to several contributions, the senior authors of citation (2) [612], J.-J. Quisquater [124], and the head of COSIC [643], Bart Preneel [124], were not in attendance. However, Joos Vandewalle [511, 644, 645], a key figure at KU Leuven [430] / COSIC [643], and their other researcher like Bart van Rompay [646], Joris Claessens [647], and Jorge Nakahara [648-650] from this cryptography epicenter participated.

Individual, non-regional academics joined from Universities of Essen [651, 652] and Mannheim [653, 654] in Germany [655], and the IITP [656, 657] in Moscow [658], Russia [659]. There were also two researchers from TU Budapest [660] in Hungary [661]), who were visiting KU Leuven [430] at the time, but their research area was not cryptography. (Probably entirely coincidentally, a Satoshi Nakamoto statue has been unveiled in Budapest in 2021 [662, 663].) Remarkably, seven delegates were seconded from industry, exclusively various related departments and subsidiaries at Philips [664-667], six and one from their facilities in Eindhoven and Leuven, respectively.

Attendees most likely deposited hardcopies of the symposium proceedings [628] in the libraries of their research groups or universities. Thus, for instance, The British Library [668] lists the proceedings as available in their reading rooms, but electronic copies do not appear to have been available online for convenient worldwide trackability and download before 2020, or at least prior to the release of the Bitcoin whitepaper [175] in 2008.

Hence, it looks like Satoshi Nakamoto could not have just accidentally "stumbled across" citation (2) [612]; either himself, or at least someone close to him, must have personally attended this symposium that took place on 27-28 May 1999 in Haasrode [669], Belgium, a location within the strong research ecosystem of KU Leuven [430], UC Louvain [637], UCLL [670], COSIC [643] and IMEC [671]. Note that Leuven [672] is located in the Flemish Region of Belgium; its French name is Louvain [672].

The only alternative explanation of having attended the Benelux meeting is that the same authors from the Quisquater [124, 125] group presented a paper [673] with a similar content, and also referencing (2) [612], shortly after at a meeting [674] held in June 1999 in Stanford [413], i.e., the San Francisco Bay Area [675]. The main difference is that (5) [615] from the Bitcoin whitepaper, which was from an ACM conference in 1997, is "substituted" by a hard-to-retrieve Cryptobytes paper from a meeting that took place in autumn 1995 [676].

Also note that while the Bitcoin whitepaper correctly cites (3) [613] with the last page as 111, while both, (2) [612] and [673], have it as 112. In addition, (4) [614] which was published at a 1991 conference, is cited in the Bitcoin whitepaper is referenced as 1993 (the year of publishing), while (2) [612] (and [676]) cite it with the year 1992, as quoted in the printed article [614]. These subtle deviations suggest that Satoshi Nakamoto must have taken a closer look at the Haber and Stornetta papers (3) [613] and (4) [614], and not just copy & pasted them to beef up his stack of references.

Besides citation (2) [612], the table of contents of the Benelux symposium [628] exhibits other intriguing papers from the same group led the highly accomplished researchers Joos Vandewalle [511, 644, 645] and Bart Preneel [365, 494, 505, 508, 677-679] at COSIC [643] / KU Leuven [430]. One is entitled "Anonymity controlled electronic payment systems" [680], which would have an obvious connection to Bitcoin as an electronic payment network.

Another contribution from the same senior researchers [681] was first authored by "Jorge Nakahara" [648-650], which would be a link to the Japanese last name Nakamoto. Note that his given name might indicate a relation to the Japanese minority across Latin America [682], e.g., Brazil [683, 684]. Still, the list of participants of the 1999 symposium neither recorded Bart Preneel, nor the senior author of (2) [612], Jean-Jacques Quisquater [124, 125], in personal attendance [685]. Overall, none of the commonly quoted big shots associated with Bitcoin / blockchain / electronic cash in academia or the cypherpunk [165] scene were in attendance of the 1999 symposium, further emphasizing that its global reach was very small.

In a later section, we investigate whether the exceptionally gifted American cryptographer Leonard Harris ("Len") Sassaman [46, 47, 77, 686], who was a researcher at KU Leuven's [430] Computer Security and Industrial Cryptography (COSIC) [643] group headed by Bart Preneel [505], could have gotten interested in cryptocurrencies, and been motivated to create Bitcoin after having read the paper (2) [612] through the hardcopy of the symposium proceedings [628] that was probably available in the local library. Suspiciously, the co-supervisors of Sassaman's PhD project were Bart Preneel [505] and David Chaum [73, 486, 687, 688], who pioneered blind signatures [485], and conceived the anonymous cryptographic electronic money and electronic cash system "ecash" [271] back in 1983.

Also the renowned cryptographers Hendrik Lenstra [113, 434], his former PhD student, Daniel Bernstein [122, 431], and the PGP [689] inventor Phil Zimmermann [74] held positions in, or entertained strong links to the Netherlands [400] or Belgium [401]. However, a direct link to the 1999 Benelux symposium [628] could not be established. However, even though neither attendant nor author in 1999, Henk van Tilborg [402, 690], the founder of Ei/Psi [432] who retired from TU Eindhoven [431] in 2011, and is a board member of the organizing WIC [630], and might thus well had access to the proceedings before 2008.

### Benelux Epicenter

Due to the eminence of the citation (2) [612, 628] in the evolution of the Bitcoin whitepaper [175] (Figure 1), we first investigate the Benelux [400, 401, 618, 632] environment. In addition to the UC Louvain [637] group of senior author Jean-Jacques Quisquater [124, 125], who was a major player in the scene at the time, Bart Preneel [505, 677-679] from the neighboring KU Leuven [430] as the head or COSIC [643], (still) one of the most prestigious research centers in cryptography on the planet. Both groups, as well as their partners and academic descendants, are key stakeholders in IACR [494], e.g., serving on their board of directors [508], and as session chairs and speakers.

Well-renowned cryptography researchers like Joos Vandewalle [511, 644, 645], Frederik Vercauteren [691, 692], Yvo G. Desmedt [506, 693], and Joan Daemen [694] were / are affiliated with this Belgian cryptography beacon. COSIC members at the time (now mostly alumni) [688] also entertained vivid relations with industry, and collaborated with several academic and corporate groups internationally, including Japan [490, 650, 688, 695-700].

High-caliber cryptography activities are / were also conducted with other top Benelux-based [401] institutions, for instance, at TU Eindhoven [122, 402, 431, 432, 690, 701, 702], Leiden University [113, 434, 703], TU Delft [74, 639], and the University of Amsterdam [704, 705] in the Netherlands [400], and the Université Libre de Bruxelles [433, 706] in Belgium [618]. It is hard to envision a Satoshi Nakamoto would not even have had, at minimum, secondary links to this world-class cryptography landscape, and its academic members, staff, students, partners, or affiliates in the Benelux [401].

### Blockchain Pioneers (3-5)

Out of the (only) eight citations in the Bitcoin whitepaper [175], three (3-5) are directly from the team around the “fathers of blockchain” [425, 426, 613-615], and one (2) [612] is largely based on their work on timestamping in the early 1990s, which is foundational for Bitcoin’s distributed ledger technology (DLT) [707]. Astonishingly, none of these four references (2-5) appears in (6) on hashcash [269, 619]. Note also that Satoshi Nakamoto updated the number of the last page in (3) with respect to its (false) representation in (2), revealing again his diligent research approach.

Remarkably, the citations (3-5) in the Bitcoin whitepaper [175] are listed in their entirety, with (4) slightly deviating, but in a different order, than in (2), which would support a “(2)&(6)-first” hypothesis (Figure 1): while certainly a gifted cryptographer and programmer, Satoshi Nakamoto might have only become aware of digital timestamping (3-5), as an essential ingredient of blockchain technology, by somehow getting his fingers on the Benelux conference paper (2), and prior art on electronic cash systems, first through (6) and then (1).

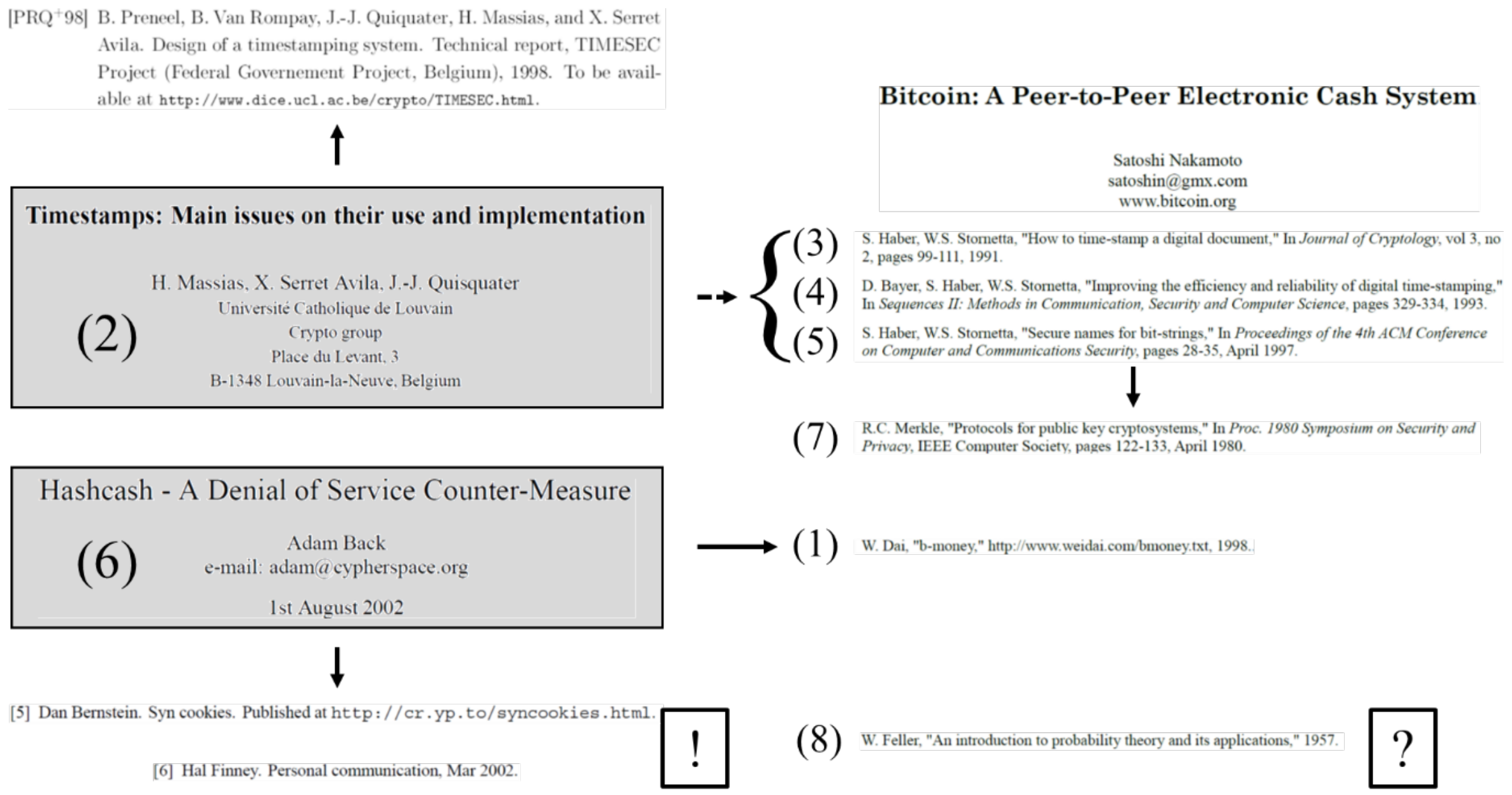

Figure 1 Evolution of references following the so-called (2)&(6)-first hypothesis, i.e., that Satoshi Nakamoto's Bitcoin initiative was principally inspired by two sources: (2) - a paper from a Benelux symposium [673] which, in essence, provided (3-5), and, in turn, referenced (7), a foundational 1980-conference paper [616] by Ralph Merkle [419]; and, secondly, (6), a paper on "hashcash" [269] by cypherpunk Adam Back [69], which cited (1), an 1998-online article on b-money [270] by Wei Dai [72]. Anecdotally, Satoshi Nakamoto approached Adam Back by personal email in August 2008 [266, 708], who guided him to Wei Dai's [267] work. Notably, in (6), Adam Back also listed a personal communication with another cypherpunk Hal Finney [70], who became an early adopter of Bitcoin, and Daniel Bernstein [122, 123, 527, 709, 710], a world-class cryptographer and freedom-of-speech advocate. The odd one out among the references in the Bitcoin whitepaper [175] is (8), the first, 1957-edition of a standard textbook on probability theory [557, 711] by William "Vilim" Feller [712] who passed away in 1970. Its age, i.e., more than half a century in 2008, indicates that the book was purchased by a person who might have been at least a college student in his twenties in 1957, i.e., likely pertaining from the parent / uncle or grandparent cohort of relatives of a generation-X [713] Satoshi Nakamoto. The vintage book or his author probably had a special meaning in Satoshi Nakamoto's family to be bequeathed and used through the family. Remarkably, other relevant precursors, such as Nick Szabo's [71, 262] Bit Gold from 2005, and David Chaum's [73] eCash [271] from the later 1990s, are missing. Interestingly, as referenced in [612], citation (2) seems to have arisen from a collaborative project TIMESEC [625, 626] between the Quisquater [124, 125] at UC Louvain [637] and the (nearby) Preneel [505, 677, 678] group at KU Leuven [430] in Belgium [618, 672].

Only then he identified their key shortcomings, such as exhibiting a single point of failure, and conceived the missing element for their solution through decentralization by a distributed public ledger [707] file. This rather plausible "(2)&(6)-first" assumption would thus underpin Satoshi Nakamoto's ingenuity, and that his Bitcoin project did not directly emerge from the highly intertwined cypherpunk movement [165, 166], possibly even nearly evolved "from scratch" in 2007 (Figure 1).

## Merkle Trees (7)

Also the seventh citation [616], a contribution to a 1980 symposium held in Sunnyvale (San Francisco Bay Area [675], California, USA), would have been rarely included 28 years later by an author in the year 2008. This work on so-called "Merkle trees" [714] roots in a patent [715] that was filed in 1979 and granted in 1982. The technology has presumably been outlined best in a 1987 book chapter [716], which may have been the most appropriate reference for the readership of the Bitcoin whitepaper [175].

The method is named after Ralph Merkle [419], who is one of the inventors of public-key cryptography [717], and the architect of cryptographic hashing [718]. Note that the very 1980-conference symposium citation (7) [616], which is rather hard to retrieve, was also referenced in (4) [614] and (5) [615], where it might have been spotted by Satoshi Nakamoto; this scenario would further support his discovery sequence (2)()↦(3,4,5)()↦(7) according to the "(2)&(6)-first theory (Figure 1).

## 1957 Edition of Textbook (8)

The last citation [557] in (8) pops into the eye as it refers to a standard (2-volume) textbook "An Introduction to Probability Theory and Its Applications"; the Bitcoin whitepaper remarkably cites its 1st edition from 1957. In Satoshi Nakamoto's (proposed) lifetime, the mid-20th century hardcopy must have been mostly available for purchase in antiquarian bookstores. If the content of this book was essential in assembling Bitcoin for Satoshi Nakamoto, who supposedly read this source not earlier than in the 1990s, he would have normally referenced its way more recent editions from around 1970 [719], or even a newer, alternative source. In contrast to (3-5) and (7), citation (8) does not appear to be retrieved from another reference in the Bitcoin whitepaper. There are a few possible explanations for the peculiar choice of a 1957 edition.

The simplest, and most plausible one points to the libraries of universities that were already founded before the second edition was issued in the mid-1960s. The author conducted exemplary searches through the catalogues of academic institutions that play a role in the later parts on this article. While the title is still ubiquitously available (as print-only version), the 1957 edition is (nowadays) only stocked at KU Leuven [77, 430, 436, 505, 645], UC Louvain [124, 637] in Belgium, and UC Berkeley [81, 122, 412, 720] and Stanford [74, 413, 509, 721] in the Bay Area, as well as MIT [414] and Harvard [421] on the East Coast [722], but not at TU Eindhoven [122, 402, 431, 702], UI Chicago [122, 720], UC Santa Barbara [73]. However, the libraires that carried to 1957 book also had the newer editions in their portfolio, so it would be unusual that someone select the most ancient edition.

Of course, there are usually photocopies ↦ post, digital scans ↦ email, or even remote hardcopy loans offered by such libraries; but it would be extremely unusual that a client would remotely request an outdated copy of a book if a newer edition that was locally available. Also, these services would normally only (manually) copy a few pages, but not an entire book. Overall, this likely scenario would mean Satoshi Nakamoto was, as a (postgraduate) student or academic (around 2007/2008), immersed in a university setting where he just went into the nearby library to collect (8) [557].

Less likely, the historic, 1957 standard textbook (8) [557] on probability theory might have been part of a personal collection, e.g., handed as an heirloom through generations; so close, elder family members, like his parents and grand- or godparents, uncles or aunts, might have had studied, or been professionally involved in mathematics, to make (8) precious enough not to be discarded over several decades. Reference (8) might hence be attributed to a special relation, or gesture of respect of Satoshi Nakamoto himself (or his mentors) to the book author; William "Vilim" Feller [712] (1906-1970), a Croatian-born professor who eventually retired from Princeton [723], is widely remembered for pioneering probability theory [711] as a branch of mathematical analysis.

This peculiar age of citation (8), more than half a century in 2008, might also hint to the significance of the year 1957 [724], maybe referring to the beginning of the space age with the Soviet Sputnik [725] satellite, the start of the S&P 500 stock market index [726], or IBM releasing the first Fortran [727] compiler; or was the choice of the 1957 edition of the book (8) just a (hard to spot) hint to Satoshi Nakamoto's year of "birth" 1975 by swapping of the last two digits to 195 ↔ 7?

## Dated References

Overall, except for two citations (1) and (6) on digital currency projects [269, 270, 619] that were, reportedly, added, *post scriptum*, to the draft of the whitepaper in late summer 2008, most references precede the Bitcoin whitepaper by more than 10 years, and literally none of them is younger than 6 years; their average year of publication is situated before 1990, i.e., about 18 years prior to the release of the Bitcoin whitepaper [175].

Such a short and antique mix of citations, and its diverse caliber, are unquestionably conspicuous, thus indicating that Satoshi Nakamoto tried to mimic scientific publishing, but did not possess an extended track record of active authoring in this arena; alternatively, he was simply not paying much attention to academic etiquette, and just dropped a brief, "quick-and-dirty" list; this impression resembles the footprint of a junior-level stint in academia, for instance, as a gifted postgraduate [728] or postdoctoral researcher [729].

Alternatively, Satoshi Nakamoto could have also opted to omit references. It is therefore more likely that the mastermind of Bitcoin came out of a relevant public agency or corporate background, where he routinely studied techno-scientific papers from the academic space, but had seldomly spearheaded authorship of publications in top-quartile journals. In addition, it can also be firmly assumed that Satoshi Nakamoto avoided self-citations (under his real name), as this would have quickly jolted him on the list of suspects for having created Bitcoin.

# Bitcoin Dates

Satoshi Nakamoto could evidently not have rewritten factual history, so we should also consider the wider context of the constituents of his self-fabricated date of birth.

## Bitcoin Whitepaper on 31 October 2008

### Day

The day Satoshi Nakamoto chose for the release the Bitcoin whitepaper [175] in 2008 was 31 October [279], i.e., Halloween [560] in many Western countries. This may indicate awareness of this ancient tradition, or superstition. Yet, depending on the time zone, the publication date is sometimes falls on, 01/11/2008 [561] (a Christian holiday [730]), which is only composed of the binary numbers 0 and 1 – the number set of digital math. While this might be coincidental, the 1st of November was a Saturday, which resonated with "Satoshi"; both theories hint towards a work schedule peaking towards the weekend.

### Course of Year 2008

In January 2008, Satoshi Nakamoto's potential chess hero Bobby Fischer [555], who was famous for his "21-move brilliancy" [558], passed away, the movie "21" [731] and the Mystery-Jets [732] album "Twenty-One" [733, 734] were released in March, and the song "42" [735, 736] by Coldplay [737] in June. The global financial crisis [520, 738] had its first doomsday with the collapse of Lehman Brothers [524], one of the largest US investment banks [525], on the 15th of September [526] that same year, i.e., just a few weeks prior to the release of the Bitcoin whitepaper [175].

## Genesis Block 0 on 03 January 2009

### The Times

#### *Headline*

These dramatic happenings of the along the unfolding global financial crisis [520, 738] must have certainly been on the radar of Satoshi Nakamoto, as impressively recorded by the message he engraved into the genesis block 0 [177, 302] of the Bitcoin blockchain [177]; issued on Saturday, 03 January 2009 [739], he included the (exact) headline of the London-based British daily newspaper "The Times" (of London) [303, 304]: "Chancellor on the brink of second bailout for banks" [176] of that very day. The inclusion of newspaper content remarkably resembles the timestamping methods [425, 426, 612-615] developed in the 1990s, which are extensively referenced (2-5) in the Bitcoin whitepaper [175].

#### *Editions*

Note that the exact wording of the headline referenced in the genesis block 0 [176, 177, 302] exclusively showed on the front page of the UK print version of "The Times" [303, 304], of which nearly half of its readers are based in London; crucially for this investigation, the corresponding article in globally accessible online edition [740] was titled “Chancellor Alistair Darling on brink of second bail-out for banks” [741-746]. Since its start in 1996 up to July 2010, i.e., before the minding of the genesis block, the website operated without a paywall [740].

In the international (print-only at the time) edition of “The Times”, in the (inner) section (on page 21) [747] titled “Britain”, the corresponding punchline also, even only subtly, but decisively for the purpose of this study, deviated (in its ending) from the quote in Bitcoin’s genesis block: “Chancellor on brink of second bank bailout to boost lending” [747].

Since 2006, “The Times” [303, 304] has been issuing a designated US (print-only version) [748], which is primarily disseminated in the capitol area of Washington D.C. [749], and New York City [750]. This US edition is / was based on the international edition; it also did not display the headline cited in the Bitcoin genesis block [751] on its front page at all; the “bailout” story was identical to the international version [752], and buried on an inner page, i.e., invisible when briefly glancing the papers, while passing a newsstand.

“The Times” is ranking high among the English-language newspapers preferred by rather well-educated and internationally minded readers, chiefly in the UK [599] and the Commonwealth [565], and outside North America. This correlation obviously indicates Satoshi Nakamoto’s keen interest in the current affairs in Britain, which would be very uncommon for a non-British “techie” residing in countries outside the UK and Commonwealth [565].

As, most likely, a digital native, it is surprising that Satoshi Nakamoto resorted to the UK print version. Since the headline was included in the genesis block [176, 177, 753], the print version must have been available to Satoshi Nakamoto on the same Saturday, 03 January 2009 [739], meaning that he must have been located in the UK, or have had purchased, or walked past one of the few newsagents abroad, e.g., at international traffic hubs like airports, railway or bus stations, which would stock the latest UK issue of “The Times”.

The latest print edition of this prestigious newspaper was most probably also well available in and around Brussels [706], the headquarter of the European Union [597], especially in pre-Brexit [598] 2009 when British affairs where highly relevant for both, UK and EU politicians, commissioners, offers, and bureaucrats were in the capital of Belgium [643]. Note that the cryptographer epicenter Leuven / Louvain [430, 637, 643, 670, 672], e.g., the team around the second citation (2), is only 30 km away from Brussels [706], so some of their researchers might even live there. For timely quotation in the genesis block, Satoshi Nakamoto was unlikely to be travelling, but rather keeping his computer stably connected to the internet in a fixed place for assuring continuous running and monitoring the embryonic Bitcoin blockchain.

Of course, also a colleague or friend based in the UK might have taken a screenshot and sent it to Satoshi Nakamoto, for instance, as he would have known about his interest in the bailout; yet, this came at the risk of another person being able to unmask his identity.

Therefore, in addition to a timestamp, in a sense that the genesis block cannot have been mined before 03 January 2009, the citation of the (literal) headline also provides a “geostamp” to places where the UK print version was available sufficiently early to be embossed in the genesis block. Not, however, that since Satoshi Nakamoto was the only person running a node / the nodes, he could of course have forged the timestamp, e.g., to just before block 1.

### *Time of the Day*

Block 0 [302] was mined at 6:15 pm GMT [239], probably derived the time of the date, time of the day, and time zone [231] setting on his own computer (which he could have manipulated). This timing fits to Satoshi Nakamoto’s main posting hours [754] starting around 3 pm GMT, as further discussed later in this article. Such schedule may have allowed that Satoshi Nakamoto had gotten up on that very Saturday at some hour in the morning, saw or picked up a copy of “The Times” displayed at a local store while passing by on a walk, noticed the daunting headline, returned home, rushed to give the code some final touches, and then, in hindsight, possibly prematurely kicked off the Bitcoin blockchain in the late afternoon of the same day.

The launch of the blockchain occurred around midday in continental American time zones [280, 595, 755, 756]. However, newsagents only carried the printed US issue of “The Times” featuring the proper headline in on the US-East Coast [722]. So the genesis block would have been mined at 1:15 pm ET [280], thus leaving a rather short interval for Satoshi Nakamoto between picking up a printed “The Times” from an East-Coast international newsagent and initiating the Bitcoin blockchain. However, I this putative scenario, Satoshi Nakamoto would have not known, and thus been able to quote, verbatim, the exact headline of the UK print issue.

This genesis-block timestamp corresponds to very early Sunday morning [757] in the Far East [758], e.g., about 3 am in Japan [454, 757]. Factoring in non-uniform DST [233] during summer on the Southern Hemisphere [235] in Australia, while Britain is on GMT [239] in January, 6:15 pm GMT spreads into the period of 2 am to 5 am in Australia [457, 759]. However, it would be unlikely that Satoshi Nakamoto got hold of a hardcopy of “The Times” during the late Saturday in these countries that are far ahead of GMT [239], and then started Bitcoin shortly after 3 am JST [562] in, for instance, Tokyo [562, 760], or 4 am in Brisbane [761] located in AEST [759].

This hybrid, analog-digital timestamp analysis makes it very likely that Satoshi Nakamoto resided in the UK, or a metropolitan area in Europe, on Saturday, 03 January 2009 [739]. As the, at the time, sole lifeline of Bitcoin, he would probably not have operated his freshly initiated blockchain from an intermittently connected computer, so he may have settled at the same place for a longer period of time.

### Historical Context

On this very day in January 2009 [739], which was a Saturday, also Sir Alan Arthur Walters [762], a prominent British economist, passed away. He was best known as the Chief Economic Adviser to the Prime Minister Margaret Thatcher [763], and fought for Britain to maintain strict monetary targets, and that the money supply should not be manipulated for political reasons. These attitudes of rigorous fiscal policies might have appealed to Satoshi Nakamoto. However, given that his Alan Walter’s activities in his political role dated as far back as the 1980s, it is less likely that this gentleman was on Satoshi Nakamoto’s agenda.

## Block 1 on 09 January 2009

### Technical Reason

Upon seeing these events, Satoshi Nakamoto might have expedited issuing the genesis block [177, 302], but it then took him more time to mine block 1 [305] after a six-day gap on Friday, 09/01/2009 [764] at 2:54 am GMT. Experts believe that the remarkable mining hiatus may have simply been caused by the instability of the frequently crashing Bitcoin v0.1 code. Again, the lack of independent nodes on the P2P network of neonate Bitcoin made timestamping prone to manipulation, e.g., to retrofit the headline in “The Times” on 03 January 2009.

### Symbolism

The historical events on this day do not seem to have particular significance to Bitcoin, except that its cross sum $0 + 9 + 0 + 1 + 2 + 0 + 0 + 9$ yields 21, a figure which assumes a central role in Bitcoin, most strikingly in the strictly limited total supply of BTC.

### Origin Myth

This way, in addition to obfuscating his own identity, Satoshi Nakamoto might have wanted to enhance Bitcoin’s origin myth [765]. As for many world religions, their founders [766-769] were typically not particularly keen on depositing formal data of their lives; in absence, or extreme paucity of primary sources, historians can often not even agree on much more than that they lived within a loosely defined time span in a certain region. While it is highly doubtful whether Satoshi Nakamoto wanted to plant a reference to such religious context, he, at least intuitively, knew well about the power of an origin myth preserved by his pseudonymity (in addition to protecting his privacy, and to deflecting potential legal issues [496]).

In a rather fancy, entirely speculative, but hopefully enjoyable theory (which is not at all meant to hurt religious feelings of any kind), Satoshi Nakamoto wanted to fog Bitcoin with an *ex nihilo* [770] ("out of nothing") type of founding narrative; for this, in the context of major monotheistic [142], Abrahamic religions [771] believing in a single deity [772] ("Satoshi Nakamoto"), the mining gap ought to evoke the six-days of creation of the world [773] in the Book of Genesis (1:3–2:3) [774], which constitutes the first book of the Hebrew Bible [775] and the Christian Old Testament [776]. Continuing the alignment with biblical happenings, his later departure might refer to The Exodus [777] or Ascension [347], and prophets [778] will spread divine messages.

In Christianity [779], Jesus sacrificed himself ("burning Satoshi's BTC") [780], so humanity was no longer bound in sin ("dysfunctional banking system"). There are equivalents to the disciples [781] ("early Bitcoin contributors" [70, 296, 353]), the (12) apostles / ("early messengers of Bitcoin story") [29, 69, 782], Saint Peter ("Petrus") [783] (aka "Gavin Andresen" [75]), the first pope [784] appointed by Jesus ("Satoshi Nakamoto"), the (4) evangelists [785] ("promotional story writers"), and present-day missionaries [387, 786-793].

On the other hand, and the exegetes [794] critically analyzing story of Bitcoin, and impactful heretics [795, 796] [797, 798], derogatively termed "gerontrocacy" [799] by the believers, who emphatically deny the economic value of Bitcoin. First-century Christians ("present-day Bitcoiners") believed Jesus would have a Second Coming [350] during their lifetime [800]. On doomsday [801, 802], only faithful Bitcoin "hodlers" [803] will pass the final judgement and ascend to heaven ("wealth").

Of course, the concrete biblical dimension in the story above may, most likely, be a fictional product exclusive to the author's overactive imagination; however, from the striking parallels, it may still be inferred that Satoshi Nakamoto wanted to plant a founding myth [348] into Bitcoin, resorting to a similar plot, and ingredients that critically contributed to the success of various religions and dynasties.

In the aftermath of the origin, a religion tends to split [804-806] into different churches, schools, branches, denominations, and faiths [807]. This pattern resembles the path of Bitcoin's hardforks [808], and the emergence of new religions ("next-generation cryptocurrencies"), which still benefit, to some extent, from the origin myth and technological backbone ("blockchain") of their common ancestor.

There is a range of evidence that Satoshi Nakamoto was aware of the supreme authority endowed by a foundational legend, and deliberately deployed it for bolstering Bitcoin, and assure its sustainability. So the allegoric [208] saga presented here, even if not scoped in this (specific) way, hopefully illustrates the abstract design, purpose, and power of a founding myth; Bitcoiners and neutral analysis as well as crypto-sceptics ("nocoiners") can thus understand the major driver for Satoshi Nakamoto's [214] pseudonymity, his inconspicuous departure, his ongoing silence, his reasoning / self-commitment to abstain from cashing out, and then draw their qualified conclusions for the future.

## Personality & Bio

### Pseudonym

The inventor of Bitcoin was evidently free to mint his online name, which is broadly acknowledged to represent an alias; otherwise, the tremendous accompanying measures to mask his identity would not have made any sense at all.

## Name

His given (male) name "Satoshi" [809], as well as his last name "Nakamoto" [810], are quite common, predominantly in Japan, and its emigrant subcultures abroad; (rather debatable) connection of this pseudonym have been fielded by Craig Wright [96, 811] to underpin his claimed inventorship of Bitcoin: Wright claimed [812, 813] that the first name is connected to in a semi-fictional book "The House of Morgan: An American Banking Dynasty and the Rise of Modern Finance" [814] published in 1990. In the story "Satoshi Sugiyama", was adopted by an American, and given the name David Phillips, possibly alluding to his junior partner Dave Kleiman [117]; the last name is declared to refer to a historical figure, "Tominaga Nakamoto". However, these connections seem to be too far-fetched to be noticeable for any reader, even with some effort. To the author's best knowledge, the pseudonym is thus not inspired by a renowned historic figure or fictional character in the Japanese or global culture.

Satoshi Nakamoto selected a rather generic name and nationality having a vivid cryptographer scene, rather than just a fantasy, gender- and stateless internet handle; this specific choice might indicate his intent to tailor a message through this artificial character. Note that in the Japanese and other Asian cultures, the order of the two names is reversed, i.e., he would be called Nakamoto Satoshi, and his initials would read as "NS". These first letters were part of a Bitcoin address [815] generated by Satoshi Nakamoto, which he highlighted in an email to Hal Finney [70] on 12/01/2009 [299] for its potential name matching. "NS" would evidently also go with Nick Szabo [71], the author of Bit Gold [608], as discussed later in this article.

## Group?

The singular of the pseudonym suggests an individual, not a team; the immense effort to coordinate and eternally protect a cover among multiple conspirators also suggests the hypothesis of a single actor. Yet, it has been purported [816] that an agent from the Department of Homeland Security (DHS) [817, 818] interviewed Satoshi Nakamoto in the context of Bitcoin and illicit online markets, and reportedly met a group of four men in California [493].

The credibility of this second source cannot be verified, and also the names of the group members have not been revealed in this feed. Some speculation spawned around the Bay-Area [675] based team around Jed McCaleb [130, 819, 820], David Schwartz [132], Arthur Britto [821], Ryan Fugger [134], Brad Garlinghouse [822], and Chris Larsen [131] involved in setting up what later became as the payment protocol / cryptocurrency Ripple (XRP) [133, 138, 823-828].

If Satoshi Nakamoto was indeed a moniker for a group, the durability of their bullet-proof collusion pact must have been fortified by a vigorously sanctioned agreement or self-commitment, e.g., to protect against the threat of prosecution by a federal agency, or to shield significant financial interests. Thus, for the sake of this paper, we continue considering Satoshi Nakamoto a single male (he/his), even though it could also be part of a "red herring" [829] strategy to mislead about gender, or to conceal a party of contributors. Notably, to the author's best knowledge, all candidates pitched as Satoshi Nakamoto (so far) are men.

## Translation & Interpretation

The pseudonym also leaves room an interesting aspect. There is some significant scope for its translation from Japanese into English; "Satoshi" may be translated as intelligent history [830], knowledge, or wisdom [831, 832]. The initial syllable "Naka–" of his last name can be rendered as "being within", "between or in the middle of something" [833]. The final part "-moto" constitutes a common terminus of Japanese family names; "Nakamoto" may stand for "the origin, the cause, the foundation, the basis" [834]. There is some, probably random, similarity to the Japanese computer scientists Satoshi Obana [835], who published at CRYPTO 2001 [836], and co-authored with Shizuo

Sakamoto [837]. As a faint hint, reversal (and minor deletion / insertion) yields “Ottom~~ak~~an”, possible hinting to the historic Ottoman Empire [838] (1299-1922), and thereby a connection to modern Turkey (Republic of Türkiye) [82, 839].

## Intelligence

In absence of reference to a known figure, and in the context of Bitcoin and ideological spirit, the chosen pseudonym Satoshi Nakamoto might thus reflect a self-description, which may be construed as “a genius who is highly focused on work to disrupt the foundations (of the established banking system)”. The entire stage name may also be creatively understood as “central intelligence”.

Moreover, it is worth pointing out that in Chinese, Satoshi Nakamoto may be read as “the wise one of Chinese currency” [840, 841]. It is indeed reported that three-letter agencies like the CIA [554], DIA [461], NSA [460], or other entities pertaining to the United States Intelligence Community [842], the FBI [544], or their international equivalents [438, 843-850] and hybrids [851, 852], have, by the secretive nature of their own service, the vital importance to military communication [853], or by eavesdropping on their targets, a keen interest in cryptography [250], and are (nowadays) likely to be specifically involved, via various modes, in cryptocurrencies [854].

However, it is widely believed to be unlikely that such governmental institutions have directly instigated Bitcoin [855], at least not as part of their remit as a public agency. Nevertheless, their former or present (renegade) staff, or their external service providers, might have gained sufficient knowledge of technologies and methods to pull off Bitcoin in a self-driven after-work or post-contract initiative. This hypothesis was also advocated by insiders the author communicated with under the condition of their anonymity.

Overall, given these semantics and their fit to a personality required to realize Bitcoin, the choice of the stage name Satoshi Nakamoto does not seem to be pure coincidence; in fact, it made the author curious about further revelations that he might have encoded. Just to mention that the Bitcoin and cryptocurrency advocate Elon Musk [15], presumably as a good joke he adopted from a 2013 forum post [856], considered Satoshi Nakamoto as an acronym for large Korean, Japanese, Chinese, and US-owned companies / conglomerates SAmsung [857], TOSHIba [858], NAKAm[~~in~~]ichi [859], and MOTOrola [860] – as a highly gifted individual, Musk certainly enjoys spotting contextual patterns.

## Date of Birth

Satoshi Nakamoto populated a profile on his personal page at the P2P Foundation [861, 862], intriguingly in 2012, i.e., after he disappeared under his online alias. Apart from claiming his Japanese nationality or residency, he provided 05 April 1975 [863] as his date of birth. The real Satoshi Nakamoto is unlikely to be born on this very day, as this would resemble publishing most digits of his passport number. The reason is that there are 365 options for the day of the year, and – say – 100 cryptographers globally who could possibly have developed Bitcoin. So given that dates of birth are rather overt biodata which commonly reside in several unsafe online databases, especially considering that federal agencies possess a backdoor to public administration, it would most likely lead to only one match per year of birth.

Furthermore, as Satoshi Nakamoto most likely wanted to convey a message through his alias, he probably also opted to put a meaning into his day of birth. It is observed that the cross sum $0 + 5 + +0 + 4 + 7 + 5 = 21$ of the birth date in its short format DD/MM/YY [230] result in 21, the most ubiquitous number in Bitcoin (see date of birth above, and in the following). Other than the arrival of the creator of Bitcoin on the planet, there are not many associated happenings that caught the attention of the global news on that very day [863], except that it was a Saturday in 1975, possible referring to his chosen given name.

Still, there was a sequence of potentially relevant events in the two days prior to Satoshi Nakamoto's birthday [864] (see also Appendix).

## Preceding Days

On 03 April 1975, the chess genius Bobby Fischer [555] very unexpectedly declined to resume competing for the World-Chess title [865] against Anatoly Karpov [866] in Manila [867] (Philippines [868]), thereby turning down a chance to clench a premium of more than USD 1.5 million. He thus became the first world chess champion [869] to willingly forfeit this most prestigious trophy. Bobby Fischer's rather odd behavior shows amazing parallels to Satoshi Nakamoto's sudden withdrawal in 2010/2011, and his ongoing abstinence from reaping the future of his 1-million BTC fortune.

On 04 April 1975, Bill Gates [13] and Paul Allen [870] founded "Micro-Soft" [871], a sea-change moment for the modern age of information. Notably, Satoshi Nakamoto used Microsoft's Windows (XP) [591] operating system (but OpenOffice.org [579], instead of Microsoft Word [583], as text editor for typesetting the Bitcoin whitepaper [175], see above).

## Age

This self-claimed birthday features two constituents. The year 1975 tethers him to the so-called Generation X [713], and infers a "credible", not too young or old age of 33 years around the time he was issuing the Bitcoin whitepaper [175] and the genesis block [177] in later 2008 and early 2009, respectively. A lot of great inventors were at the prime of their creativity in the 30-55 age envelope [872], and that at this stage in life people tend to have garnered sufficient personal life experiences and professional competences to form strong opinions, and to execute on effective solutions.

Like Satoshi Nakamoto, also the English computer scientist Tim Berners-Lee ("TimBL") [212] was 33 years old when he proposed an information management system in 1989 via the Hypertext Transfer Protocol (HTTP) [873]. His invention seeded the World Wide Web ("WWW", "Web1.0") [213], the technology boosted the "information age" [874], and has most drastically changed lives of people, societies and economies forever, which might have been on Satoshi Nakamoto's agenda with Bitcoin heralding the "Internet of Value" [875].

Satoshi Nakamoto could have certainly shifted the (final one or two digits of) the year he claimed to have been born, so maybe somewhere between the mid-1950s and about 1990, but probably not much farther. Combining his assumed age and his gradual, seemingly long-haul choreographed exodus from the Bitcoin community, it is unlikely that Satoshi Nakamoto was mentally incapacitated, or has deceased in 2011.

## Year

The year 1975 [876] was marked by a few potentially related events (see also Appendix), such as the Watergate scandal [877], [877], which featured the abuse of government agencies the FBI [544], the CIA [554], and the IRS [878] as political weapons, and the end phase of the Vietnam war [879], both shattering the US government [880] in Washington DC [749]. Some less likely connections were that the Rockefeller Commission issued its report on CIA abuses [881], the signing of the Helsinki Accords [882], which officially recognize Europe's national borders and respect for human rights, the approval of the bailout of 6.9 billion total USD for New York City [876], and the coining of the term "fractal" [883] by the Polish-born French-American mathematician and polymath [884] Benoit Mandelbrot [885].

Also in 1975, the so-called Monty-Hall problem [886] was solved biostatistician Steve Selvin [887] from the University of California, Berkeley [412] (which will play a role in a subsequent section). Finally, the number 1957195↔7 results from swapping the last two digits on the publication date of the final citation (8) [557] in the Bitcoin whitepaper [175].

### Day of the Month

The situation is much different for the particular day in the past [888]. Blaise de Vigenère [889], a French cryptographer, diplomat, translator and alchemist was born on April 05 in 1523. Kurt Cobain [890], the enigmatic, anti-establishment artist, singer and angst-fueled songwriter of the American grunge rock band Nirvana [891] from the Seattle music scene [892], an iconic figurehead of Generation-X [713], committed suicide on 05 April 1994.

Most conspicuously for Bitcoin, during the Great Depression [33, 893] on April 05 in 1933, US-President Franklin D. Roosevelt [894] signed the Executive Order 6102 [895, 896], which made it illegal for American citizens to privately own gold; this ban was not lifted for more than 40 years later by US-President Gerald Ford [897], effective on 31 December 1974. By virtue of his (surmised) ideology, Satoshi Nakamoto certainly had a keen interest in the evolution and practices around money [898] / currencies [899], and was concerned how individuals could preserve their personal wealth, and, as an avowed libertarian [378], keep it safe from state intervention [900]. His deliberate choice of April 05 is likely to be related to this event.

### Sakura

In Japanese, a cherry blossom, also known as Japanese cherry, translates to “Sakura” [901]. Bloom typically occurs between March and April in the Northern Hemisphere; the exact date depends on factors like the species, climate, weather and geographical location, but a date around April 05 is not uncommon. In Japanese Shinto-influenced mythology [902], cherry blossoms symbolize clouds due to their nature of blooming *en masse*, besides being an enduring metaphor for the ephemeral [903], or transitory nature of life. Satoshi Nakamoto might have reinforced his (adopted) attachment to the Japanese culture through his date of birth.

### First Contact Day

In the Star Trek sequel [904], the first interactions between humans and Vulcans “took” place on 05 April 2063 [905]. Sci-fi aficionados celebrate this “historic” event on this day of the year [906]. So is Satoshi Nakamoto a “Trekkie” [907]?

## Nationality & Country of Residence

The choice of his pseudonym would initially propose a Japanese person. People from other Asian backgrounds, like Koreans or Chinese, owing to certain historic events, some deeply rooted animosities, and clashes of culture, are less likely to voluntarily adopt a Japanese nationality.

### Britain and the Commonwealth

Actually, there is a train of evidence that Satoshi Nakamoto possesses a Western, presumably a British / Commonwealth or North-American nationality, or at least spent a good part of his formative years in these cultural habitats.

In fact, virtually all (so far) seriously considered nominees originate from such countries where all or most people master the English language with native-level fluency, and might be more aware of the British culture and current affairs than other nationals. Registering “The Times” [303, 304] headline in the genesis block [176, 177, 302, 753] hints towards his state of mind and objective for Bitcoin, and possibly towards his cultural embedding within the realm of the United Kingdom of Great Britain [599], or the Commonwealth [565]. The expression “Chancellor” (of the Exchequer) [908], would be very uncommon in the USA, where the role is referred to as the “Secretary of the Treasury” [909].

This very pattern certainly matches Satoshi Nakamoto’s flawless mastering of the English language with a peculiar mélange of British [563] and American [571] (and possible Canadian [910]) spelling and idioms, e.g., expressions like the expletive "bloody (hard)" [572-574], terms such as "flat" (Am.: apartment), “mobile” (Am.: cell phone), “defen<u>c</u>e” (Am.: defen<u>s</u>e), and "math<u>s</u>" (Am.: math), and the spellings "grey" and "colo<u>u</u>r") in the Bitcoin whitepaper [175], code base [911], and related online communication [912], and preference for the European date format [230] DD/MM/YYYY [913], and reference to the Euro currency [353], all very uncommon for US-natives.

This ostentatious blend might be characteristic of an excellent non-native speaker, as typical for, e.g., Dutch [914] / Flemish (dialect) [915] speakers from the Benelux [401], or Scandinavians [916], who are widely exposed to a mixture of British and US language variants through un-dubbed, original-language broadcasts, such as movies and TV series, in their mainstream media. These nationals are often indistinguishable from native-speakers in writing, with some accent in speaking, and possibly flipping between American and British English. Alternatively, a native American English [571] speaker living for some formative period might have partially assimilated to a British English [563] influenced environment, e.g., as a postgraduate student or visiting scholar in Europe, or vice versa.

### Daily Schedule

Yet, systematic timestamp and metadata analysis of his numerous online postings [66, 172, 364, 917] indicate that Satoshi Nakamoto did not post between the hours of 5 am and 11 am Greenwich Mean Time (GMT) [239] in the UK, which would translate to 2 pm and 8 pm Japanese Standard Time (JST) [562], making it unlikely that he resided in Japan, or another Far-East (Asian) [758] location, assuming rather normal (say midnight to 6 am) sleeping schedules (which is not necessarily applicable to coders and hackers having daytime college lectures, social activities, or jobs to pay their way).

### IP Address

In the very early days of Bitcoin, IP addresses [174] were used in transferring BTC. On 10 January 2009, a systematic analysis of the two IP addresses [59, 918-921] (arguably [922]) of a debug log [923] suggested that Satoshi Nakamoto was located in the Van Nuys [924] neighborhood in the central San Fernando Valley [925] region of Los Angeles [926], California [493].

### Multi-Location

Therefore, Satoshi Nakamoto was not residing in Japan, but moving across US (Eastern to Pacific) [172, 280, 595, 755, 756] or British (most likely: London [742, 743]) time zones [231] during his post-2008 interactions with the online fora. Alternatively, the messages might have originated from a geographically spread company of authors, each of them potentially with their own writing style and spelling patterns.

### Clock Settings

Satoshi Nakamoto also appears to have (deliberately?) changed his onboard computer clock between different American and British / GMT / Far-East time zones, implying frequent intercontinental travel, or deliberate manipulation. We have already concluded from PDF metadata that that the author of the Bitcoin whitepaper was in Mountain Time zone in mid-March 2009.

A post by Satoshi Nakamoto dated 'February 15, 2010, 06:28:38 AM' (UTC [238]) refers to an event that took place on 14$^{th}$ of February (presumably in his own time zone) as "yesterday" [927]. This means that, on that day [928], Satoshi Nakamoto could have not been farther West than UTC-06:00 [929], corresponding to (standard) Central Time (CT/CST) [756], e.g., the state of Illinois [720, 930].

In addition, some inconsistencies in sent and receive times in personal emails Satoshi Nakamoto exchanged with Hal Finney regarding early code development in January 2009 were pointed out [293].

### Connectivity

At 8:41 am (UTC [238]/ GMT? [239]) on Monday, 12/01/2009 [318], Satoshi Nakamoto mentioned "Unfortunately, I can't receive incoming connections from where I am" [299], directing to travel, (remote) rural areas, or government-restricted blockage of select internet services.

## Links to Japan

### *Manga / Anime*

Still, if this is not a decoy, an evidently fabricated Japanese [454] identity might be rooted in a link to this unique, Far-East [758] country, like family, friends, or fondness of the rich, ancient, or modern East-Asian culture; for instance, Mark Karpelès [120], the CEO of the notorious Bitcoin exchange Mt.Gox [120, 819, 820] in 2010, moved from his native country of France [931] to Japan [454], as he was fascinated by Manga [932] / Anime [933].

In this context, a (faint) link might be constructed with the monthly manga magazine "Nakayoshi" [934] ("Good Friends"), and its popular title "Cardcaptor Sakura" [935]; however, the target readership for Nakayoshi is teenage girls. Another magna series titled "Naruto" [936, 937] is written and illustrated by Masashi Kishimoto [938]. While some syllables may have inspired the pseudonym of the creator of Bitcoin, they are also quite common in the Japanese language, and this connection thus seems to be quite unspecific.

### *Internet Service Provider*

There is actually another interesting pointer towards Japan, specifically to its capital city of Tokyo [61, 760]. Satoshi Nakamoto frequently used, in addition to his account with the (German [655]) provider gmx.com [175, 251, 939], the email domains anonymousspeech.com [252, 253, 940] (shut down on 30/09/2021) and vistomail.com [254-256, 941, 942] (defunct as per time of writing), for instance, on the "The Cryptography and Cryptography Policy Mailing List" [283] when announcing the release of Bitcoin v0.1 [943].

These providers offered highly privacy-preserving email services, and anonymousspeech.com temporarily accepted anonymous payment in e-Gold [384, 385, 944], and, anecdotally, later Bitcoin [61]. In a story published in 2014 [118], several links, e.g., via domain registration addresses for email providers, bitcoin.org [257, 258, 328, 356], and a Swiss bank account, were construed via the elusive, (presumably) by the elusive Swiss internet entrepreneur "Michael Weber" [941, 942, 945] (wwwmichi@gmx.ch), notably to Sakura House [61, 941, 946, 947] and Lounge [947]. (Note that several insiders coherently commented that "Micheal Weber" might have just been a commonly known fake identify for covering up identifies.)

### *Sakura House*

These places in Tokyo [760], which are obviously named after the Japanese cherry blossoming, offer apartments and facilities that are preferentially rented by foreigners. Interestingly, the Sakura House and Lounge are situated near in the Nakano-sakaue station [948] for the Tokyo districts of Nakano [949] and Shinjuku [950], which are notorious amusement quarters that are well-known as gay hotspots (Shinjuku Ni-chōme [951]), and for sex-related establishments (Kabukichō [952]). There was also several branches of the American fast fashion retailer "Forever 21" at the time in Tokyo and other parts of Japan [953, 954].

### *Stint in Japan*

This intriguing plot constitutes one of the very few direct connections that could be quite convincingly established between Satoshi Nakamoto and Japan [454]. (Other than that one of the founding fathers of blockchain [5, 955], who co-authored 3/8 papers cited in the Bitcoin whitepaper, speaks Japanese.)

In a conceivable scenario, Satoshi Nakamoto had a stint in Tokyo around the year 2006, staying at the Sakura guest house, possibly for studies, work, fun, or a combination thereof; during this period, Satoshi Nakamoto (re-)connected to the owner of the highly privacy focused email services anonymousspeech.com and vistomail.com who resided, or had his web portals registered to these facilities. Having enjoyed his time, Satoshi Nakamoto got inspired by the locality, i.e., specifically Sakura, Nakano and Nakano-sakaue, and the "Nakano Sakaue gang" [118] that was anecdotally reported to have been around in these locations the year 2006, to adopt his pseudonym from the first syllables of the local ambient.

#### *Matching Story*

With all these cues at hand, it can be speculated that Satoshi Nakamoto likely has a Western educational and cultural background, either British / European or North-American, possibly pertaining to an ethnic minority, such as Japanese-American. In the years immediately prior to developing Bitcoin in 2007/2008, Satoshi Nakamoto might have stayed for a certain period, or at least immersed with an expat community in Tokyo that inspired his alias (possibly in addition to its "intelligence" related interpretation, see above).

#### *Japanese Candidates*

Japanese candidates were proposed in a 2017 presentation [124-126]. The senior author, who published a conference contribution [612, 673] referenced in the Bitcoin whitepaper (2) [175], scrutinizes a 2001 publication entitled "The Security Evaluation of Time Stamping Schemes: The Present Situation and Studies" [129, 956], which is authored by a Japanese person affiliated with the Institute for Monetary and Economic Studies (IMES) [957] from the Bank of Japan [958].

The source finds stunning intersection between some core technological elements that later entered Bitcoin, as well as the references cited in their report and the Bitcoin whitepaper [175]. Moreover, the personal homepage of the author of the report [956] was last updated 09/02/2007, i.e., roughly around the time when Satoshi Nakamoto must have started conceiving and developing Bitcoin. Another Japanese candidate was proposed to have created Bitcoin in a rather quirky YouTube video [127] in 2013 by the American computer scientist and co-inventor of hypertext [257, 258]. However, these leads never seemed to have received significant traction in the community.

## Privacy & Anonymity

### Highly Refined Armor

As a mere fact that, even until today, i.e., nearly 1½ decades after the release of the Bitcoin whitepaper in fall 2008, and despite major efforts to track him, Satoshi Nakamoto's true identity has not been unraveled; this impressively proves that he was able to employ sophisticated methods to persistently protect his anonymity. Using a pseudonym was only one part of his cleverly and skillfully devised, and vigorously executed privacy shield, for which he effectuate(d) a very strict separation between his online and real personality.

As Satoshi Nakamoto, he never communicated through analog channels like handwriting, voice, video, or personal meetings; instead, he exclusively messaged electronically via email, and postings to online newsgroups, essentially "sterilized" ASCII [575] text. Potential "eye-witnesses", whether involved in Bitcoin, or just accidentally noticing related activities, do not exist, or are at least not willing "testify" (yet).

Direct clues, for instance, to his bio, family, friends, socializing, cultural background, (inter-)nationality, political attitudes, physical appearance, (dis-)abilities, coining / traumatic experiences, hobbies, sense of humor, education, and professional vita, are widely missing. Even worse, the scant, and rather shaky hints Satoshi Nakamoto left in his messages might be a part of his intelligently and diligently orchestrated cover-up. Thus, pretty much all digital evidence available from and about the father of Bitcoin has to be taken with caution.

Moreover, Satoshi Nakamoto employed technologies like the Tor [407] ("The Onion Router" [408], a free and open-source software for enabling anonymous communication) network [408, 959] and I2P [409] ("The Invisible Internet Project"), he somehow found a way to register bitcoin.org [257, 258, 328, 960] incognito with a prepaid credit card [118, 960], established email accounts at highly privacy-focused domains like vistomail.com [254-256] (defunct as per time of writing), anonymousspeech.com [252, 253, 940] (shut down on 30/09/2021), and gmx.com [251, 939, 961] for concealing his identity and location. These capabilities unambiguously display the signature of an experienced, very privacy-conscious character versed in skills which are actually quite common among cryptographers, but very rare in beyond these small circles.

### Disciplined Character

As per the whole setup of his persona, per his strict adherence to online communication and his avoidance of live, in-person interaction, and his sudden disappearance from the community, it is obvious that Satoshi Nakamoto was extremely clear and consequent in his decision making, and very solitary; he carefully avoided the limelight by expressing that others, like fellow developers from the steadily growing Bitcoin community, should represent the emerging cryptocurrency project in public. It might also be inferred from the continuing success in maintaining his privacy and anonymity over long periods of time, and by his prolonged abstinence from reaping the fruits of his incredible wealth, fame and impact, that Bitcoin was unlikely to have been created by a group.

### Forgoing Pseudonymous Communication

Note that, in principle, Satoshi Nakamoto could have, and of course still can, include short texts in transactions (through a 40 Byte metadata by the Bitcoin script command OP_RETURN [962]); if they are issued from Bitcoin addresses the community widely accepts to be owned by him [216, 374, 375], such messages would surely send some ripples to developers and markets. But Satoshi Nakamoto has (so far) not once tapped this pseudonymous communication channel, that is incorporated into his own Bitcoin (node) software.

### Techniques

Yet, it might well be argued that, as an undoubtedly exceptionally sophisticated computer- and internet-literate who is evidently proficient in deploying advanced privacy technology [404-410, 963], Satoshi Nakamoto might have easily run message texts past spell and grammar checkers set to (occasionally intersperse) British English, insert double-spacing after a full stop into ASCII ([964]) texts, tune computer clocks and time zone [231] settings of operating systems, and manipulate sent times and forge metadata in email headers.

### Institutional Cover-Up?

While there are serious speculations, e.g., by virtue of the interpretation of his pseudonym (see above), that an intelligence organization might conceived and implemented Bitcoin. Other than certain “rogue states” [965], e.g., to circumvent internationally enforced economic sanctions [966], the author could not come up with a solid reasoning, and convincing plot.

As also discussed in other sections of this article, it is known that the CIA [554] had an eye on Bitcoin, at latest when they invited Gavin Andresen [75] for a presentation that eventually took place in mid-2011. Nowadays, they are even “deeply involved” in the cryptocurrency space [854, 855]. In addition, the DHS [817] seems to have arranged a meeting with “Satoshi Nakamoto” [816], 2022 #7071}.

The author could only speculate that there might have been so government interest to stop, or at least control Bitcoin, but the genie may have already been out of the bottle at that stage due to decentralization. So, the best they could do at that point was to put pressure on Satoshi Nakamoto, e.g., by threatening to report him/her/them for potential tax evasion to the IRS [878], issuing non-compliant issuing of a security to the SEC [534], or infringing with the “coinage power” of congress engraved in the US Constitution [535].

Under such serious intimidation, Satoshi Nakamoto might have budged and transferred his private keys, and possibly access to his online accounts, to a federal agency, e.g., to be able to wield some over Bitcoin in case it became a menace to national interest. In this entirely hypothetical constellation, the governmental organizations involved would surely not have any interest, neither in the disclosure of the deal, nor in unmasking Satoshi Nakamoto.

Furnished with the full playbook of professionally managed and organizationally sanctioned techniques at hand, such as identity replacement technology [963], e.g., adopted from witness [967], whistleblower protection [968], or espionage [969] programs, Satoshi Nakamoto would evade most of the attribute mapping and exclusion filtering methods developed in this article; the investigation would then go way beyond the reach of the author of this article.

### Impersonation

An element of a wide-ranging identity spoof [963] would be to install an impersonator [970] to direction attention away from him/her/them. Among the many possible constellations, Satoshi Nakamoto might have (even pseudonymously) contracted a sufficiently competent person with a suitable vita, and coordinated a reasonably plausible narrative with him/her. Equipped with a carefully scoped, bullet-proof story and tactical plan with Satoshi Nakamoto as advisor and insinuator in the backhand, this "bait" would step into the public to mislead his "chasers" onto the wrong trail, and thus keep them busy, and leaving them discouraged. However, the author of this article does not hold any tangible evidence that Satoshi Nakamoto deployed such shrewd tactics.

## Work Pattern

### Day of the Week

There is a markedly high frequency of Saturdays in the events that Satoshi Nakamoto had some room to maneuver with. These are his date of birth on Saturday, 05/04/1975 [863], and the Genesis block 0 [302] mined on Saturday, 03/01/2009 [739]. In Japanese time zone, also the release of the seminal Bitcoin whitepaper was dated on Saturday, 01/11/2008 [561]. Note that in GMT / UTC or US time zones, the announcement of Bitcoin v0.1 [306] was on Thursday, 08/01/2009 [308], and the mining of block 1 [305] on 09/01/2009 [764], as well as the publication of the Bitcoin whitepaper [175] on 31/10/2008 [279] fell on Fridays. So, in general, a trend for the build-up of outcomes can be observed towards the end of the work week, and work day, when referenced to GMT [239].

However, a comprehensive analysis of Satoshi Nakamoto's activities on Bitcoin-related online fora discloses marked dips on Tuesdays and Saturdays [66].

### Peak Activity Months

Analysis of Satoshi Nakamoto's postings to newsgroups and commits to the Bitcoin subversion repository in the final months of 2009 and 2010 [172] reveals activity peaks, involving "night shifts", in February, a noticeable gap between March and May, resuming intensive activity over the following summer months [66].

### Activity and Time Zones

In the GMT [239] time zone [231], which applies to the United Kingdom [599] (roughly) between November and March. While activities start building up from 11 am, Satoshi Nakamoto mostly posted between 3 pm and 2 am, regardless of weekdays or weekends. Silence is recorded between 6 am and 11 am, likely indicating his sleeping schedule. This would correspond to rather normal hours from 1 am to 6 am in US-Eastern time zone (ET [280]), which is (for the most part of the year [236, 237]) 5 hours behind GMT [239].

If he had resided in the US-Mountain (MT [595]) or Pacific time zone (PT [755]), Satoshi Nakamoto would have gone to sleep rather early, i.e., way before midnight, which would be atypical, especially for (younger) coders. Residence in Far-Eastern [758] time zones, e.g., Japanese Standard Time (JST) [562], may only apply to daylight sleepers, a highly unnatural schedule which would mainly apply to people on night-shift (labor) job rosters.

### Job Type

This fluctuating pattern may correspond to work obligations that are not compatible with a regular office job, but more for a student or academic who is busy on campus during lecture months in college, or a contractor, consultant or freelancer working along tight project completion schedules; even a seasonal schedule of a farmer has been brought forward [971].

Due to the brevity of Satoshi Nakamoto's online life, and his likely capability to forge metadata to disguise his identity and whereabouts, it cannot be ultimately concluded whether this calendrical, work-hour and time-zone [172, 231] consistencies were just specific to the (post Bitcoin-whitepaper) period of observation, or recurrent over preceding and consecutive years.

## Professional Embedding

### Secret Service

While there is some solid indication of Satoshi Nakamoto having participated in academic research, e.g., as a postgraduate student [728] or postdoctoral [729] scholar at a university, Satoshi Nakamoto might have (also) had a (possibly even brief) career in, or related to, cryptography in the corporate sector, or in a secret service institution. These organizations regularly recruit top (mathematical) talent, that they identify based on their educational scores, to their relevant departments.

Before embarking him on a dedicated cryptography training, these three-letter intelligence agencies [460, 461, 544, 554, 842-849, 851, 852] would have rigorously committed Satoshi Nakamoto to strict confidentiality over extended periods of time, even extending way beyond the actual duration of his employment or contract [972].

Such thoroughly monitored obligations might explain Satoshi Nakamoto's apparent fear towards exposure of the nascent Bitcoin project to the CIA [554] in 2011. Personally, he might have even been scared to be in breach of grave commitments to his confidentiality, which actually would prevent him from coming out for a long time, if not forever.

### Freelancer or Contractor

It is also known that highly capable cryptographers would have generously paid options as freelancers or contractors in industry, or even on the dark side, which would elucidate why Satoshi Nakamoto might have had sufficient money to be able to abstain from monetizing his BTC assets after his sudden disappearance; more likely, his staunch ideals might have directed him towards leaving his (early) mined coins (so far mostly) untouched, in order to bolster a catchy narrative of a mysterious origin, and intrinsically endowing innate decentralization; this approach indeed still deeply appreciated by "his" Bitcoin community.

## Computing Setup

### Computer Operating System

Through his communication and digital fingerprint, it can be deduced that Satoshi Nakamoto initially ran a Microsoft Windows (XP) [591] operating system, only later he installed Linux [973], and asked others to port and test Bitcoin on Apple's Mac OS [974], which he evidently never ran on his own machine. As already pointed out earlier, in the section of the Bitcoin whitepaper [175], the (easy to massage) metadata, and other evidence, indicate that it was produced by OpenOffice.org [579], version 2.4, with PDF [286] version 1.4 formatted (print) output.

### Programming Language

Bitcoin was coded in C++ [975], working with the "Feemium" [976], integrated development environment (IDE) [977] for multiple programming languages, Microsoft Visual Studio [978]. The C++ language is rather popular with elder programmers, typically born before the 1980s; younger coders might have opted for modern languages like JavaScript [979] or Ruby [980]. Experts opined that Satoshi Nakamoto's C++ skills revealed in the Bitcoin source code was equivalent to 5-10 years of specific programming experience. With these advanced capabilities at hand, it can be well contended that C++ offers more control than more modern languages towards specifically customizing the cryptographic security for the Bitcoin blockchain.

#### Hardware

Satoshi Nakamoto also seemed to have run somewhat mediocre computer hardware, neither equipped with an Intel Core i5 [981] or AMD [982] processor, thus exhibiting rather mediocre hash rate. This comparatively inferior computing [983] power (CPUs [983] were sufficient for mining at the time) is uncharacteristic for a high-caliber programmer might rule out opulent financial resources of Satoshi Nakamoto around the birth and infancy of Bitcoin.

The distinct blend of Microsoft Windows [591], Microsoft Visual Studio [978] for coding, OpenOffice [579] for publishing, and fair CPU performance for mining would be uncommon among academic cryptographers and cypherpunks, and thus represent a unique recognition feature of Satoshi Nakamoto (at the time).

### Interdisciplinary Skill Set

#### Computing & Cryptography

In addition to his unquestionably outstanding (C++ [975]) programming abilities [366, 367, 369], as exhibited in the first version of the Bitcoin code [306], Satoshi Nakamoto must have, at least, had some advanced knowledge and passion for a strikingly rare combination of techno-scientific fields; this unusually wide interdisciplinary repertoire encompasses elements of cryptography [250], (digital) timestamping [171] (2-5) [612-615] as the precursor of blockchain [5], proof-of-work (POW) [316, 984], computer science (e.g., distributed computing [985], peer-to-peer (P2P) networks [986], decentralization [987, 988]), probability theory (8) [557], economics (e.g., game theory, financial systems, banking, money, market psychology), and law (e.g., financial regulation, freedom of speech).

#### Electronic Cash

Regarding digital currencies, evidence stays inconclusive to which extent Satoshi Nakamoto was involved (under his real identity or another alias) in the cypherpunk [165, 166] movement that spawned in the early 1990s from the Bay Area [675] in California [493]; his alleged (or mock) lack of awareness of (at least part of) related precursor projects [262, 268, 271, 272, 463, 465, 467, 468, 470, 472-475, 478, 481, 611, 621, 989, 990], to some extent including (1) [269] (1) and (6) [270], might be owing to the absence of a specialized scientific community and publishing scene on this cryptocurrencies at the time, but also linked to his (young) age in the 1990s. Furthermore, this apparent unfamiliarity with important prior art reveals that intellectually, Satoshi Nakamoto may have even independently developed, or re-invented, key pillars of Bitcoin, while he was cognizant of the foundations of timestamping and blockchain [613-615, 955, 991, 992].

#### Main Inventory Step

From a formal, intellectual property perspective, it is widely accepted [1] that Satoshi Nakamoto has authored the Bitcoin whitepaper [175] in 2008, and was part of Bitcoin's original reference implementtation. In the course of this, Satoshi Nakamoto devised the first blockchain database, through which he was the first to solve the double-spending problem for digital currencies using a (decentralized) distributed ledger [707] via a peer-to-peer (DLT) [986] network. By combination with cryptographic proof-of-work (PoW) [316, 993], he eliminated a central point of failure, which constituted a decisive weakness in previous attempts.

## Genius

According to the various statements [366, 367, 369] of independent, highly recognized experts, an ingenious intellectual virtuoso [609, 994] [994] must have cleverly designed and successfully delivered Bitcoin [491]. A genius stands out by undeniable talent, resolve, and success, but he often does not meet societally widely accepted behavioral schemes. Or in the words of the impactful, 19$^{th}$ century German philosopher Arthur Schopenhauer [995]: “Talent hits a target no one else can hit. Genius hits a target no one else can see”. Satoshi Nakamoto’s uniqueness is impressively underpinned by his far-ranging spectrum of, in parts presumably self-taught, interdisciplinary competences, tremendous work ethic, and volatile biography.

### Common Traits

While there is no “one-fits-all template” for a genius, their outstanding talent is normally confined to a select area; their major commonality is that they are atypical, their overall character and psychology [994] regularly falls out-of-the-norm. Many of such extraordinarily intelligent personalities clash with generic societal templates in terms of common behavioral traits, such as pronounced modesty, rather poor sociability, unusual sense of humor, and major volatility in their biographies.

A genius tends to have a peculiar sense of humor which is mostly inapt for small-talk, he might be inclined to understate (introvert) or oversell (extrovert) their sophisticated ideas, he shows major discontinuities in his private and professional life, with departures from projects that he considers delivered earlier than others, and may resurface with new, disruptive or unexpected initiatives that are primarily motivated by his portfolio of ideals.

Such erratically perceived decision making may reflect signs of abnormal psychological conditions, or even a degree of mental disorder [996-999]. These rare species can intensively work long hours with extreme focus and persistence on solving extremely challenging problems they feel passionate about. They usually do not retire in a classical sense, their brain needs to offload its energy, for instance, via in interest-, or ideals-driven activities in business or leisure.

### Cryptographers

In particular highly gifted mathematicians in the field of cryptographers often detest races for fame and for showing off their fortune; instead, they frequently show a preference for privacy, and even complete reclusion. They may have an apathy against testosterone-driven power games and hierarchies, e.g., within federal agencies, or the corporate sector, and get easily frustrated when discovering that the main focus of initiatives they care about is not on the delivery of factual solutions, but on personal career objectives.

### Strategy Games

Furthermore, especially math-affine geniuses [613-615] are frequently bestowed with an extraordinary talent to quickly recognize patterns in arrangements of numbers, letters, or geometries, and they often enjoy tackling and creating puzzles. This particular type of intellectual high performers sometimes tends to have a penchant for strategic games involving statistics and memory, e.g., involving card counting, pattern recognition, and complex strategies, such as the banking games Poker [1000, 1001], Blackjack [1002-1004], as well as the table and board games Backgammon [1005], and Chess [1006] (or its Japanese variant Shogi [1007]).

For instance, as pointed out further above, the prodigy chess player Bobby Fischer [555], who blended genius in the game as, for instance, documented by his world-famous 21-move brilliancy [558], with a rather turbulent professional and private vita, might have been one of Satoshi Nakamoto’s childhood heroes. It is also known that a few such gambling-affine “superbrains” were banned at some stage from casinos in Atlantic City or Las Vegas, e.g., the notorious “MIT Blackjack Team” [1008], which operated successfully from about 1980 to the beginning of the 21st century.

### Celebrities

There are numerous, historical and present-day ingenious celebrities, e.g., artists, actors, filmmakers, entrepreneurs, and scholars, who are (sometimes arguably) reported to have displayed some (episodes of) mental abnormalities [999, 1009], including bipolar disorder [1010], schizophrenia [1011], Asperger syndrome [1012], and autism [1013, 1014], or an imposter syndrome [1015]. Amongst the most famous personalities in science and technology are Isaac Newton [1016, 1017], Charles Darwin [1018], Ludwig Boltzmann [1019], Albert Einstein [1020], Paul Dirac [1021], Nikola Tesla [1022], John Nash [90], Richard Feynman [1023], Grigori Perelman [1024], Steve Jobs [14], Terry A. Davis [1025], John McAfee [1026], and Elon Musk [15, 1027].

# Bitcoin Parameters

On their journey into Bitcoin, readers usually stumble across mysteries that they cannot make immediate sense of. These frequently concern the technical or practical reasoning for the choices of the (targetted) block time $T_{\mathrm{B}} \approx 10\ \mathrm{min}$, the (maximum) block size $S_{\mathrm{B}} \approx 1\ \mathrm{MB}$, the halving period of $N_{\mathrm{H}} = 210{,}000$ blocks, roughly corresponding to $T_{\mathrm{H}} \approx 4$ years, the initial coinbase transaction $C_0 = 50\ \mathrm{BTC}$, and the (ensuing) total supply $M_{\mathrm{tot}} = 21 \times 10^6\ \mathrm{BTC}$. Table 1 summarizes the “ambient”, fixed, free, and resulting design parameters of the Bitcoin blockchain as detailed in the following sections.

The values of these parameters sculpturing Bitcoin are not fully explained in its whitepaper [175]; yet, email correspondence between Satoshi Nakamoto and early adopters [355, 1028, 1029] details some objectives. The following considerations are unavoidably quite technical and mathematical, so the reader might fast-forward to the last subsection “Take-Home Messages” if not interested in the specifics.

## Practically “Fixed” Parameters

### Block Time

The block time $T_{\mathrm{B}} = 10\ \mathrm{min}$ draws its technical rational from the latency for synchronizing the common ledger by message propagation time $T_{\mathrm{P}} \ll T_{\mathrm{B}}$ between nodes on an open and decentralized peer-to-peer (P2P) network [354]; it is already reasoned in optimizing the mining efficiency, i.e., that the nodes should not spend too much time on futile mining efforts when the next block has already been found [1030].

For instance, with the $T_{\mathrm{P}} \approx 1\ \mathrm{min} = \mathrm{const.}$ propagation time (which might actually not properly represent current internet and future internet speeds) assumed in Satoshi Nakamoto’s back-of-the-envelope calculations, about $T_{\mathrm{P}}/T_{\mathrm{B}} = 10\%$ of the mining efforts would be deterministically ”wasted” due to delayed notification about the generation of a new, valid block during the final interval $T_{\mathrm{P}}$ of $T_{\mathrm{B}}$. The block time $T_{\mathrm{B}}$ thus constitutes a compromise between fast transaction processing on the one hand, and optimized power usage favoring longer periods $T_{\mathrm{B}}$ for mining on the other.

There is also a further, important practical aspect of $T_{\mathrm{B}}$ associated with the quasi-finality of transactions, which are only deemed safely confirmed after a certain number of blocks were mined, for instance, $60\ \mathrm{min} = 6 \times T_{\mathrm{B}}$ for six blocks. Hence, the block time $T_{\mathrm{B}}$ cannot be randomly set when factoring in the requirements of Bitcoin as a payment system running on a distributed network. Note that $T_{\mathrm{B}} = 10\ \mathrm{min}$ just represents a target value for the average block time measured over many periods; significant deviations have been recorded for individual block periods in the past [1031].

### Block Size

A ($S_{\mathrm{B}} \leq 1\ \mathrm{MB}$) block-size limit, in conjunction with $T_{\mathrm{B}} = 10\ \mathrm{min}$, restricts the write speed $W = S_{\mathrm{B}}/T_{\mathrm{B}}$ of the Bitcoin blockchain’s distributed ledger to . Hence, with $B_{\mathrm{Y}} = (365.25 \times 24 \times 60)\ \mathrm{min}/T_{\mathrm{B}} \approx 52.596$ blocks in a calendar year, the storage requirement of the Bitcoin ledger increases by $W = B_\mathrm{Y} \times S_\mathrm{B} \approx 50$ GB per annum. Faster write speeds $W$ would thus, over the course of time, oust a significant portion of non-professional contributors who volunteer to operate (full) Bitcoin nodes, and consequently, severely undermine Bitcoin's pivotal concept of decentralization through prolific and unrestricted network participation.

Such *de-facto* marginalization of contributors to the distributed ledger is averted under the condition that the storage space of commodity hard drives grows faster than the write speed . This issue constituted one of the main disputes in the Bitcoin "civil war" in the mid-2010s, associated with the (eventually unsuccessful) Bitcoin XT [9] fork, and follow-ups like the controversial Bitcoin SV [1032]. At the time of writing (mid 2022), i.e., about 13 years after its launch in 2009, the full size of the Bitcoin's distributed ledger file ranges in the vicinity of 350 GB.

### Transaction Throughput

As the peer-to-peer payment system alluded to in the very title of its whitepaper [175], speed, measured in transactions per second (TPS), is a common benchmark. While each block has a maximum size of $S_\mathrm{B} \leq 1$ MB, as engraved in the code, the storage space for each of its transactions varies, for instance, as they can contain different amounts of Bitcoin addresses and script commands, called opcodes [1033]. Using an average size of $S_\mathrm{T} \approx 435$ Byte, a ($S_\mathrm{B} = 1$ MB) block typically contains about $S_\mathrm{B}/S_\mathrm{T} \approx 2300$ transactions.

This coarse estimate provides a transaction speed $W_\mathrm{T} = (S_\mathrm{B}/S_\mathrm{T})/T_\mathrm{B} \approx 4$ TPS; the actual transaction speed is reported to be in the range between 3.3 and 7 TPS [1034, 1035]. Importantly, the throughput $W_\mathrm{T}$ cannot be freely adjusted, but results from the fixed design parameters $S_\mathrm{B}$ and $T_\mathrm{B}$, and the space needed for scripting a transaction on the Bitcoin blockchain $S_\mathrm{T}$.

For comparison, the VISA network is capable of , which might be the reason why layer-2 solutions, such as the Lightening [1036] or Liquid [1037-1039] networks, are, or have already been, developed and implemented for Bitcoin as fast, consumer-level payment solutions headlined in its whitepaper [1040].

### Halving

Satoshi Nakamoto's plot for Bitcoin was obviously to issue a finite supply of BTC. Mathematically, this somewhat clever and altruistic objective requires a special distribution function that converges to a fixed number over time. Amongst the many options for the BTC minting policy, he chose a stepwise adjustment which periodically halves the initial reward for the miner, called coinbase transaction.

### Period

The logic behind the halving interval $T_\mathrm{H} \approx 4$ years is not immediately evident, other than its potential correspondence to the leap year schedule, or typical legislative periods of democratically elected governments. This peculiar choice might furthermore indicate Satoshi Nakamoto's intent to maneuver in the binary domain, given that $4 = 2 \times 2 = 2^2$, i.e., doubling, and halving, or division by 2, relating to the fundamental multiplication operators (left- and right shifting) in the binary domain.

### Block Count

The ratio between the aimed halving period $T_\mathrm{H} \approx 4$ years and the (target) block time $T_\mathrm{B} \approx 10$ min yields a block count $N_\mathrm{H} \approx T_\mathrm{H}/T_\mathrm{B} = 210{,}384$, which is roughly represented by the more catchy number $N_\mathrm{H} = 210{,}000$ blocks, as hardwired in Bitcoin's design parameters [355]. Note that since the (maximum) block time $T_\mathrm{B}$ may substantially deviate from its 10-min target, Satoshi Nakamoto may have also rounded to a 200,000-block interval, so there may be some additional message in the choice of $N_\mathrm{H}$.

| **Environmental Parameters** | |
|---|---|
| Propagation time (estimate): $T_\mathrm{P} \approx 1$ min | Calendar: 1 year $\approx 525{,}960$ min |
| Transaction size (mean): $S_\mathrm{T} \approx 435\ \mathrm{B}$ | Global M1 in 2008: $M_1 \approx 21 \times 10^{12}\ \mathrm{USD}$ |
| | |
| **Fixed** | **Free** |
| Block time (target): $T_\mathrm{B} = 10\ \mathrm{min}$ | Initial coinbase transaction: $C_0 = 50\ \mathrm{BTC}$ |
| ⇓ | Adjustment period: $T_\mathrm{H} \approx 4$ years |
| ↦ Transaction "finality" after 6 blocks: 1 hour | ↦ In blocks: $H_\mathrm{B} = T_\mathrm{H}/T_\mathrm{B} \approx 210{,}000$ |
| ↦ Mining efficiency: $\mathrm{T_P}/T_\mathrm{B} = 10\%$ | Adjustment algorithm: $C_h = C_0/2^h$ |
| | with $h = \mathrm{Integer}(b/H_\mathrm{B})$ |
| Block size (maximum): $S_\mathrm{B} = 1\ \mathrm{MB}$ | Smallest subunit: $1\ \mathrm{Sat} = 10^{-8}\ \mathrm{BTC}$ |
| ⇓ | ⇓ |
| ↦ Write speed: $W = S_\mathrm{B}/T_\mathrm{B} \approx 50\ \mathrm{GB/year}$ | ↦ $M_\mathrm{tot} = 2C_0 \times H_\mathrm{B} = 21 \times 10^6\ \mathrm{BTC}$ (1) |
| ↦ Transactions / block (mean): $S_\mathrm{B}/S_\mathrm{T} \approx 2300$ | ↦ $M_1 = 10^{14}\ \mathrm{cent} \cdots 10^{14}\ \mathrm{Sat} = M_\mathrm{tot}$ |
| ↦ Speed: $W_\mathrm{T} = (S_\mathrm{B}/S_\mathrm{T})/T_\mathrm{B} \approx 4\ \mathrm{TPS}$ | ↦ $Y_\mathrm{fin} = Y_\mathrm{ini} + T_\mathrm{H} \cdot (h_\mathrm{fin} + 1) \approx 2142$ (2) |
| | ↦ Inflation at block number $b$: $I(b)$ (3) |

Table 1 Overview of Bitcoin design parameters which are reasoned by technical and practical aspects of a payment network on a decentralized, globally distributed public ledger in a real-world user environment. The values of the fixed parameters (target) block time $T_\mathrm{B} = 10\ \mathrm{min}$ and block size (limit) $S_\mathrm{B} = 1\ \mathrm{MB}$ determine the throughput in terms of transactions per second (TPS), the write speed $W$ in GB per year, and thus the storage needs of full network nodes. The initial coinbase transaction $C_0 = 50\ \mathrm{BTC}$, the adjustment (halving) method $C_h = C_0/2^h$ and the period of $H_\mathrm{B} \times T_\mathrm{B} \approx 4$ years, and the smallest currency unit that can be captured on the blockchain of $1\ \mathrm{Sat} = 10^{-8}\ \mathrm{BTC}$ are freely chosen to adjust the resulting total supply of $M_\mathrm{tot} = 21 \times 10^6\ \mathrm{BTC}$ (1), which has been speculated to relate to the M1 money in 2008 of $M_1 = 21 \times 10^{12}\ \mathrm{USD}$ in terms of their indivisible subunits, and the (theoretical) year of the last mint $Y_\mathrm{fin} \approx 2142$ (2); the "inflation rate" $I$ as defined in (3).

## Initial Coinbase Transaction

For the first, ($T_\mathrm{H} \approx 4$ year)-interval starting with block 0 [177] and 1 [305] on 03 and 09 of January 2009, respectively, the initial coinbase reward was set to $C_0 = 50\ \mathrm{BTC}$. This particular value of $C_0$ constitutes, at first glance, a rather arbitrary pick, in a sense that it does not hinge on practical arguments, like the block time $T_\mathrm{B}$ and the (maximum) block size $S_\mathrm{B}$. Given that $C_0 = 50\ \mathrm{BTC} = 50 \times 10^8\ \mathrm{Sat}$ is halved at intervals $T_\mathrm{H}$, a natural option might have been to express $C_0$ by an amount to $C_h = 2^h\ \mathrm{Sat}$, with an unsigned integer exponent $h \in \mathbb{N}$, so that the final coinbase transaction would cleanly turn out as $2^0\ \mathrm{Sat} = 1\ \mathrm{Sat} = 10^{-8}\ \mathrm{BTC}$, before it drops below 1 Sat, which cannot be represented by the corresponding integer (token) variable on the Bitcoin blockchain anymore.

## Total BTC Supply

Nowadays, the limit of 21 million BTC that will ever be mined is almost faithfully cited among Bitcoin maximalists, and even a membership in the "21 million club" for those owning at least one BTC is popularized. The reasoning behind this rather mystical choice has rarely been discussed. Figure 2 displays the (approximate) evolution of the total supply available at a specific point in time after mining commenced in January 2009.

The total BTC supply is calculated by

$$M_\mathrm{tot} = N_\mathrm{H}(T_\mathrm{Y}) \times C_0 \times \sum_{h=0}^{\infty} 2^{-h} = 2 \times C_0 \times N_\mathrm{H}(T_\mathrm{Y}) = 21 \times 10^6\ \mathrm{BTC} \tag{1}$$

using the identity $1/2 + 1/4 + 1/8 + \cdots = 1$ [1041]; $M_\mathrm{tot}$ (1) is therefore completely determined by the product $N_\mathrm{H} \times C_0$ of the block cycle $N_\mathrm{H} = 210{,}000$ (determined by $T_\mathrm{Y}$, and the "fixed" block time $T_\mathrm{B}$), and $C_0 = 50\ \mathrm{BTC}$. Yet, notice that the sum only converges for $h \mapsto \infty$, so $M_\mathrm{tot}$ will actually fall slightly short of $21 \times 10^6\ \mathrm{BTC}$. (Note that Satoshi Nakamoto refrained from pre-launch mining, which might have tainted the public-good image of Bitcoin for fostering its wide-scale acceptance.)

Consequently, and especially given that $T_{\mathrm{Y}}$ cannot accurately correspond to full years anyway, it would have been very straight-forward to alter the total BTC supply from $M_{\mathrm{tot}} = 21 \times 10^6$ BTC to a more catchy number, like 20, 10, or a 100 million, or even 1 million or a billion to comply with the common scale systems [1042].

The fixed total supply $M_{\mathrm{tot}}$ (1) of BTC (not the chosen amount) was reasoned by Satoshi Nakamoto [339]:

“In this sense, it's [Bitcoin is] more typical of a precious metal. Instead of the supply changing to keep the value the same, the supply is predetermined and the value changes. As the number of users grows, the value per coin increases. It has the potential for a positive feedback loop; as users increase, the value goes up, which could attract more users to take advantage of the increasing value.”

on 11/02/2009.

Only a second source is cited in an email [1028, 1029] exchanged in January 2011 with an early adopter, who directly conversed with, and received BTC from Satoshi Nakamoto in spring 2009. It was surmised [353, 1028, 1029] that Satoshi Nakamoto’s choice of the 21 million of BTC was inferred from the (global) money supply [1043] of fiat currencies $M_1$, which (reportedly, but not accurately) amounted to roughly 21 trillion USD at the time of releasing the Bitcoin whitepaper in fall 2008. M1 money comprises of all the currencies and assets that can easily be converted into cash.

Satoshi Nakamoto is quoted [353]: “I wanted to pick something that would make prices similar to existing currencies, but without knowing the future, that’s very hard. I ended up picking something in the middle. If Bitcoin remains a small niche, it’ll be worth less per unit than existing currencies. If you imagine it being used for some fraction of world commerce, then there’s only going to be 21 million coins for the whole world, so it would be worth much more per unit.”

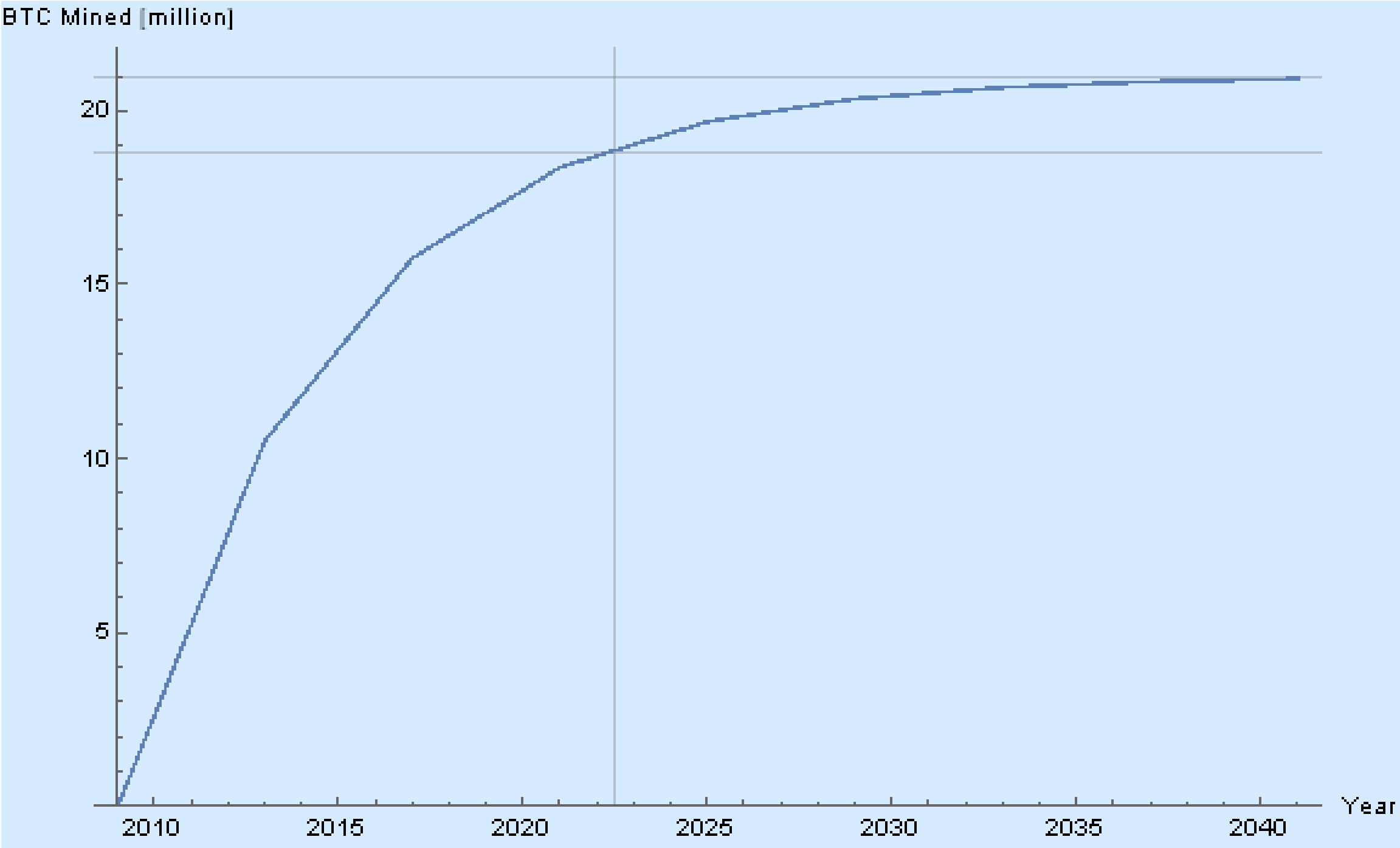

Figure 2 Total supply $M_{\mathrm{tot}}$ (1) in terms of BTC minted as a function of the calendar year, assuming an initial coinbase transaction of $C_0 = 50$ BTC, a block time $T_{\mathrm{B}} = 10$ min of, and a halving period of $N_{\mathrm{H}} = 210{,}000$ blocks.

From a pure quantitative point of view, this makes sense as the two smallest (integer) units, i.e., cents ($1$ ¢ $= 10^{-2}$ USD) and $10^{-8}$ BTC $= 1$ Sat, we mathematically obtain $M_1 = 21 \times 10^{12}$ USD $= 21 \times 10^{14}$ ¢, and $M_{\mathrm{tot}} = 21 \times 10^6$ BTC $= 21 \times 10^{14}$ Sat for Bitcoin. However, the correlation between the two inherent quantities is very fuzzy, and might not align well with the other brilliant logic and expertise demonstrated by Satoshi Nakamoto. (Another article [1044] tabled that a total supply in the region or 21 million BTC would be favorable towards minimizing rounding errors when operating in the realm of a 64-bit floating-point arithmetic [1045]. The author of this work does not follow this argument.)

However, in a different email [355], Satoshi Nakamoto disclosed to have played a bit with this number, initially considering 42: "I thought about 100 and $42 \times 10^6$ BTC $= 2 \times M_{\text{tot}}$, but 42 million seemed high." The notion "seems high" does not make much sense, if 21 is eventually chosen. Also note that the adjustment by the factor ½ refers, similar to the halving period, to the binary system. In fact, following formula (1) for $M_{\text{tot}}$, the first coinbase transaction $C_0$ could have been shifted over a wide range without compromising the fundamental mechanisms underpinning the Bitcoin blockchain.

In addition, revisiting equation (1), the total supply $M_{\text{tot}} = 21 \times 10^6$ BTC could have also been preserved by a more "intuitive" annual cycle, i.e., $T_{\text{Y}} = 1$ year and $N_{\text{H}}(1\text{ year}) \approx N_{\text{H}}(4\text{ years})/4 =$ 52,500 blocks, by multiplying $C_0$ by the same factor, i.e., $4 \times C_0 = 200$ BTC. Hence, the latter communication sheds doubt of the exclusivity of the (alleged) reasoning of $M_{\text{tot}}$ by the global M1 money supply $M_1$; it rather indicates that there may be (additional) motivations for the prominence of the number 21. This circumstance will be further expanded in the following sections.

## Final Block

Starting with the initial mining reward of $C_0 = 50$ BTC $= 50 \times 10^8$ Sat, we can extrapolate the number of halving periods $h$ required to mint (nearly) all BTC, terminating when the coinbase transaction $C_h = C_0 \times 2^{-h}$ slips below 1 Sat; this final period $h_{\text{F}}$ is obtained by solving

(2)

and with a (roughly) ($T_{\text{H}}$ =4-year) halving interval (more accurately $N_{\text{H}} = 210{,}000$ blocks) starting with the genesis (or first) block in January of the year $Y_{\text{ini}} = 2009$, i.e., in the 21st century, we approximate the final non-vanishing coinbase transaction (note that miners will continue to receive transaction fees) to be issued in $Y_{\text{fin}} = Y_{\text{ini}} + T_{\text{H}} \times (h_{\text{fin}} + 1) \approx 2141.88 \approx 2142$, which is in the same region as quoted elsewhere [1046]. Intriguingly, the digits of this iconic year start with a 21, and end with a $42 = 21 \times 2$, just like Satoshi Nakamoto's final choice and initial idea for $M_{\text{tot}}$, respectively.

Trivially, choosing a "more natural" annual halving period $T'_{\text{H}} = 1$ year would have adjusted the final year to $Y'_{\text{fin}} = Y_{\text{ini}} + T'_{\text{H}} \times (h_{\text{fin}} + 1) \approx 2042$, so about a century earlier than with the original Bitcoin parameter $T_{\text{H}} = 4$ years. However, retention of the same year $Y_{\text{fin}}$ for the last coinbase transaction $C_h \geq 1$ Sat around 2142 could have been arranged by annual halving cycles $h'$ calculated by $Y_{\text{ini}} + T'_{\text{H}} \times (h_{\text{fin}} + 1) + h' = Y_{\text{fin}} \Rightarrow h'_{\text{fin}} = 132$; however, this choice of $T'_{\text{H}} = 1$ year would have resulted in a huge number for the adjusted initial coinbase transaction $C'_0 = C_0^{h'_{\text{fin}} - h_{\text{fin}}} \approx C_0^{100}$, which would fall far outside the scope of integer-type variables that are common in computer storage.

## Inflation Rate

For multiple reasons, a classical inflation rate is hard to define. First, Bitcoin started without exclusive pre-launch mining (except the genesis block) by its creator, i.e., from an initial BTC amount of literally zero in January 2009. Second, the BTC supply is not continuously increased by a government to implement certain fiscal policies and to satisfy consumer psychology, but by the need to secure a decentralized ledger on a blockchain.

Furthermore, it is not clear how the transaction fees, which will likely rise in their fiat value over time, should be factored in. Finally, with increased minting of new BTC and halving cycles, the inflation rate decreases until it deterministically merges to zero (i.e., below BTC's non-divisible subunit of 1 Sat).

Note, however, that fading rewards for participating in the crucial proof-of-work [993] competition will have to be compensated, to some degree, by transaction fees to still incentivize miners in the PoW-secured [993] public ledger.

Mathematically, we can define an (annual) inflation rate at block number $b$ in the halving interval numbered by $h(b)$,

$$I(b) = \frac{C(h(b)) \times \frac{T_{\mathrm{H}}}{4.25\ \mathrm{year}}}{M_{\mathrm{tot}}(b)} \tag{3}$$

with $M_{\mathrm{tot}}(b)$ minted so far at block number $b$, starting with $C(h) = C_0 \times 2^{-h}$ and $C(h = 0) = C_0 = 50$ BTC with the Bitcoin genesis block $b = 0$ (or $b = 1$) in January 2009. This ratio (3) may be understood as the inflation rate $I(b)$ at block number $b$. The (logarithmic) plot in Figure 3 shows the development of the Bitcoin "inflation rate" $I$ (3) over the ($T_{\mathrm{H}} \approx 4$ years) halving periods, commencing with $C_0 = 50$ BTC. The calculations are based on a constant block time of $T_{\mathrm{B}} \approx 10$ min; this is just a target value which deviates in practice from the actual block time [1031]. The calendar years on the horizontal axis are thus not in exact sync with the present state of the Bitcoin blockchain.

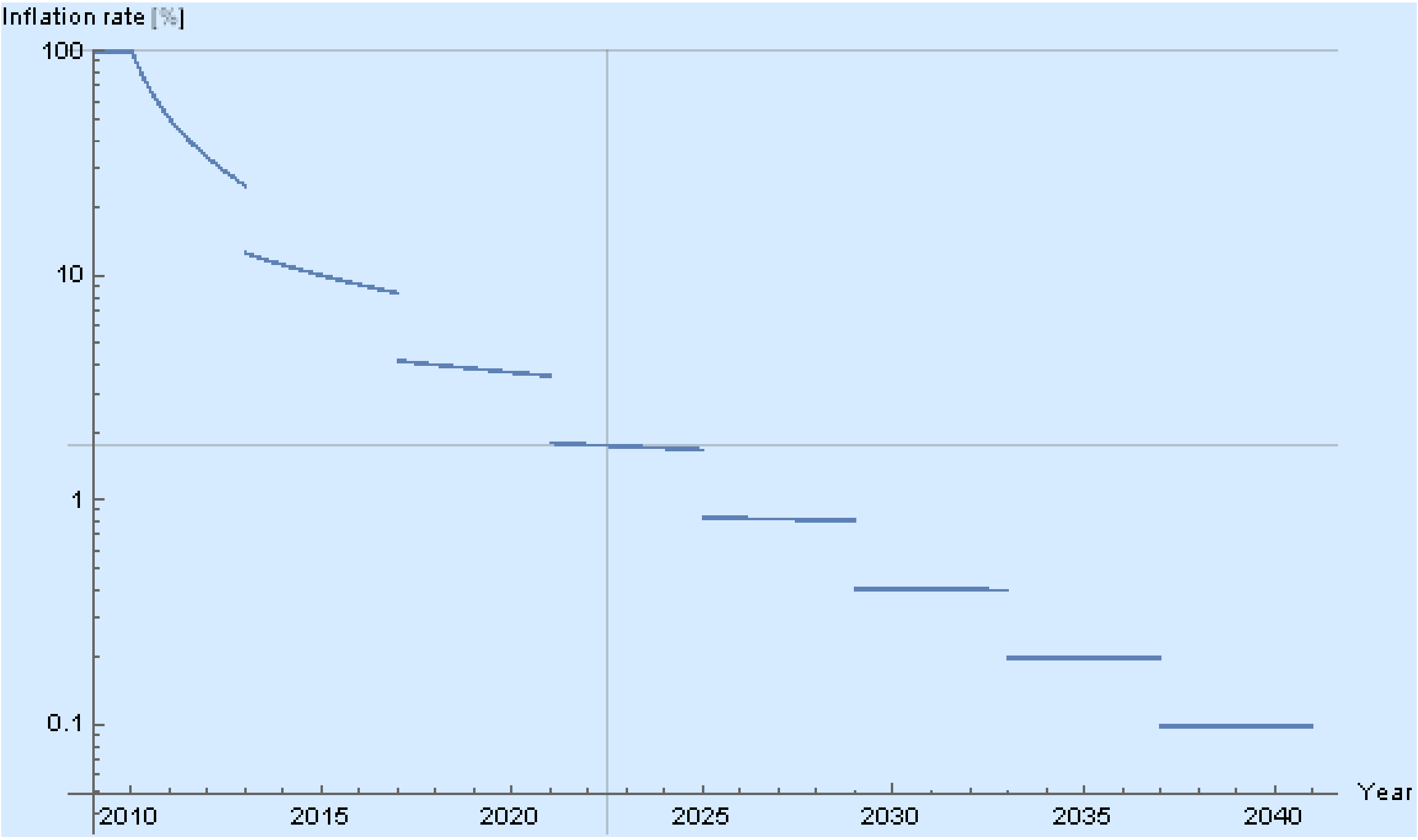

Figure 3 BTC inflation rate $I$ (logarithmic scale) over calendar time calculated according to the method quantified in (3).

Just as an example, at the time of writing (in 2022), and assuming a constant block time, we are within the 4th halving period (see vertical line) when about 19 out of the a total $M_{\mathrm{tot}}$ (1) of 21 million BTC have already been issued, and the inflation rate (3) amounts to about 1.75%, dropping below 1% in the second half of the 2020s, and completely vanishing after the final Sat token had been issued around the year 2142. An alternative choice with an annual adjustment of the coinbase transaction while maintaining the year 2142 for its last, non-vanishing value, is displayed in Figure 4.

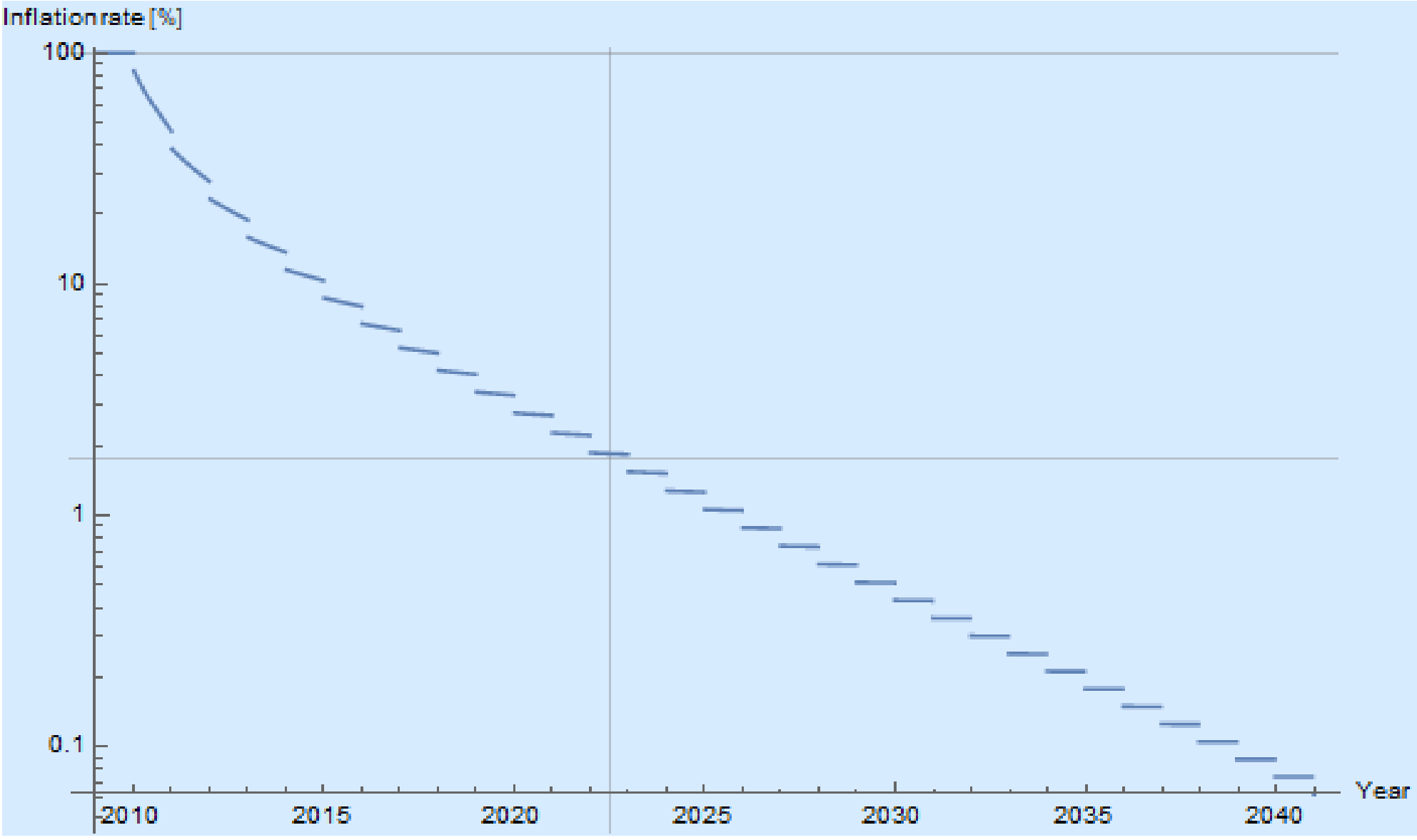

Figure 4 Evolution of (annual) inflation rate $I$ (logarithmic scale) if the coinbase transaction $C_h$ was adjusted by a factor $2^{-1/4}$ each year ($T_\mathrm{H} = 1$ year) with respect to equation (3) and, thus effectively producing a halving every $T_\mathrm{H} = 4$ years as in Figure 3.

## Parameter Analysis

The set of Bitcoin parameters is composed of the practically justified choices for the (target) block time $T_\mathrm{B} = 10$ min and the (maximum) block size $S_\mathrm{B} = 1$ MB, plus a subset where Satoshi Nakamoto had a greater degree of discretion. A key design decision was to implement "halving", i.e., a binary right-shift operation, rather than a floating-point factor for the reduction of the coinbase transaction every interval lasting $T_\mathrm{H} \approx 4$ years; nevertheless, the initial value $C_0 = 50$ BTC $= 50 \times 10^8$ Sat was not chosen as an integer power of 2, i.e., $C_h = C_0 \times 2^{-h}$, leading to rounding errors, and even a slight shortfall of the total BTC supply (1) often quoted as $M_\mathrm{tot} = 21 \times 10^6$ BTC.

Assuming the halving operation for periodic adjustment of the coinbase transaction as given, an alternative annual cycle, i.e., $T'_\mathrm{H} = 1$ year containing $N'_\mathrm{H} = N_\mathrm{H}/4$ blocks, the total BTC supply $M_\mathrm{tot}$ (1) could have been preserved by scaling the initial coinbase transaction by the same factor, i.e., $C'_0 = 4 \times C_0$. Yet, if the last non-vanishing coinbase transaction was to be kept around the year 2142, the annual halving would imply to a tremendous increase of the required initial coinbase transaction $C_0$ to a huge number outside the storage space reserved for typical integer variables.

## Long-Term Dilemma

The often-quoted finite BTC supply $M_\mathrm{tot}$ (1) faces a major sustainability issue. A basic principle for network security of the distributed ledger requires that mining remains profitable, while attackers need to be economically disincentivized. In Bitcoin terms, this implies that a certain hash rate needs to be maintained, which corresponds to investment in equipment and energy. Since resulting bills can be assumed to be paid in fiat (money) [411], the reward for mining a block needs to have a certain minimum value in traditional currency.

At the time of writing in 2022, the lion share of the reward originates from the coinbase transaction $C_n = C_0 \times 2^{-n}$, which amounts to 6.25 BTC for $n = 3$. In an order of magnitude calculation, this equals to about 100,000 USD, and, for simplicity, we assume 2,000 transactions per block. This back-of-the-envelope calculation essentially suggests that, currently, a cost per transaction of about 50 USD.

As long as this amount is mainly covered by the coinbase reward, the actual transaction fees paid by the user remain much lower. However, this necessitates that the ratio of demand-to-supply for BTC on exchanges stays sufficiently high; a dropping USD-BTC pair also lowers the hash rate, and thus network security.

Let's consider two examples while neglecting inflation, which will most likely affect the BTC price and mining expenses in a similar manner. For $n \geq 6$, i.e., by the early 2030s, the coinbase transaction $C_h$ will have fallen below 1 BTC. So, by then, one BTC must be traded at approximately 100,000 USD to keep transaction costs fairly constant. After 10 halving cycles, i.e., at $n = 10$ in the middle of the 21st century, less than 0.05 BTC will be newly issued per block, meaning that one BTC would have to sell at about 2 million USD at exchanges in order to restrict transaction fees to present levels.

Therefore, and in addition to the limited transaction throughput, Bitcoin can only meet the economic requirements of a micropayment system if the number of transactions that can be crammed into a single block is substantially increased, e.g., by layer-2 solutions such as the Lightening [1036] or Liquid [1037-1039]. As layer 1 with the current block size of $S_\mathrm{B} \leq 1\ \mathrm{MB}$ and clock rate $T_\mathrm{B} \approx 10\ \mathrm{min}$, Bitcoin will, otherwise, inevitably become "Digital Gold" where high transaction fees (and comparatively low throughputs $W_\mathrm{T}$) are acceptable in the market.

## Take-Home Messages

The detailed scrutiny of Bitcoin's design parameters underpins the following hypothesis: The target values for the block time $T_\mathrm{B} = 10\ \mathrm{min}$ and the maximum block size $S_\mathrm{B} = 1\ \mathrm{MB}$ are prescribed by real-world boundary conditions of a deliberately decentralized peer-to-peer [986] payment system relying on a distributed public ledger [1047] file. The evidence in this section suggests that Satoshi Nakamoto then needed to choose a proper finite total supply $M_\mathrm{tot}$ of his new cryptocurrency, and a commonly accepted mechanism to deal out $M_\mathrm{tot}$ for incentivizing widescale participation in Proof-of-Work [984] mining for securing Bitcoin.

A fixed total supply $M_\mathrm{tot}$ could only be achieved by continuous slashing of the initial coinbase transaction $C_0$, for which Satoshi Nakamoto initially picked a catchy amount of the order of 100 BTC; he also opted for the (generic) binary operation of halving. Satoshi Nakamoto then played around with the numbers and found that only a ($T_\mathrm{H} = 4 = 2^2$)-year interval would sufficiently spread out to a "fairer" distribution of Bitcoin over several generations well into the next century.

Satoshi Nakamoto calculated from the ratio of $T_\mathrm{H}$ and $T_\mathrm{B}$ that a 4-year period corresponds to roughly 210,000 blocks, which would have resulted in a total supply $M_\mathrm{tot}$ of 42 million BTC. It is theorized that Satoshi Nakamoto then compared this $M_\mathrm{tot}$ to the M1 money supply in 2008 of $21 \times 10^{12}$ USD, and discovered that halving the value of his originally planned initial coinbase transaction to $C_0 = 50\ \mathrm{BTC}$ produces $M_\mathrm{tot} = 21 \times 10^6\ \mathrm{BTC}$; additionally, the issuance date of the last BTC around the year 2142, which is linked a final mint of about 1 Sat, remains untouched. By equating $1\ \mathrm{BTC} = 10^8$ Sat, numerical parity between the M1 money supply and $M_\mathrm{tot}$ is achieved in their indivisible subunits Cents and Sats (tokens), respectively.

In a similar way as shown in the preceding section, Satoshi Nakamoto probably experimented with a few alternative parameter sets, but was content to have found a sweet spot in the parameter space where the binary system for alluding to his new, computer-based digital currency, quantitatively "matches" the legacy monetary system. As we will see in the following, and to the possible delight of Satoshi Nakamoto, the resulting, rather "weird" numbers 21 and 42 nicely lent themselves to further analogies; these figures convey messages reflecting historical and societal happenings as well as a portion of "nerdy" fun relevant to both, the creator and the creation of Bitcoin.

## Features in Code

The original 0.1.0 Bitcoin code also has some intriguing details [1048], such as an IRC [1049] client, a P2P [986] marketplace, and a virtual Poker [1000] game. Especially the latter feature might be picked up by (rather speculative) interpretations in the following sections may indicate a penchant for casino banking games like Blackjack [1002] ("21"), and the board game chess [1050].

## Digits & Numeral Systems – Vires in Numeris

By constructing a new persona, Satoshi Nakamoto had a broad range of options in fabricating his pseudonym, age, and nationality. It could be observed that such (widely) self-definable biodata and dates of Bitcoin, while protecting his privacy, may contain messages that fit very well with the psychogram of a genius, the historic and current events at the time, and his socio-economic attitudes. It is also quite striking that the "odd" number 21 repeatedly occurs in the rather free design parameters within the setup of Bitcoin, often without conclusive technical reasoning.

It can be conjectured that these digits, as well as further peculiarities excavated in the following, may convey one, or potentially more, messages conceived by the elusive mastermind of Bitcoin. This section identifies and delivers potential interpretations of such "suspicious" numbers to eventually extract some information about Satoshi Nakamoto's attitudes and sense of humor around Bitcoin. A plaque stating "I am … and live in …" is highly unlikely to have been deployed by Satoshi Nakamoto.

Note that, to the best of the author's knowledge, Satoshi Nakamoto did not suggest looking for any hidden stories; yet, they might have been a joke for his self-amusement, and not for the public, and thus reflect an aspect of his state-of-mind at the time. But in the end, there is a non-negligible probability that (even all of) the following interpretations are unintended, possibly putting a bright smile on Satoshi Nakamoto's face when reading them. Yet, there is also a chance they might form essential artefacts of the mysterious riddle he left for the world to untangle. Nevertheless, a proposed candidate passing all more vigorous exclusion criteria presented before, and then also meeting a few of the following criteria certainly reinforces his/her/their "Satoshiness".

### Why 21?

What strikes most Bitcoin enthusiasts is the choice of the often almost religiously deemed total supply $M_{\text{tot}}$ of 21 million BTC. Following the previous considerations (1), this BTC limit could have been freely selected, e.g., by adjusting the original coinbase transaction $C_0$ accordingly; in fact, most would have either expected a quantity in the decimal system, such as ten, a hundred, or multiples of 1,000, to arrive at convenient names like million, billion, trillion [1042], or in the binary system [1051], i.e., a power of 2, or even a byte-based [1052] amount, that can be expressed in the commonly used units kB, MB, GB or TB, in the quasi-decimal, octal domain.

We have already seen that 21 also pops up in other contexts, like the $21 \times 10^4$ blocks per halving interval $N_{\text{H}} \approx T_{\text{H}}/T_{\text{B}}$, the first two digits of the year when the final coinbase transaction will be issued, and the cross sum of the block 1's (somewhat freely set) date of mining. Further occurrences and potential meanings of 21 (and 42) will be unveiled in the next sections.

As almost any clue derived in this article, each of them might, *per se*, be fully coincidental, unintended, and even outrageously overinterpreted. Moreover, it is also very likely that the author has blatantly overlooked other correlations further supporting or dismissing decisive ingredients of the subsequent story. But for now, let us take it a step further before making a final judgement.

### Binary

#### Operators

If we play around with the numbers, all revolves about the binary "alphabet" $0,1,2$, the number 21, which may also be regarded as compounding the digits 2 and 1, and Satoshi Nakamoto's indirectly, by repeatedly stated intention to navigate the binary space by left- and right shifting operators, i.e., multiplication by 2 and halving, to produce the numbers $4 = 2 \times 2 = 2^2$ and $42 = 2 \times 21$ within the base-2 "grammar".

### Patterns

Regularly alternating binary patterns when converting from these decimal numbers:

- $1 = 2^0$: I
- $2 = 2^1$: I0
- $21 = 2^4 + 2^2 + 2^0$: I0I0I
- $42 = 2 \times 21 = (2^5 + 2^3 + 2^1)$: I0I0I0

These oscillating sequences might be symbolizing blocks ("I" ↦ ■) and their links ("0" ↦ ○), i.e., inclusion of the previous block hash in the new block, which may be a numeral depiction of a blockchain ■○■○■○■○■○■.

## Decimals

There is also some sprinkle of the decimal system in the block time and the subunits of BTC in the setup, specifically in Satoshi Nakamoto's choice of subdividing $1\ \mathrm{BTC}$ into $10^8$ tokens, i.e., atomic subunits of "Satoshis", or merely "Sats". Here, a typical value linked to signed or unsigned integer variable would lead to ranges between $-2{,}147{,}483{,}648 = -2^{31}$ to $+2{,}147{,}483{,}647 = 2^{31} - 1 \approx 2.1 \times 10^9 = 21 \times 10^8$, and $0$ to $+4{,}294{,}967{,}295 = 2^{32} - 1 \approx 4.2 \times 10^9 = 42 \times 10^8$ in common 32-bit systems, respectively. What strikes most is that the 21 resurfaces again if expressed in units of $10^8$, which is the same order of magnitude between BTC and Sats.

As previously elaborated, it also surprises that the initial mining reward $C_0$ was set to 50 BTC, which was initially intended to be 100 BTC [355]. Given its halving cycle, the "logical" choice for $C_0$ would have been $2^h$ (Sat), or expressed as something like "I000…0000" (with a long "chain" of "0s") in the binary system, so that the last (finite) coinbase transaction would have been exactly 1 Sat.

## Other Numeral and Binary-to-Text Systems

The author could not scavenge any evidence of further messages, e.g., be expressing "Satoshi Nakamoto" in other encoding systems, such as hexadecimal [1053], or any piece yielding Bitcoin addresses in its base-58 encoding [1054]. Also ASCII code [964], or Roman numerals [1055] did not yield any compelling outcomes.

### Bitcoin Keys and Addresses

Maybe one can juggle with some elements, like the pseudonym Nakamoto, and applying Bitcoin's Elliptic Curve Cryptography [1056] ("ECC", specifically the secp256k standard) for generating public keys, and then hashing by SHA256 followed by RIPEMD160 algorithms may disclose some further cues, e.g., when stated in the common alphanumeric base-58 system [1054].

In an email to early adopter Hal Finney dated Monday, 12 January 2009 [1057] at 8:41 AM [299, 300, 1058], Satoshi Nakamoto observed that the Bitcoin address

'1NSwywA5Dvuyw89sfs3oLPvLiDNGf48cPD' [815]

which he had (fortuitously) generated (but which received its first BTC transaction only on 31/08/2014 [1059]) contained his initials (in the reverse Japanese order NS, i.e., last name followed by given name), and elaborated on how users could intentionally generate such vanity Bitcoin addresses containing words or names through computational brute-force efforts.

Note that the (Japanese ordered) initials N.S. would evidently fit well to a cryptocurrency pioneer, Nick Szabo [71], who conceived Bit Gold [262] in the 2000s, arguably deemed the technologically closest precursor to Bitcoin. While this reclusive person is quite concordantly acknowledged by the expert community as one of the few candidates to have had the know-how to develop Bitcoin in 2008, he adamantly denies being Satoshi Nakamoto [65].

### Alphabet

Numbers can be associated with letters according to their ordering in the alphabet, and possibly also with other meanings in a mathematical context. A few connections may be construed, e.g., $0 \mapsto$ 0: Neutral element of addition, $1 \mapsto$ A: Algebra, Neutral element of multiplication, $2 \mapsto$ B: Bitcoin, Blockchain, Binary, Bay Area, Berkeley, $4 \mapsto$ D: Decentralization, Distributed Ledge Technology (DLT), $21 \mapsto$ U: UK, USA. Yet, the author would rank these correlations to Satoshi Nakamoto and Bitcoin as very weak.

## Geometrical Symbolism

### Dice

Counting the spots on the six faces of a dice $1+2+3+4+5+6$ yields 21. For similar geometrical reasoning, 21 is also referred to as a triangle number [1060]. Moreover, a common dice is arranged so that each pair of opposite sides adds up to 7, so offering the alternative calculation $3 \times 7 = 21$. These geometrical associations, which a math genius recognizes quasi instantaneously, allude to "rolling the dice", which is a common theme in the mathematical discipline of probability (calculus); this syntax may point to the topic of the 1957-textbook citation (8) [557] in the Bitcoin whitepaper [175], and further reference the popular table game of Craps [1061] in the realm of gambling.

Geometrically, the dice is shaped like a cube or block, which then evokes the notion of blockchain in the context of Bitcoin. Consequently, there exists a rather astonishing correspondence between the number 21, statistics, the distributed ledger technology (DLT) blockchain, and the casino mentality that is often affiliated with the (investment) banking system that was presumably scorned by Satoshi Nakamoto considering the message he embedded into the genesis block [177].

### Magic Cube

Even more, if we "left-shift" 21 in the binary space, we arrive at the number $42 = 2 \times 21$, which was Satoshi Nakamoto's original pick for the first coinbase transaction $C_0' = 42$ BTC. This number, at least for a math-driven mind, associates with a (simple) "magic cube" [1062, 1063]; all its lines parallel to the faces, and all four triagonals, sum up to $m \cdot (m^3 + 1)/2$; for the case $m = 3$, the formula results in 42. The potential relevance of this peculiar number to the topic of this article will be outlined in a later section.

Geometrically, a cube can evidently be interpreted (again) as a block. As stated in a personal email [355], Satoshi Nakamoto originally considered setting a finite supply $M_{\text{tot}}$ (1) of 42 million BTC, but this "seemed high", so he halved it. Note that, as already stated above, this context also highlights a contradiction, or another variation of the interpretation, to a secondary source [355], stating that the 21 million BTC limit $M_{\text{tot}}$ (1) was pegged to the M1 global money supply measured in USD [1028].

### Blockchain

By investigating Bitcoin parameters like the total supply of 21 million BTC, i.e., $M_{\text{tot}} = 21 \times 10^{14}$ Sat, we observe that they contain a long chain of 0s; in contrast, common computer based measurements are usually expressed in the octal (Byte = by eight or 8 bit [1052, 1064]) system; for instance, $1 \text{ MB} = 10^{20} \text{ Byte} = 1{,}048{,}576$ Byte, which is only close to 1 (decimal) million (=mega) Byte, but not ending with a longer sequence of 0s.

Hence, by interpreting 21 (and 42) as standing for "block" denoted by the symbol ■, and the many zeros as a chain-link (technically speaking, the inclusion of the previous block hash into the next block) by ○, we may, again, "depict" the number $21 \times 10^{14}$, representing $M_{\text{tot}}$ in terms of the basic supply of smallest subunits of Bitcoin, as a block ■ chain ○○○○○...○○.

## Writing Systems

Such (surely notional) symbolic interpretations may relate to the Japanese writing systems [1065] composed of the syllabic [1066] kana [1067] / Hiragana [1068], and the logographic [1069] kanji [1070], which are logograms [1069] adopted from Chinese characters [1071]. Amongst their important types are pictograms [1072], which are highly stylized and simplified pictures of material objects, like 人 for person, or 木 for tree / wood, and ideograms [1073], that attempt to visualize abstract concepts, such as 上 'up' and 下 'down'. (In Japanese, Satoshi Nakamoto is represented by サトシ・ナカモト [1074].) Hence, assuming that he showed signs of a cultural affinity to Japan, Satoshi Nakamoto probably knew at least the basics principles of kanji [1070], and possibly tried to introduce some (fun) symbolism not the fabric of Bitcoin.

The methods for deciphering his (potential) messages have similarity to translating the writings of (temporarily) forgotten, ancient cultures, such as the Egyptian hieroglyphs [1075] which were eventually decoded by the help of the famed Rosetta Stone [1076], or the manifold, often still not, or only fractionally understood, Mesoamerican writing systems [1077-1079], some of them featuring pictorial elements similar to the above case of “blockchain”.

# Media, Sports & Games

## 21

### *21 (Film)*

On 28 March 2008, i.e., presumably during a decisive phase for Satoshi Nakamoto in creating Bitcoin, the Columbia Pictures heist drama film “21” was released [731]. This movie, as well as others [1080], are based on the book “Bringing Down the House: The Inside Story of Six MIT Students Who Took Vegas for Millions” [1081]. Allegedly spiced with some dramatic exaggerations, this Hollywood production describes the story the MIT Blackjack Team [1008, 1082], which was believed to be founded around 1980. In the context of Satoshi Nakamoto and Bitcoin, the following snippets might be of interest: MIT [414], Harvard [421], Las Vegas, financial (tuition) problems, scholarships, casinos, gambling, banking games, and Blackjack [1002], where the number 21 is of pivotal importance.

The movie also features the so-called “Monty Hall Problem” [886], which is a brain teaser, in the form of a probability puzzle, loosely based on the American television game show “Let's Make a Deal” [1083], and named after its original host [1084]. The underlying mathematical problem was originally posed (and solved) in a 1975 letter [1085], and an article [1086] submitted to the American Statistician [1087].

Born in 1941, the author Steve Selvin [1088] is a professor emeritus of biostatistics at the University of California, Berkeley [412] in the San Francisco Bay Area [675], which is known as the birthplace and hotspot of the cypherpunk [165, 166] movement in the early 1990s. Note that these articles have been published in 1975 [876], i.e., the same year Satoshi Nakamoto claims to have been born on his P2P Foundation profile [861].

In the aftermath of its box office success, a racial “whitewashing” controversy emerged over the decision to star the majority of the characters by white [1089] (“Caucasian” [1090]) Americans, even though the main players in the book template were Asian-Americans [1091]; its only actor from this minority had one-note designations as the squad's kleptomaniac and a slot-playing loser. Moreover, intra-Asian conflicts arose as one (former) Chinese-American [1092] member of the MIT Blackjack team [1008] uttered "I would have been a lot more insulted if they had chosen someone who was Japanese [454] or Korean [1093], just to have an Asian playing me."

### *A Beautiful Mind (Film)*

As it most likely attracted a very similar audience, it is also noteworthy to look the Universal / Dreamworks movie “A Beautiful Mind that started screening (in US cinemas) on 21/12/2001 (note the exclusive composition of the date by the numbers 0,1,2,21, and 2001 ↦ 2~~00~~1↦ 21?) [1094]. It is a Hollywood adapted biopic of John Nash [90] (1928–2015), an American mathematician who made fundamental contributions to game theory [1095], which evidently plays a major role in Bitcoin. John Nash's work shed light onto the factors that govern chance and decision-making inside complex systems found in everyday life.

In 1994, Nash was co-awarded the Nobel (Memorial) Prize in Economic Sciences [1096] "for their pioneering analysis of equilibria in the theory of non-cooperative games" [1097], and the prestigious Abel Prize [1098] in 2015. In 1959, Nash began showing clear signs of mental illness, and spent several years at psychiatric hospitals being treated for schizophrenia [1011]. After 1970, his condition slowly improved, allowing him to return to academic work by the mid-1980s. The case of John Nash may exhibit some (partial) overlap with Satoshi Nakamoto, e.g., around the buzz words genius [994], game theory [1095], encryption (ECC [1056]), mental disorder [1009], MIT [414], and Princeton [723].

#### *Music Album*

"Twenty One" [418] is the second studio album by the English indie rock band Mystery Jets [419]. It was released in the UK on 24 March 2008, i.e., about half a year prior to the publication of the Bitcoin whitepaper [175]; it is explain [734]: "Twenty One is the perfect name for this album on more than one level. Identities are still being shaped during the first half of most people's 20s." Protecting his "identity" evidently also paramount for the staging the "mystery" of Satoshi Nakamoto.

Note that the "21" [1099] by the British artist Adele [1100] was world's best-selling album of the year for both 2011 and 2012, but obviously released after the start of Bitcoin. The title refers to her (approximate) age at the time of composing and recording, so most likely no connection to Bitcoin's choices for 21.

### 42

Due to the role of the binary operators that occurred already in the context of Bitcoin parameters for doubling and halving, we explore "left-shifting", i.e., doubling 21, to obtain $42 = 2 \times 21$. There are a couple of intriguing correlations between Bitcoin and this number; the "Magic Cube" [1062, 1063] was already mentioned above in the context of geometries. It can be found in his emails with an early adopter [355] that the number 21 emerged from halving Satoshi Nakamoto's initial pitch of 42 million total supply $M_{\text{tot}}$ (1) of BTC.

#### *The Hitchhiker's Guide to the Galaxy*

At least in "nerdy" academic and techie circles around mathematicians, physicists, or computer scientists, 42 is known as the "Answer to the Ultimate Question of Life, The Universe, and Everything" [1101]; this absurdly simple number to a complex philosophical question is calculated after 7.5 million years by the humanoid supercomputer "Deep Thought" [1102] in "The Hitchhiker's Guide to the Galaxy" [1103]; this science-fiction comedy franchise was created by Douglas Adams [1104], who died in 2001 ( 2~~00~~1?). Originally a 1978 radio broadcast on BBC Radio 4, it was later adapted to other formats, including stage shows, novels, comic books, a 1981 TV series, a 1984 text-based computer game, and a 2005 feature film [1103].

#### *Level 42*

The same "Answer to the Ultimate Question of Life, The Universe, and Everything" [1101-1103] also inspired the name of the English jazz-funk band "Level 42" [1105]. Formed on the Isle of Wight [1106] in 1979, the band disbanded in 1994, and reformed in 2001; they had a number of UK and worldwide hits during the 1980s and 1990s. One of Level 42's legendary songs is "Running in the Family" [1107] from their homonymous 1987-album [1108].

This circumstance (weakly) implies that Bitcoin might have been developed by a multigeneration group of "relatives" (in blood or mind); such a peculiar multi-author constellation would explain some conundrums regarding locations, time zones, spelling, typesetting, competences and historical embedding. Furthermore, strong family bonds offer enhanced cohesion for robust and sustained collusion compared to a team of fairly independent players which are merely forged by complementary competences and joined interests. But this aspect is very speculative.

### *Coldplay Song*

"42" [735] is a song by the British rock band Coldplay [737] (released on 12 June 2008); it is called "42" after the favorite number of (some of) its composers [736]. Its lyrics go:

> Those who are dead are not dead
> They are just living in my head
> And since I fell for that spell
> I am living there as well, oh
> Time is so short
> And I'm sure
> There must be something more

Whether this poetry was Satoshi Nakamoto's mind cannot be verified.

### *42 Laws of Cricket*

The beginnings of the of the sport of cricket [1109] date back to the beginning in the late 16th century in south-east England. Its rules are specified in a code called "The Laws of Cricket" which has a global remit [1110]. There are 42 Laws (always written with a capital "L"). The version of the code has been owned and maintained by its custodian, the London-based Marylebone Cricket Club (MCC) since 1788. While it has significantly moved on by now as an open international sport including female competitions, cricket still strongly roots in a deep tradition of a "Gentlemen" sport, with an air of upper class and elite education in its home country of England [1111].

Cricket is predominantly popular in the United Kingdom [599, 1111, 1112], Australia [457], South Africa [1113], New Zealand [1114], Ireland [1115], India [452], Pakistan [1116], Sri Lanka [1117], West Indies [1118], Bangladesh [1119], Namibia [1120], and Zimbabwe [1121]. These countries are / or were, in some form, affiliated with the former British Empire [1122] and the Commonwealth [565], where English, with a preference for British spelling [563], is either the native language, or acts as a (quasi) lingua franca [1123].

### *Texas 42*

There is also a trick-taking domino game known as Texas 42 [1124]. It is played with a standard set of double six dominoes. 42 is often referred to as the "national game of Texas", where it was designated the official "State Domino Game".

### *Jackie Robinson*

The number 42 is tightly affiliated with the legendary baseball player Jackie Robinson (1919 – 1972) [1125], the first black athlete to play in Major League Baseball (MLB) during the modern era. Robinson played for the Brooklyn Dodgers [1126]. Since 1977, his famous jersey number 42 has been retired by his alma mater UCLA, and by all MLB teams since 1977; commencing in 2004, each MLB player wears No. 42 on "Jackie Robinson Day" [1127], which is celebrated on April 15 [1128].

Robinson's outstanding role for sports and society is underpinned by the 2013 (i.e., after the 2008-release of the Bitcoin whitepaper [175]) American biographical film "42" [1129]. His story may give cues to racial discrimination, for instance, felt by Satoshi Nakamoto as member of a (possibly Japanese-American [1130]) minority, and to locations like New York (Brooklyn) and Los Angeles (UCLA [1131]), the Dodgers this franchise was much controversially, moved to Los Angeles in 1957 [1126, 1132]. This metropolitan area also was where the strong "Satoshi Nakamoto candidate" Hal Finney [70, 299, 300, 1058, 1133], and the Japanese-American [1130] "Dorian Nakamoto" [1134], who, presumably by sheer coincidence, shares the last (and middle) name with Satoshi Nakamoto [7, 371], resided at the time of the release of Bitcoin.

#### *Computer Science School*

As a direct reference to this science-fiction franchise, the private, non-profit and tuition-free computer science school "42" was opened in 2013 [1135], i.e., post-launch of Bitcoin. It was created and funded by a French billionaire [1136] with several businesses in the telecommunications and technology industry. This unconventional name was an explicit reference to "The Hitchhiker's Guide to the Galaxy" [1103].

## Science Trivia

### Geography

The possible meaning of the numbers 21 and 42 might also be scrutinized from a geographical perspective. Yet, neither the 21st nor 42nd meridians east or west provided any locations that related to Bitcoin, nor did their corresponding latitudes on the Southern hemisphere. In the North, the 21st latitude [1137] crosses Hawaii, and the 42nd parallel [1138] Massachusetts [1139] (Boston [1140]), Connecticut [1141], New York (state) [1142], and Illinois [930] (Chicago [1143]) – nothing super compelling overall, with the potential exception of Massachusetts hosting MIT [414] and Harvard [421], as related to the film "21" [731], and the University of Illinois Chicago [122, 720], and the New York University [750, 1144] and state [1142] related to a Satoshi Nakamoto candidate [122]. At the 42nd parallel, the rotational speed of the earth approximately matches the speed of sound.

### Biology

As one of the 23 pairs of chromosomes in humans [1145], chromosome 21 is both the smallest human autosome and chromosome. Most people have two copies, while those with three copies of chromosome 21 have Down syndrome [1146], also called "trisomy 21". Chromosome 21 is associated with diseases, disorders and conditions, such as cancers, intellectual disability, or early onset familial Alzheimer's disease. A likely connection of chromosome 21 to Satoshi Nakamoto could not be recognized.

# Sorting Circumstantial Evidence

## Numbers and Numeral Systems

This section groups the pointers derived for various numbers in the context of Bitcoin. Note again that the interpretations are to be taken with a grain of salt, and only those that are surfacing in various reasonable, and tightly linked scenarios, might actually have been intentionally deposited by Satoshi Nakamoto; even if this was the case, the probably do not reveal his name tag and whereabouts, but would rather reflect his penchant for puns, or even red herrings [829].

### 21

The most striking pointers gather around the number 21, which occurs in the rather free Bitcoin design parameters: the total BTC supply capped to $M_{\mathrm{tot}} = 21 \times 10^6$ BTC $= 21 \times 10^{14}$ Sat (1) each halving period $T_{\mathrm{H}} \approx 4$ years corresponding to $H_{\mathrm{B}} = 21 \times 10^4$ blocks. Reportedly [353], this quantity, as expressed in the smallest BTC unit (token) of Sats, may have been derived from the 2008 global M1 money supply, as expressed in USD cents.

On top of the this rather weak, anecdotal correlation, copious other contexts indicate that the choice of 21 may have had additional, non-exclusive interpretations that may have been encoded by Satoshi Nakamoto. 21 also equals the cross sums of the "free digits" in his self-issued date of birth (05/04/75). i.e., $5 + 4 + 7 + 5 = 21$, and the timestamp of block 1 (09/01/2009 $\mapsto 9 + 1 + 2 + 9 = 21$) [305] after his 6-day [773] mining gap to the genesis block [177, 302]. As a reference to "block" – possibly the most frequent, non-trivial word in the Bitcoin whitepaper [175], 21 also corresponds to the sum over the faces of a dice, which geometrically resembles a cube or block. When expressing the supply limit $M_{\text{tot}}$ (1) of $21 \times 10^6$ BTC $= 2100000000000000$ Sat in the decimal space, the series of 0s certainly looks like a chain, thus potentially depicting a blockchain.

Bitcoin was also launched in the 21st century, when his potential chess idol Bobby Fischer [555], who conceived the famous "21-move brilliancy" [558], passed away (on 17 January 2008). Several films [1147], music albums[1148], and songs [1149] were themed 21 (or Twenty One).

In March of 2008 [522], the film "21" [731, 1082], and the album "Twenty One" by the Mystery(!) Jets [732] were released. The card game Twenty-One [1004], and its popular casino variant Blackjack [1002], are also named after this central number to Bitcoin. Amongst further occurrences of 21 [1150] are the American game show Catch 21 [1003], the American fast fashion retailer Forever 21 [954] which had a shop in Tokyo [953], and the "Twenty-One Demands" [1151] made during the World War I [1152] by the Empire of Japan [454] to the government of the Republic of China [453].

### 0,1,2,4

The decimal number 21 yields an alternating "I0I0I" in the binary space, which might be deemed as the elements (0) and links (I) constituting a (block-)chain. The 21 de composes into the digits 2 and 1 of the base-2 system, which has the neutral element 0, and right- and left-shifting bits implementing division by 2 (i.e., $\times 2^{-1}$) and doubling (i.e., $\times 2$); these binary operators directly occur in the halving period for the value of the initial mining reward, and $4 = 2 \times 2$ as the approximate period $T_{\text{H}}$.

Somewhat surprising, the initial mining reward of $C_0 = 50$ BTC was not chosen as a power of 2, but, in the same way as the total supply $M_{\text{tot}}$ (1), and the (target) block time $T_{\text{B}} = 10$ min (Table 1), in a quantity complying best with the decimal system. (Most likely, 50 BTC was most likely obtained after first trying 100 BTC as $C_0$, and then halving / right-shifting the value.) This also holds for $H_{\text{B}} = 210{,}000 = 21 \times 10^4$ blocks. Evidently $H_{\text{B}} \times T_{\text{B}} \approx 4$ years, which then leads to a last coinbase transaction $C_h = C_0 \times 2^{-h}$ awarded in a year starting with 21, from a theoretical, pre-launch perspective, around 2142. The octal system, which is common to express digital storage size, was utilized by Satoshi Nakamoto to express the (maximum) block size $S_{\text{B}}$ of 1 MB (Table 1) for the Bitcoin blockchain.

### 42

Doubling, or binary (left-shifting) of 21, yields the number $42 = 21 \times 2$, again corresponding the to a regularly alternating binary "chain" pattern "I0I0I0". The number 42 also defines a (simple) magic cube [1062], which evokes the symbolism of blockchain. Satoshi Nakamoto mentioned [355] that 42 million BTC was his initial consideration for its total supply $M_{\text{tot}}$ (1), as following from a $C_0$ of 100 BTC, so this notorious number was definitely on his agenda. A (simplified) calculation arrives at the year 21<u>42</u> when the final Bitcoin token will be minted.

Even more, the 42 constitites the "Answer to the Ultimate Question of Life, the Universe, and Everything" [1101] as calculated and checked by the supercomputer "Deep Thought" [1102] according to the "The Hitchhiker's Guide to the Galaxy" [1153], an entertaining sci-fi comedy that is very popular amongst techies. Even the name of the British band "Level 42" [1105], and a French computer science school [1135] (established post 2008) were inspired by this multi-layered figure.

The 42 is also a sacred number in the historic Laws of Cricket [1110], and the domino game Texas 42 [1124, 1129]. Moreover, 42 is associated with the jersey number of the famous baseballer Jackie Robinson [1125, 1129], who broke the race boundaries in professional sport; the number 42 is retired to all Major League Baseball [1154] franchises, and all players wear the number 42 once per season on "Jackie Robinson day" [1127] (April 15 [1128]). The Northern latitude of 42 [1125] cuts through the

US federal states, such as Massachusetts [414, 421, 1008, 1138], New York [122, 1142, 1155], and Illinois [122, 720, 930], which all pop up various parts of this article.

While the occurrence of the number 42 is less prominent than 21 in the final version of the Bitcoin implementation, the sheer fact that a supply of 42 million was Satoshi Nakamoto's initial choice for the total supply BTC [355] highlights that this number was definitely on Satoshi Nakamoto's radar. Therefore, the numerous associations brought forward in this work on 42 ought to be considered for drafting his mindset.

## Pseudonym

The choice of his pseudonym and the nationality in his (mock) birth certificate clearly articulate Satoshi Nakamoto's anchoring to Japan [454]. While this attribution is, most likely, a decoy, his particular choice of a rather common Japanese name without reference to a celebrity (discarding "Satoshi Sugiyama" [812-814], and "Tominaga Nakamoto" [96, 1156, 1157]), likely teaches that Satoshi Nakamoto wanted to convey a message. He might have been inspired by his connection to the Sakura House [946, 947], a popular residence for foreigners in Tokyo, the nearby Nakano district [949] and Nakano-sakaue station [948], and the "Nakano Sakaue gang" [118].

The choice his first name [809] is often translated as "intelligent history", possibly indicating a stint, e.g., as employee or contractor, in a national intelligence agency, or a corporate enterprise in the context of cryptography and cybersecurity. The syllables Naka- [833] and -moto [834] are widespread in Japanese last names, also in the academic and corporate cryptographer community [500-504, 650, 688, 695-699], some of them are even similar to "Nakamoto". The entirety of his pseudonym might be creatively interpreted as "a genius focused on a project", or "the wise one of Chinese currency" in Chinese. The initials, in their Japanese ordering NS, were even highlighted by Satoshi Nakamoto in an email on 12/01/2009 [1057] to Hal Finney [70], one of the earliest Bitcoin adopters [299, 300, 815, 1058].

## Locations

Pinning Satoshi Nakamoto to a fixed location has so far proven to be extremely difficult. While his self-defined pseudonym and birth certificate point to Japan [454], few other evidence advocates Japanese nationality. Instead, a chain of evidence [118, 946-949, 1158, 1159], mainly gained from the similarity of (bits of) his name to various places in Tokyo [760, 946, 947] [118, 948, 949], and the registration of the privacy-focused email domains [252-256], signals that the alias Satoshi Nakamoto was selected as part of a well-orchestrated endeavor to conceal his true identity and whereabouts; yet, they might indicate that he might have had a spell of some duration in Tokyo, and an attraction to (facets of) the Japanese culture, in advance of creating Bitcoin.

However, time zone analysis of his postings after the release of the Bitcoin whitepaper reveals some stays across various zones in the continental USA, which is supported by the role places like California with the greater Los Angeles [926, 1131, 1132] and the Bay Area [675] / Silicon Valley, the state of Massachusetts with the Boston area [414, 421], and Great Britain [599] / The Commonwealth [565] / Europe [281], which is further supported by Satoshi Nakamoto's (occasional) spelling preference, and the citation of the UK print edition of the London-based "The Times" [303, 741, 747] newspaper in Bitcoin's 2009-genesis block [176, 177]. The second citation (2) [612, 628] of the Bitcoin whitepaper [175], and the association of the postings of X [34] that were traced back to Dutch [400] IP addresses [174] suggest some liaison with the Benelux [401] countries.

## Sport & Games

The various contexts disclosed some relevance of strategic games that are known to appeal to mathematically highly gifted "brainiacs" [1160], like Blackjack [1002, 1003], the Monty-Hall problem [886, 1083, 1084], or chess [555, 558]. In addition, connections to the bat-and-ball games [1161]

Baseball [1125] and Cricket [1109, 1110] are shown up, which enjoy particular (but not exclusive) popularity in Anglo-America [617, 1162, 1163], Great Britain [599], many former colonies and impact zones of the British Empire [1122], respectively.

### Ethnicity

The evidence collected so far also points at minority issues, such as the whitewashing of Asian-Americans in the cast of the 2008-movie "21" [731], and the racial discrimination as epitomized by the case of the African-American baseball legend Jackie Robinson [1125, 1129] and his iconic jersey number 42 [1127]. The author would regard this evidence, without additional pointers, as weak.

### Exposure to Intelligence Organizations

As a high-caliber cryptography expert, Satoshi Nakamoto was certainly also well aware of legal issues around cryptography, like the perpetually smoldering conflict between freedom-of-speech [527] for citizens on the one side, and crime-prevention, and national security interests on the other. Professionals in this highly demanding field are eagerly coveted by departments in federal three-letter agencies [460, 461, 544, 554, 842-849, 851, 852], as well as the corporate (ICT [1164]) sector. It is also not uncommon that accomplished cryptographers in academia are approached by intelligence institutions.

Satoshi Nakamoto's unique portfolio of interdisciplinary skills might root from a designated training in such organizations, who would have committed him to long-term confidentiality on information, and refrained him from utilizing intelligence techniques acquired through them in external, or post-departure projects. Cases like the WikiLeaks [359, 538-541, 1165] disclosures, Silk Road [119, 543] and the regulatory watchdog SEC [138, 534, 1166] challenging to the legitimacy of issuing cryptocurrencies before, during his brief online life, and even after his unexpected exit from the landscape of Bitcoin might have distinctly sharpened his very pronounced sensitivity for public exposure of himself, and his electronic cash masterpiece.

## Biographical Cues

### Competences and Work Ethic

Without doubt, Satoshi Nakamoto, especially if assuming an individual actor, must cover an extraordinarily broad spectrum of techno-scientific and economical competences. His amazing C++ [975] programming skills are supplemented by a deeper understanding of distributed computing [985], peer-to-peer (P2P) [986], privacy technologies [404, 407-409, 689, 1167], operating systems [1168], mathematics [1169], probability theory, game theory [1095], the money [898], banks [1170], law [1171], current affairs [121, 359, 362, 520, 538, 738, 741, 747], federal (intelligence) agencies [554], and (likely) history [527, 710, 895, 1172, 1173]. Satoshi Nakamoto's unusually large spectrum and composition of relevant knowledge was paired with his outstanding ingenuity, creativity, resolve, and dedication, probably rooting in a very pronounced libertarian [378] streak in his personal ideology.

Such staunch determination may have grown out of a sympathy, if not a substantial immersion in the cypherpunk movement, and an irreconcilable personal aversion towards the legacy banking system; this characteristic would be untypical for a person much younger than – say – the age of 25-30, and would thus never have been personally exposed, and significantly delusioned by real-world professional environments and financial matters. Satoshi Nakamoto's work on Bitcoin may well have commenced around August 2007 with the forerunners of the upcoming global financial crisis [521], which has motivated him to crack, for the first time, the "double-spending" problem, and thereby achieving the much sought-after goal of "ultrasound digital money".

The tremendous intellectual and creative effort of creating Bitcoin certainly involved unstinted commitment to long daily working hours without (project-related) technical or social communication to avoid compromising the veil over his identity; managing such a rather extreme social and psycho-logical situation might imply a genius [994] with some (possibly mild) level of mental condition [999, 1009]. Such an extravagant, exceptionally talented individual must definitely have left highly visible marks along his private life, as well as his educational and professional career.

## Education & Career

With all this, Satoshi Nakamoto must still have had funds to sustain his everyday life and pay bills. Given his widely assumed, distinct despise for the banking establishment, it is unlikely that he was coming from a wealthy background in 2007/2008 that let him live from inherited assets and passive income. While there are examples of geniuses living a purposely austere lifestyle, e.g., by still living at advanced age and deliberately jobless with their parents [1024], such extreme modesty seems to clash with Satoshi Nakamoto's evident interest in monetary systems [898].

The previously used analysis of the work hours (on Bitcoin) and time zones [231] when he submitted postings and committed to the code repository in 2009/2010 [172, 917] indicates some interstate travel in the USA [617] or Canada [427, 1163], and some stints in the UK [599] / Europe [281], the latter is also corroborated by the preferred date format DD/MM/YYYY [913], and his (sporadic) British spelling and idioms [563]. Overall, it is likely that Satoshi Nakamoto was intercontinentally commuting between the North [1174] / Anglo [1162] America and Europe [281].

His Bitcoin-related activity pattern peaked in February, followed by a valley until resuming again towards the summer, resuming around June. Such monthly schedules may suggest an alignment with typical college schedules, for instance, as a postgraduate student, postdoc, or as a member of academic staff; an experience in scientific publication is further backed by his use of (old-style) double-spacing after the period at the end of sentences (in ASCII-format [575] postings), e.g., in his original announcements (and associated threads) of the Bitcoin whitepaper themed "Bitcoin P2P e-cash paper" on 31/10/2008 [282], and the public release of the Bitcoin source code on 08/01/2009 [307] on the cryptography mailing list [283].

However, other facts, like the unusual brevity and composition of references in his Bitcoin whitepaper [175], or his first name meaning "intelligent history", direct towards a designated training and job experience in a three-letter secret service agency [460, 461, 544, 554, 842-849, 851, 852], which must have strictly obliged him towards high levels of long-term secrecy on classified information and methods acquired. This assumption was shared by insiders that anonymously advised (parts of) this work.

A personal connection to the 2008-film "21" taking place within East-Coast [722] elite-college environments (MIT [414], Harvard [421]) also fits well into this framework. The activity peak at Saturdays / end-of-the-week revealed that in this work might not specifically underpin his academic embedding, as it is rather common in many professional environments, like external consultants who often work from their homebase on Fridays.

## Genius

The invention of Bitcoin required deep interest and expert skills in an unusual combination of multi-disciplinary skills extending substantially beyond cryptography and programming. In particular the bibliography in the Bitcoin whitepaper indicates that Satoshi Nakamoto may not have been a long-term core member of the cypherpunk and electronic cash community, supporting the characteristics of a superfast learning, highly autodidactic genius.

This document produced compounding evidence that, in addition to practical or logical explanations, Satoshi Nakamoto may have (deliberately) encoded a mixture of information and puns into a set of "free" variables, like select Bitcoin parameters, certain dates, and his alias. It is important to note that these messages do not (directly) interfere with his resolve for obfuscating his identity and locations;

instead, these witty riddles reflect the signature of a genius who cannot resist putting his mind to work to produce jokes that, at least instantaneously, only resonate with similarly endowed superbrains.

In a more creative interpretation, these messages tell about a genius who, while being capable of doing amazing things, also has a normal life, with plenty of personal frustrations, and possibly even suffering from social isolation, minority discrimination, and a (mild) form of mental disorder [1009], which has been reported for many other famous inventors and scientists.

## Biography and Psychogram – An Attempt

After a deep dive the substantial amount of existing research, and a set of new circumstantial evidence exposed in this work, it exudes that Satoshi Nakamoto used his undoubted ingenuity, multidisciplinary competences, and unstinted resolve to pull off Bitcoin in 2008, and, likewise, to methodologically, and (so far) successfully, obscure his identity and whereabouts, i.e., for far more than a decade ago by now. The author agrees with the vast majority that the Bitcoin community that we should be very grateful to Satoshi Nakamoto for his priceless gift of developing the first practically functional, decentralized and globally inclusive cryptocurrency based on a public, and distributed ledger to the world.

Yet, from a history of science and technology perspective, as well as to anticipate and gauge possible fallout for the future of Bitcoin, it is worthwhile elucidating the story, character, motivation, and mentality of its enigmatic inventor. Satoshi Nakamoto's pivotal masterwork was formed under the impact of happenings in the past prior to, and at the time of launching the whitepaper in 2008. Just only a few years after its birth, Bitcoin has swiftly turned into a seminal disruptor of the 21$^{st}$-century technology, and has laid the groundwork for adding a native value layer to the present Web2.0 [1175].

Despite that the creation of Bitcoin is undoubtedly not a felony [1176] at all, the search for Satoshi Nakamoto resembles offender profiling [1177] in a cold case [32], e.g., the infamous "Jack the Ripper" [1178], and others [181, 1179, 1180] in the recent century; such criminal investigation involves drafting a person's mental, emotional, and personality characteristics based on things done or traces left. Other parallels to the methodology pursued here are found in unraveling historical mysteries [179, 202, 203, 205-207], in reconstructing forgotten techniques[175][176][177, 178][179-188][198-201], and in deciphering the meaning of ancient writings left by ceased cultures [1075, 1077-1079].

Accordingly, the quest for Satoshi Nakamoto is largely anchored in an ever-evolving forensic collection of circumstantial evidence, which is, unavoidably, prone to notorious tunnel vision [1181], subjective conclusions, and blatant overinterpretations. Still, by filling the voids with the author's hunch, a best attempt is implemented on how the available pieces of the unavoidably incomplete puzzle might optimally assemble.

Note that most likely, the real Satoshi Nakamoto, when put on the spot, would adamantly deny any interest or involvement in cryptocurrencies, possibly refusing any comments, and declining interviews. He would presumably even be happy if other nominations and self-claims were fielded, thus distracting investigators, and defocussing public attention.

### General Profile

The author would consider Satoshi Nakamoto as a strongly ideals-driven, legally well versed, intercontinentally engaged, endurably closemouthed, extraordinarily gifted, multi-talented, interdisciplinary polymath [884], and productive modern-age cryptographer [1182] who is very confident in applying his highly advanced, hands-on (C++ [975]) coding skills; the still enduring success of shielding his real-world identity reveals a cunningly orchestrated, and wide-ranging expertise in online privacy technology [404-410] enabled by a comprehensive, long-term sustainable strategy, enforced by diligent, highly disciplined, and consistent execution. From what is known, Satoshi

Nakamoto never met in person, at least not under his pseudonym, or left any form of analog traces, such as handwriting, voice, video or face-to-face talks with other members of the Bitcoin community, a practice that no uncommon in online interest groups.

He is probably a male individual who was raised, or has spent formative years, in a British cultural background. He was physically located in early 2009 where the paper version of the London-based "The Times" was available, e.g., the UK, or metropolitan areas in Europe, like Brussels in Belgium. The peak hours of his post-whitepaper postings suggest that he either ran Bitcoin as an afterwork hobby, or mostly resided in or near the Eastern-US time zone.

Satoshi Nakamoto very likely entertained some solid link into the Benelux cryptographer powerhouse around Leuven, and probably fostered a close relation to an elder person born well before 1940, e.g., a family member, friend or tutor, with a profound passion for math. He displayed a certain familiarity, or even fondness of (elements of) the Japanese culture, and the choice of his pseudonym was possibly inspired, and assembled, by similar-sounding names of Japanese cryptographers. The ingenious architect of Bitcoin himself most likely belongs to the (extended) Generation X [713], or was a (second-half) "baby boomer" [1183]; so he might have been not much younger than $30 \pm 5$ years of age, i.e., not a "Millennial" [1184], and not much older than $50 \pm 5$ in the later 2000s, i.e., approximately born in the mid-1950s to 1980 bracket.

Satoshi Nakamoto must have had excellent formal or autodidactic, fast-tracked and award-decorated education in a rather extraordinary portfolio of disciplines, plus a creative mind for thinking outside the box, and an unsettled resolve for implementation. His CV must be unusual, a career of an early high achiever boasting outstanding accomplishments, and inexplicable discontinuities. This attributes perfectly converged for pulling off Bitcoin in 2007/2008, and fading out in 2010/2011.

His interest in "sound money" may well have been accompanied, or sparked, by events around the turn of the millennium, such as the start of the Eurozone [1185] in 1999, or the burst of the dot-com bubble [1186] in the early 2000s, when Satoshi Nakamoto may have discovered cryptocurrencies as a means to implement his libertarian ideals; since then, he might have (possibly anonymously, or via another alias) followed, posted and interacted with the cypherpunk or cryptographer scenes, or been involved in, or passively following the conception, development, discussion, or implementation of other electronic cash / micropayment systems [269-271, 608, 611, 619, 1187]. Yet, he might not have been a seasoned cypherpunk at the time.

When writing the Bitcoin whitepaper and early mining, Satoshi Nakamoto ran Microsoft Windows (XP) on comparatively low-end hardware, and, very distinctively, used the quite uncommon word processor OpenOffice. The format and writing style of his Bitcoin whitepaper also indicates some exposure to academic publishing, either through just regular screening of scientific literature, or by active involvement at a rather early stage, like postgrad, postdoc or reader / lecturer level, but not carrying the signature of an internationally reputed, highly accomplished senior faculty member.

Satoshi Nakamoto's texts featured rather impeccable, native-level English with a notable mixture of American [571] and British English [563] spelling and expressions, which would be very uncommon for people culturally firmly anchored to the USA or UK, respectively. This linguistic signature might either indicate assimilation through a longer-term stay in an environment where the alternative version is dominant, e.g., a person from regions impacted by both spellings, e.g., Canada [910] or other (present or former) Commonwealth [564] countries. Another option would be that English was not Satoshi Nakamoto's mother tongue, but he perfected his skills acquired in school while exposed to diverse private, academic or professional communities, or through original language services, e.g., in journals, books, movies and interactive online media.

Another conspicuous feature in Satoshi Nakamoto's writing (in his ASCII text postings) is his pedantic double-spacing after a full stop at the end of sentences [285]. This custom has been abandoned decades ago, especially when mechanical typewriters were increasingly substituted by mechatronic systems, and eventually computers with modern text processors and high-resolution printers. This habit would thus be characteristic for a Baby Boomer [1183] who still learnt typing on a classical machine, e.g., in the 1980s. Overall, the evidence from his writings, i.e., blending English language variants and the two spaces before a full stop, turns out to be inconsistent, and it may well surmised that embodied a clever maneuver in his sophisticated, multi-level strategy for camouflaging his identity.

Satoshi Nakamoto may have had some disappointing experiences, such as bullying by peers or teachers as a prodigy child, for instance, as a (Commonwealth [565]) expatriate, or pertaining to an ethnic minority, e.g., as a Japanese American [1130]. Possibly treated as a maverick geek, he might have been ridiculed by envious fellow students who were sent by their wealthy family to elite colleges, and showed off with prestigious cars and high-fly lifestyle, while he was socially somewhat isolated, and just getting along on a rather tight scholarship budget.

Satoshi Nakamoto might have found, early on, consolation in computer and internet technologies, strategy (banking or board) games like Blackjack, Poker or Chess, following idols with similar mental and social issues accompanying their ingenuity, like the extravagant world chess champion Bobby Fischer [555], or the game theorist John Nash [90, 1094], whom he could intellectually and emotionally emphasize with.

His exuberant motivation for working extended periods with immense emotional commitment on a cryptocurrency must have been rooted from a deep frustration, e.g., with the traditional banking system, paired with a distinct libertarian attitude, as prevalent throughout the cypherpunk movement. However, possibly due to his age, he might only have discovered, and immersed with this rather niche scene, and familiarized with predecessor electronic cash / cryptocurrency attempts at the time when he began pondering and working on Bitcoin in the second half of the 2000s.

Around the year 2006, Satoshi Nakamoto may have had a longer stay in the Tokyo where he enjoyed the local culture, and socialized with a similar-minded community; these circles may have been comprised of (possibly including himself) Japanophiles, digital nomads, free-speech advocates, and highly privacy-oriented internet natives, which helped him to systematically cover his identity from an early stage, and, in parallel, to decisively enhance his interest in the politico-economically quite radical objectives similar to those promoted by the cypherpunk scene.

Under the influence of this ambient, Satoshi Nakamoto gradually developed an increasing, eventually quasi-obsessive passion for electronic cash systems / cryptocurrencies, which offered a path to address severe issues he affiliated with the, at least from his point of view, dysfunctional banking system by leveraging his outstanding intellectual capabilities and exceptional work ethic towards what later materialized as Bitcoin.

Though the interplay of these experiences and awareness of this own remarkable skills, Satoshi Nakamoto decided to devote himself to the Bitcoin project in the second part of 2007, when forerunners of the upcoming financial crisis already crystallized. Considering his accelerated education, a dedicated training in cryptography or cybersecurity, e.g., in academia, the corporate arena, or a three-letter agency [460, 461, 544, 554, 842-849, 851, 852], Satoshi Nakamoto entered his self-chosen solitude within a 1.5-year Bitcoin development tunnel, possibly as an intensive after-work, or late-night / weekend pastime.

During this decisive period, Satoshi Nakamoto was getting progressively motivated by the concurrently escalating financial crisis [520, 738]; his concerns about free speech that he was familiar with from the preceding or emerging cases, such as Phil Zimmermann's PGP case [1173], Bernstein v. USA [527], Napster [531, 532], Megaupload [533], or Wikileaks [359, 360, 362, 538], were even amplified.

While online, he foremostly conveyed the impression of a "tech-head"; for the most part, Satoshi Nakamoto remained very rational on explanations of techno-economical aspects of Bitcoin, rather than preemptively celebrating his early-miner's opportunity for his future monetary fortunes, and economic impact; he also widely desisted from passionately advocating extreme ideological positions, e.g., those affiliated with the cypherpunk movement.

With a fondness for pattern recognition, Satoshi Nakamoto might also have enjoyed multifold overlap of his Bitcoin parameters with fun and coincidental interpretations of the number 21 in relation to math, geometry, history, politics, media and sports, and the decentralized "chain of blocks" as a keystone of his Bitcoin. Similarly enjoyable correlations also hold for the number 42, like the "The Hitchhiker's Guide to the Galaxy" [1103], the "Magic Cube" [1062, 1063], the Laws of Cricket [1109, 1110], and the legendary baseballer Jackie Robinson [1125, 1127].

Around August 2008, Satoshi Nakamoto had most of the Bitcoin package wrapped up; however, not having been a seasoned member of the cypherpunk and academic research community, he was somewhat unsure whether he properly accounted for prior art, which was very hard to pinpoint and retrieve with the traditional instruments commonplace in scientific publishing; in fact, he eventually left out some major work, and listed some rather "odd" citations to the references of the Bitcoin whitepaper, to an extent that his document would have likely been returned for major revision after peer review by a classical journal; in hindsight, Satoshi Nakamoto might even have been flabbergasted that he (independently) reinvented major elements of Bitcoin's various precursor components [485, 616, 676, 714-716, 1188-1190] and currency systems [262, 268-270, 611, 619, 825, 989, 990, 1187, 1191, 1192] that displayed technical or motivational similarity.

Satoshi Nakamoto seemed to be strong to endure the initial week of silence, and the ensuing, mostly critical and dismissive echo in the cryptography mailing list [283] following the publication of the whitepaper on 31 October 2008 [279]. After the discussion picked up, he initiated a designated "bitcoin-list" [340] on 10 December 2008 [1193]. While in the final steps of honing his work on Bitcoin's code v0.1 with the help of a few early adopters (like Hal Finney [70] and Ray Dillinger [294]), Satoshi Nakamoto came across the now legendary headline on the frontpage of (the UK print issue of) "The Times" (of London) [176]; this unique opportunity to express his objectives and dislike of the legacy monetary / banking system prompted him to expedite the mining of the Genesis block 0 [177] on 03 January 2009 [739], even though he still wanted to perform some polishing before uploading the code to a public repository almost a week later.

Over the next about two years, Satoshi Nakamoto interacted at fluctuating intensity with the continuously growing Bitcoin community. He seemed to have been quite appreciative of inputs, but also showed traits of a benevolent dictator; he seldomly merged contributions from others into the code, and had notable peaks of his postings and commitment over the course of the years. Overall, Satoshi Nakamoto focused mostly on technical issues, and remained overwhelmingly silent on ideological aspects of Bitcoin occasionally raised by the community. His business or investment attitude towards Bitcoin, as well as his desire to assume formal roles, and to wield the power arising from them, appear to have ranked at astonishingly low priority.

While nowadays many hardcore Bitcoiners tend to perceive him akin to a divine entity, the level of respect towards Satoshi Nakamoto in these early days in the online fora was occasionally appallingly small, and his natural authority even occasionally challenged. Towards the end of 2010, Satoshi Nakamoto might have gotten increasingly alienated from the budding early-adopter community, discouraged about the attitude they showed towards him, disillusioned about the feasibility of his initial objectives.

More likely, his rather odd departure might have actually been part of a long-term planned, gradual exit that he deliberately chose not to communicate in advance. Consequently, over the course of time, he began transferring the stewardship of Bitcoin development and mailing lists to trusted stakeholders. At some stage, Satoshi Nakamoto might just have considered his creative part in setting up

Bitcoin essentially delivered, and after "mission accomplished" was reached, from his perspective, he went on to new challenges that his exceptionally gifted mind could play around with and solve. He might also have faced issues balancing his personal commitments between Bitcoin and his private life and job, which was needed to settle his bills, or to deal with situations in his family.

Satoshi Nakamoto's gradual exit from Bitcoin towards end of April 2011 was also spurred by the arrival of Bitcoin in mainstream media, and in particular through discussions about Bitcoin adoption in circumventing the financial blockade of Wikileaks [359, 362, 538], and an upcoming meeting with the CIA [554, 816], 2022 #7071}.

He definitely wanted to strictly maintain his own privacy and keep the nascent Bitcoin project out of the spotlight of the press, and potential legal prosecution. This hyper-cautious attitude might have resulted from professional commitments on strict, and long-term confidentiality he signed with three-letter agencies or industry; their potential breach might lead to seriously sanctions, even reaching as far as longer-term imprisonment; Satoshi Nakamoto was certainly aware of related cases, e.g., from peers, colleagues, or via public media outlets.

After his pseudonym became inactive in the community space, he did still not want to jeopardize his brainchild, and felt a deep ethical obligation to desist pushing certain buttons. Satoshi Nakamoto believed it would be unfair to move the BTC he rather easily and copiously mined in 2009, as a lion share of them would be attributed to him. Also, the fear of legal issues, e.g., enforced by the Securities and Exchange Commission (SEC) [534] or the Internal Revenue Service (IRS) [1194], such as unauthorized issuance of Bitcoin as a security, or gigantic obligations in capital gains tax (CGT) [1195], might have led to a firm pledge to refrain from touching, or even to burn about 1 million BTC, e.g., by locking up, burning, or destructing (the physical medium storing) his private keys. (In principle, the private keys might have also been deleted or lost by mistake, but this is unlikely for an extraordinarily computer-versed cryptographer.) Such potential institutionally led investigations would have, most probably, also compromised his personally highly valued anonymity.

Conversion of BTC into fiat money [411] was also impractical before first exchanges emerged in 2010 [819, 820]; after his departure from the community, cashing out his 2009-BTC assets without leaking his real name would have been extremely difficult, other than through entering rather dubious technologies like cryptocurrency tumblers [1196] (also referred to as "mixers"), which were still to emerge in the early days of Bitcoin; such legally questionable action might have attracted the attention of federal prosecutors, who might have already blacklisted his Bitcoin addresses. A full extermination of all evidence, including his online accounts, connecting him to Bitcoin might also have been part of Satoshi Nakamoto's advanced strategy to strictly protect his privacy, and to eternally obscure the origins of his cryptocurrency project.

Satoshi Nakamoto might still enjoy operating in an unnoticed "God-mode" in the Bitcoin community, or even staying involved under his real name, or another mock identity. After all, Satoshi Nakamoto just retired his pseudonym, possibly with a self-imposed pledge to only resurface in emergency situations, such as in the resolution of the civil war about Bitcoin XT [1197], or for relieving a name-alike Japanese-American "Dorian Nakamoto" from unwanted public exposure as the (wrong) person behind Bitcoin.

Satoshi Nakamoto might nowadays be very wealthy, profiting from the Bitcoin he mined in a more competitive environment post 2009/2010, and lucrative follow-on businesses. While his focus might have shifted to other exciting, possibly even diametrically different projects, e.g., in the large spectrum of cryptography, Satoshi Nakamoto still may remain, like any parent, very proud of the breath-taking success of his now adolescent masterpiece Bitcoin. In public, Satoshi Nakamoto would surely emanate the aura of a genius, but otherwise be a normal human with his repertoire of strengths, weaknesses, concerns, flaws and vices in his character.

## Multi-Tier Filtering

The previous section described a bottom-up approach. Following a top-down perspective pursued in this section, the potential cohort of candidates is progressively narrowed [66]. Some filters feature a restrictive *conditio sine qua non*, meaning that it is hard to imagine the real Satoshi Nakamoto would not pass them, or have a very convincing excuse why they do not apply. There are also more tolerant sieving steps, but their associated criteria are unlikely to hold in their entirety.

There are about 8 billion people living on our planet. But it is reasonable to assume [425, 491] that the mastermind of Bitcoin must have been an expert cryptographer pertaining to an elite crowd of similar size as the community involved in the annual CRYPTO meetings [429, 456, 492, 494, 589], or government and other entities involved in cryptography [460, 461, 544, 554, 842-849, 851, 852]. For the sake of a back-of-the envelope calculation, let's assume about 1,000 researchers ($10^9 \mapsto 10^3$).

Only about one quarter of this selection might have had a keen interest in electronic cash / payment systems or cryptocurrencies, if considering that projects like Bitcoin are not necessarily received well in the traditional academic community of cryptographers, especially at the time ($1000 \mapsto 500$).

Of those, maybe half belonged to the Generation-X [713] or (younger range of the) Baby Boomer [1183] age group, have native-level fluency of mixed "American-British" English, and a cultural embedding in the USA, Britain, the Commonwealth or Europe ().

Another filter is the rare combination of Microsoft Windows (XP) on a rather low-performance CPU (neither Intel Core i5 series nor AMD processors), OpenOffice rather than $\TeX$ [586] / $\LaTeX$ [587], which is very widespread across the scientific community, or, for instance, common WYSIWYG [581] editors like Microsoft Word [582, 583], for typesetting. Using Open Office and double-spaces before periods (in ASCII texts), might have been part of his cover-up.

In addition to having a libertarian mindset and sophisticated capabilities in cryptography, online privacy, (C++ [975]) programming, and probability theory (8) [557, 719], which might be commonplace amongst high-caliber cryptographers, Satoshi Nakamoto also exhibited an unusual portfolio of complementary skills in distributed computing, financial systems / banking / money, economics / game theory, law / freedom-of-speech, and history, which further slashes the number of likely candidates offering this entire gamut ($50 \mapsto 20$).

Satoshi Nakamoto's psychogram exhibited an introvert, "techie" streak, occasionally showing flippant decision making, possibly also an Asperger-like [1012] syndrome, or other mental condition, perhaps flanked by depressive episodes [1198]. Additionally, he must have had the time, determination, and financial prerequisites between 2007/2008, and up to 2010, to deeply delve into the subject of electronic cash while not necessarily having been a core member of the cypherpunk community.

He must also have had the emotional persistence and social discipline to keep it for himself throughout conception, implementation, and even post departure and "eternity". So, either the circumstances of his job permitted this tremendous time commitment, e.g., in academia, or Satoshi Nakamoto must have literally lived like a socially and professionally secluded "crypto-monk" when developing Bitcoin; this psychologically stressful situation might rule out that he was in a close (but occasionally fragile) personal relationship, or that he was surrounded by an intense family life, at least not at the time of concocting Bitcoin ($20 \mapsto 10$).

His Bitcoin whitepaper reveals a genius who displayed dazzling capability for brilliantly explaining the technical framework of Bitcoin without much ideological overhead. While extrapolating the future potential of Bitcoin, Satoshi Nakamoto assumes a project perspective, rather than boasting about his tightly related (potential) own entrepreneurial and financial success story. Satoshi Nakamoto is a person with great intellectual achievements, with a propensity to watch his accomplishments taking on a life of their own once they are firmly anchored in their community. Yet, while having left Bitcoin, and even cryptocurrencies, at least as Satoshi Nakamoto, it is highly probable that he is still maneuvering in and impacting subspaces of his hobbyhorse cryptography ($10 \mapsto 5$).

The bibliography of the whitepaper discloses a stunning paucity of knowledge about historic precursors [262, 271, 485, 608] at the time of writing the Bitcoin whitepaper; this may either indicate a lack of interest in compliance with good academic practices, a disrespect or personal aversion against certain pioneers, or a sheer unfamiliarity with the state of the art ($5 \mapsto 2$).

Often overlooked, but critical in the chain of evidence, the arguably most compelling "physical" evidence of Satoshi Nakamoto are represented by the citations (2) [612] and (8) [557] that were only available as printed documents at the time of authoring the Bitcoin whitepaper [175]. Reference (8) points to a 1957 edition of a textbook on probability theory [557, 711, 719], which would have hardly been the generic choice of a generation-X [713] or Baby Boomer [1183] cryptographer born after the mid-1950s.

Even more intriguing, the second citation (2) [612], which assumes a central role in the Bitcoin whitepaper [175] (Figure 1), was only available as part of the proceedings [628] of a local meeting symposium on information theory in the Benelux [401]. According to its editor [634], the book of proceedings was only available as a in printed form at the time of the Bitcoin whitepaper [175], and it was only distributed to the (registered) attendees, and sent out to select libraries, like the "Royal Library of the Netherlands" [635] in The Hague [636].

Satoshi Nakamoto also left a trace of locality as mined he quoted a headline from the frontpage of "The Times" (of London) in the genesis block of the Bitcoin blockchain on 03 January 2009. He probably overlooked that only the print issue was exclusively available in the UK, and most likely in certain metropolitan areas in continental Europe, like Brussels as the headquarter of the EU. In the international print and online versions of the same article, the title slightly varied.

The chances of a non-academic without tight connections into the cryptography hot spot of the Benelux [401], e.g., to the attendees of the 1999 symposium, the organizing WIC [629], or the IEEE Benelux Chapter on Information Theory [631], the KU Leuven [77, 430, 506, 643, 645, 677, 692, 693, 1199], UC Leuven [124, 125, 637], TU Eindhoven [122, 431, 432, 701, 702, 1200], or the Leiden University [113, 434], possibly paired with tight links to the UC Berkeley and Californian cryptographer ecosystem [73, 113, 412, 415, 416], IACR [494, 508] organized in the CRYPTO conferences series [456, 589, 1201], and the cypherpunk scene [69-71, 74, 77, 165, 166, 168, 288, 360, 428, 1202], which emanated from the San Francisco Bay Area [675], and its go-to mailing lists (e.g., cryptography [283], cypherpunks [167], coderpunks [1203, 1204], Bluesky [1205, 1206]), where digital money was a hot topic [266], appear to be negligible. Hence, Satoshi Nakamoto was / is most conceivably part of this cross-fertilizing biosphere. Note, however, that the intersection between the academic / traditional cryptographer community [1182], at least under their real-world identity, and cypherpunk scene [168] was rather scant [74, 77, 428, 721, 1207-1213] ($2 \mapsto 1$).

After this collection of rather hard facts, there is a heap of circumstantial evidence, which might be coincidental on their own, but about half of them might apply to the real Satoshi Nakamoto. There are some penchants, namely: Japanese culture (Tokyo [118, 760, 947, 949, 953], Manga [932] / Anime [933]), strategic games (e.g., Blackjack [1002, 1008, 1082], Poker [1000], Chess [556, 558, 1050]), pattern recognition (e.g., 21: dice $\mapsto$ cube $\mapsto$ block, tokens $\mapsto$ eight 0s $\mapsto$ 00000000 $\mapsto$ chain).

In addition, there Satoshi Nakamoto might have employed his free design parameters (e.g., Bitcoin blockchain, pseudonym, date of birth, nationality) to point to various contexts that he cared about, or found amusing, e.g., political (gold ownership in the US [895-897]), political ("chancellor" message Bitcoin's in genesis block [176, 177, 753]), societal (minority issues [1008, 1127]), celebrity geniuses (e.g., Bobby Fischer [555], John Nash [89, 90], Steve Selvin [887], [712]), Sci-Fi ("Hitchhiker" [1101, 1103], "Star Trek" [904, 905]), music [735-737, 1105, 1107], or sports (e.g., Jacky Robinson [1125], Cricket [1110]) context. In addition, there are hints to certain, mostly US-locations where Satoshi Nakamoto might have resided or transiently stayed, like the San Francisco Bay Area [675], or Greater

Los Angeles [926, 1131, 1132] in California [493], Chicago [1143], Illinois [720, 930], or New York [750, 1142].

Finally, the real Satoshi Nakamoto must have had very compelling reasons for his, from an external perspective, rather unglamorous and unanticipated parting with the Bitcoin community (at least under his pseudonym), his subsequent online silence, and for desisting to touch his crypto-wealth (so far). It can only be guessed that this exceptional behavior is related to voluntary, accidental, or forced destruction / loss of private keys / passwords for account access, pure altruism to endow a beneficial founding myth, legal or tax issues, family situations, or some medical or mental condition.

By virtue of his elaborate, cutting-edge masking of identity, and by knowing that, most likely, only himself could ultimately resolve the mystery, the real Satoshi Nakamoto would unsurprisingly dismiss his involvement in the creation of Bitcoin; he might even deny any past or present interest in cryptocurrencies, and be fully focused on other research or corporate adventures.

## Shortlist of Questions

For the majority of Bitcoiners, a transaction from early mined BTC ascribed with high certainty to the creator of Bitcoin remains the most compelling, if not ultimate proof of his/her/their identity. for. As long as this is not demonstrated, a Satoshi Nakamoto claim would hinge upon plausible clarifications about key pieces of the conundrums, and the resolution of the (apparent) paradoxes he left behind.

The preceding sections of this article brought up a large number of issues that are relevant to the origins of Bitcoin. In the following lists are sorted and prioritized into essential prerequisites Satoshi Nakamoto must know, may answer to strengthen support for "Satoshiness", and more curiosity driven questions.

### *Hard facts*

1. How was it possible for you to engrave the headline of The Times UK print edition [176] into the genesis block [177] of the Bitcoin blockchain on 03 January 2009?
2. How did you rummage and retrieve the ostensibly pivotal citation (2) [612] from the print-only version of a local meeting on information theory in the Benelux?
3. How did you get hold of citation (8) [557], a first, 1957-edition of a textbook?
4. Why do so many time zones [231] occur in your online life, seemingly encompassing, and hence suggesting stays all over the (continental) US, the UK, and Central Europe?
5. Why do your texts blend American-British-Commonwealth spelling and idioms [563, 564, 570, 571], while you seemed to have implemented a spell checker?
6. Where does the rather outdated double-spacing [285] after full stops of sentences come from?
7. Why were there statistically sound gaps in your hour-of-the-day (between 3 am and 11 am GMT), day-of-the-week (Tuesdays and Saturdays), and month-of-the-year (March to May) schedules of your Bitcoin activities?
8. Why did you use OpenOffice [578, 579] for authoring the Bitcoin whitepaper?
9. Why did you run Microsoft Windows XP [591] on a rather slow machine?

### *Somewhat Important*

1. Was citation (2) [612] the eye-opener following the (2)&(6)-first hypothesis (Figure 1) on your journey to Bitcoin?
2. What was the reasoning behind the 6-day mining gap between the blocks 0 [177, 302] and 1 [305]?
3. Why did you cite (8) [557], the first, 1957 edition of a textbook, in the Bitcoin whitepaper?
4. What happened when you mentioned in an email [299] to Hal Finney [70] that "I cannot receive incoming connections from where I am" on 12 January 2009?

5. Why did you include code for virtual Poker [1000], IRC [1049], and P2P [986] marketplace client in an early version of the Bitcoin code?

### *Curiosity*

- Did you conceive and implement Bitcoin completely autonomously prior to the Bitcoin white-paper?
- Did you ever tell anybody that you were about to develop a "peer-to-peer electronic cash system" that later became Bitcoin?
- Did you ever tell anybody, post-launch, that you are, or part of, "Satoshi Nakamoto"?
- Are / were you a cypherpunk?
- Did you attend the meeting on e-Gold [384, 385] in February 2000 [386, 387] at Anguilla [388]?
- Were you the user "X", or read his daunting postings [397-399] in 2001?
- When did you start developing Bitcoin?
- Why did you start developing Bitcoin?
- How did you find the time to develop Bitcoin?
- How did you acquire the extraordinarily wide, expert-level skill set required to instigate Bitcoin?
- What did you learn through formal education, your professional environment, or your interests / hobbies?
- Are you still interested, or even involved in Bitcoin / blockchain?
- Was your gradual and inconspicuous fading as Satoshi Nakamoto long planned?
- Were the upcoming CIA presentation, the Wikileaks discussions regarding Bitcoin, and / or the dynamics on the online groups a decisive factor in your decision to disappear as Satoshi Nakamoto?
- Why did you only cite the two cypherpunk projects (1) [270] and (6) [269], but not, for instance, the work of David Chaum [73, 486], Hal Finney [70], or Nick Szabo [71]?
- Did citations (1) [270] and (6) [269] play a major role in scoping Bitcoin, or did you mostly learn about them in hindsight when you were approaching Adam Back [69] for prior art in summer 2008?
- Why is the bibliography in the Bitcoin whitepaper so outdated and short?
- Were you "shocked" when Hal Finney [70] retrieved the debug.log [923] disclosing a / your Los-Angeles based IP address on 10 January 2009?
- Did you regret to have released on 12 January 2009 the Bitcoin vanity address 1NSwywA5Dvuyw89sfs3oLPvLiDNGf48cPD [299]?
- What happened to your private keys?
- What happened to your online accounts?
- Did you intend to push many important dates towards the end of the work week, mostly Saturdays?
- Did / do you have a link to, or a penchant for Japan [454], e.g., its culture or language?
- Did you just wanted to throw a red herring [829], or are there messages behind your Japanese nationality, pseudonym, and date of birth?
- In addition to their mathematical reasoning, is there any magic implanted in numbers like 21 (or 42)?

## The Usual Suspects: Finney, Back, Dai, Szabo

These four individuals are often quoted as likely fathers of Bitcoin, as they meet many of the obvious, previously established canon of attributes associated with Satoshi Nakamoto, such as their unquestionable expertise in cryptography, cryptocurrency, or their participation in the cypherpunk commu-nity. Needless to mention that all four adamantly denied having invented Bitcoin at some stage. An excellent analysis of this quadruple eventually leading to Nick Szabo has been published by [65]; in its section "Who is Satoshi Nakamoto", this article focusses on criteria that shed major doubt on their identity with Satoshi Nakamoto.

First, being based and embedded in the USA, it is unlikely that cypherpunk Hal Finney [70, 1214, 1215], Nick Szabo [71, 609, 1216, 1217], or Wei Dai [72, 261] would have gotten notice, and have been able to retrieve citation (2) [612] of the Bitcoin whitepaper [175]. However, a similar version [673] on timestamping (3-5) [613-615], and containing a similar reference list highlighting the work by Stuart Haber [425] and Scott Stornetta [426] (Figure 1), was also presented by the same Quisquater group in 1999 at Stanford [413], though citing [676] "instead of" (5) [615] as in the Bitcoin whitepaper.

If this paper (2) [612] played a central role in the road to Bitcoin (Figure 1), it must have been obtained by Satoshi Nakamoto as a hardcopy that was mainly available from colleagues or select libraries in Benelux / Europe. It is thus hard to believe, that Hal Finney [70], Wei Dai [72], or Nick Szabo [71] would have found, and then included (2) [612] in the Bitcoin whitepaper.

Given their immersion in the US-American culture, occasional slippage to British spelling, and quoting the headline [176] of (the print version of) the London-based "The Times" [303] in the genesis block [177] mentioning "the chancellor", also seems to disqualify Finney, Dai, and Szabo.

Third, for not compromising the shield of his anonymity, Satoshi Nakamoto would have probably avoided circulating his real name in citations, early messages and postings, which would mostly eliminate Hal Finney [299, 300, 1058], who had a track record as a brilliant coder [289, 689], but would also have communicated with himself [299, 1058, 1133] on the cryptography mailing list [283], Wei Dai (1) [72, 270], and also UK-based Adam Back [69] with hashcash (6) [269, 619].

While Back used to insert double-space before full stop, his 2002 paper (6) [269] quotes a personal communication with the other cypherpunk Hal Finney (and Thomas Boschloo [1218]) [1219] on an "improvement suggestion 5 years into hashcash to simplify the verification cost", references Wei Dai's [72] b-money [270] (1), and also Daniel Bernstein's [122, 123] SYN cookies [620], but completely omits any 1990s publications on timestamping (2-5) [612-614] by Haber [425] and Stornetta [426]. In addition to citing himself, and the lack of reasoning why he even mentioned having been personally contacted through email by the creator of Bitcoin in summer 2008, this constellation makes it unlikely that Adam Back is Satoshi Nakamoto, despite his range of relevant competences, and long-term membership of the cypherpunk scene [1220, 1221].

## Candidates

Just to stress that this article does not want to, and simply cannot credibly dox a particular person (or group) as "Satoshi Nakamoto", simply because the author simply does not know. (However, the article might empower ruling out certain nominations.) Instead, this section is mainly intended to demonstrate how the rather comprehensive collection of initially unrelated sources compiled in this article can make sense in terms of being fit to a person in the real world. First, a shortlist of cryptographers is scrutinized who showed best overlap with the Satoshi Nakamoto profile elaborated in this work. The author still believes that Satoshi Nakamoto could well be a person "like them". Then, other possible candidates are investigated.

### Best Matches

This section presents certain a range of candidates who do not show major inconsistencies with the exclusion filters. However, their passing of negative selection criteria might be reasoned in missing information available to the author. Note that while the following individuals show gripping match to the hypothetical Satoshi Nakamoto profile sketched in the previous section, their individual connec-tions to the origins of Bitcoin are all unconfirmed, and will possibly even remain unconfirmable without their (unlikely) own input.

### *Daniel Bernstein*

The polymath Daniel J. Bernstein [122, 123] (born 29 October 1971 in East Patchogue, New York, USA) has very rarely been associated with the context of Bitcoin [88, 710, 1222]. Bernstein's (publicly known) core research focus has not been on cryptocurrency [2], but he might have suddenly been drawn into a rabbit hole when, out of sheer curiosity, he was taking a glance at the state-of-the-art in the mid-2000s. So, while staying active in his main daytime job academia, he created Bitcoin as a transient, "hobbyist"-type excursion to challenge his brilliant intellect. Preferring to stay incognito, he filed under the stage name "Satoshi Nakamoto". After his exodus, Bernstein successfully "moved on" to advance exciting new ideas revolving about his mainstay interests in cryptography [709].

Bernstein graduated from high school in Long Island [1142, 1223], and top-ranked in the highly prestigious Putnam competition [1224] in 1987. He completed a B.A. in mathematics at New York University [750, 1142, 1144] in 1991. Bernstein received his PhD in 1995 from the University of California, Berkeley [412], i.e., before turning 25, thus convincingly displaying the key attributes of a highly gifted genius. Curiously, Berkeley resides within the Bay Area [675], the birthplace of the cypherpunk [165, 166] movement in the first half of the 1990s, and a lot of his prior academic vita in the USA took place near the 42$^{nd}$ parallel North [720, 1138, 1144, 1223].

In his postgraduate studies, Bernstein was supervised by Hendrik Lenstra [246], who, principally, worked in computational number theory, and is well known as the discoverer of the elliptic curve factorization method, later at Leiden University [434] in the Netherlands [400], i.e., Benelux [401]. Elliptic Curve Cryptography (ECC) [1056] methods are used in Bitcoin, and Lenstra was even briefly suggested as "Satoshi Nakamoto" [34] based on the Dutch IP address [174] as the origin of the previously mentioned, mysterious postings of "X" titled "Virtual peer to peer banking" on the Usenet [392] groups alt.internet.p2p and uk.finance [393]. Could "Satoshi Nakamoto" be embodied by a junior-senior academic dream team?

Bernstein is a research professor of Computer Science at the University of Illinois Chicago [720], and a visiting professor at CASA ("Cyber Security in the Age of Large-Scale Adversaries") [445] of the Ruhr University Bochum [1225], Germany. He is also a professor in TU Eindhoven [431, 432, 690, 1200] in the Netherlands [400], and still closely collaborates with their department of mathematics and computer science, especially with Tanja Lange [432, 701, 702, 1226], since 2007.

The founder of another cryptography stronghold in the Benelux [401], Ei/Psi [432], was Henk van Tilborg [402, 690], who retired in 2011. He is a board member of WIC [629, 1227], the organizer of the Benelux symposium published [628] where the central citation (2) [612] to the whitepaper (Figure 1) was published. Tilborg was the PhD advisor of Stefan Brands [403, 1228-1230] in 1999, who published on electronic cash systems [478], and also collaborated with David Chaum [479] as early as 1993.

Lange posted on the Cryptography Mailing List [283] in November 2008 [1231], even though not on the topic of the Bitcoin whitepaper [175] released only days before. She was also an invited speaker at the ECC 2003 (The 7th Workshop on Elliptic Curve Cryptography (ECC) [1056] on 11-13 August 2003 at Waterloo, Ontario, Canada) [1232] along J.J. Quisquater [124, 125], the senior author of the second citation (2) [612] from the 1999 Benelux symposium [628, 631] in the Bitcoin whitepaper [175].

Hence, there is a tangible link to the "odd" second citation (2) [612] from a "networking" meeting in Benelux [401] region (which seemed to have also been presented only days later at Stanford [413, 673]). Bernstein's mid-1990s work [1233] is also cited as (5) in the sixth reference (6) [269] of the Bitcoin whitepaper [175]. For the last three decades, Bernstein has been giving talks all over the planet [1234], including in 2006 on "High-speed cryptographic functions" [1159] at Chuo [1235], a private flagship research university in Tokyo [760], Japan [454]. His densely packed travel schedule [1236] might also explain Satoshi Nakamoto's comment on his temporary loss of online connectivity [299], and access to the UK print version of "The Times".

Already a student at University of California, Berkeley [412], Bernstein wanted to publish a paper including the source code on his "Snuffle" [1237] encryption system. Starting in 1995, he ran a multi-year, multi-stage series of court cases on whether software source code was protected by "freedom of speech" guaranteed under the First Amendment of the US Constitution [528]; these disputes addressed whether that the government's regulations preventing his publication, and thus the export of cryptography from the United States, were unconstitutional [527]. Bernstein thus definitely had intensive exposure to legal matters and intelligence organizations.

His name also appeared in the framework of the above-described Keccak-SHA-3 controversy [547-551]. In its aftermath, an article on "Sakura tree coding" was published [1158], acknowledging him, and the NIST [422, 546] hash team, for "useful discussions". Note the astonishing relation of Satoshi Nakamoto's day of birth [888] around the Japanese cherry blossom "Sakura" [901], the striking name matching with the previously occurring "Sakura House" [946-948] accommodation and its Tokyo neighborhoods [949, 950], and the registration of the privacy web services [118, 252-256, 940] to the same facility.

Remarkably, but probably coincidental, Bernstein's Google Scholar profile [709] displays one 2012-publication [1238] in Japanese kanji [1070] writing; its senior author is the accomplished Kouichi Sakurai [1239], a professor of mathematical informatics at Kyushu University [1240] in Fukuoka [1241], Japan. Amazingly, the generation-X [713] cryptologist is not listed as an author, but only cited with his famous 2009 book entitled "Post Quantum Cryptography" [1242, 1243]. This may underpin his exceptional academic reputation, and possible links to Japan.

One of the fathers of blockchain [425] commented in an early article on the origins of Bitcoin [491] that "the community of cryptographers is very small: about three hundred people a year attend the most important conference, the annual gathering in Santa Barbara [492, 589]. In all likelihood, Nakamoto belonged to this insular world. If I wanted to find him, the Crypto 2011 [1244] conference would be the place to start."

Daniel Bernstein chaired the "Dinner and Rump" sessions [1245-1250] at IACR's [494] CRYPTO conference series [589] during the build-up to the Bitcoin whitepaper in the early and mid-2000s; among co-chairs and presenters were J.-J. Quisquater [124, 125, 612, 625, 627] (1), Stuart Haber [425, 613-615, 676] (2-4), Ralph Merkle [419, 616, 714-716] (7), David Chaum [73, 485, 486, 611], Bart Preneel [505, 677, 678], Tanja Lange [701, 702, 1251], Jan Boneh, [365, 626, 627, 680, 1252], Phil Zimmermann [74, 689, 1253, 1254], Cynthia Dwork [420], and Moni Naor [450].

Bernstein himself presented on "Smaller Decoding Exponents: Ball-Collision Decoding" with two co-authors [701, 702, 1255, 1256] from TU Eindhoven [431] in the Netherlands [400] (Benelux [401, 628]), at this very meeting on 18 August 2011, i.e., probably coincidentally, exactly three years after registering bitcoin.org [257, 258, 356], and well after Satoshi Nakamoto finally retired from Bitcoin, uttering that he "moved on to other things" [351] in later April 2011. With some "tongue-in-cheek", Daniel Bernstein's initials provide "The Answer to the Ultimate Question of … Everything: $(\mathrm{D} \mapsto 4, \mathrm{B} \mapsto 2) \Rightarrow 42$.

Note that while Daniel Bernstein ticks a lot of the boxes, his interest in cryptocurrencies and the financial world is entirely unverified. The author of this article approached him, and his Ph.D. supervisor, Hendrik Lenstra, who stated: "I can categorically and honestly declare that I never had any interest or expertise in cryptocurrencies." Daniel Bernstein did not reply upon multiple requests.

### *Len Sassaman*

Born in 1980, the American technologist and cypherpunk Leonard ("Len") Sassaman [46, 47, 77, 686, 1257-1259] (aka "Rabbi", <rabbi at quickie.net>) graduated from a Pennsylvania-based boarding school [1260, 1261] in 1998. Already as a teenager, he was diagnosed with a major depressive disorder [1198]. Reportedly [47], he suffered from traumatic experiences at the hands of "borderline sadistic" psychiatric practitioners, experiences which presumably left him distrustful of purported authority figures.

Sassaman was an amazing autodidact who taught himself in cryptography [250], and protocol development, which the mainstay of his career revolved about. While only at the age of 18, he joined the Internet Engineering Task Force (IETF) [1262] overlooking the TCP/IP [1263-1265] protocol underlying the internet, and later the Bitcoin network. In 1999, Sassaman relocated to the Bay Area [675], and moved in with Bram Cohen [288, 1266], who is best known for the peer-to-peer (P2P) [986] BitTorrent [287] protocol released in 2001. There are even rumors that the creation of Bitcoin emerged from a Sassaman-Finney combo (potentially including Bram Cohen, if he is a good actor [289, 1267-1271]), as Finney superior coding would have nicely complemented Sassaman's talent [289].

He became a frequent, often pseudonymous contributor to the cypherpunk [47, 165, 166, 1272] and the cryptographer mailing list [1273], where Satoshi Nakamoto eventually launched the Bitcoin whitepaper on 31 October 2008. Besides his extraordinary contributions to computing, Sassaman also had a keen interest in biohacking [1274], also known as human augmentation or human enhancement, a do-it-yourself biology aimed at improving performance, health, and wellbeing through strategic interventions.

Based on his distinguished expertise, he founded a start-up with the open-source activist Bruce Perens [1275] in the field of public-key cryptography, which went belly-up in the wake of the implosion of the dot-com bubble [1276]. Sassaman co-developed the GNU Privacy Guard implementation of OpenPGP [1262]; he designed, with PGP [689] inventor Phil Zimmerman [74], the "Zimmermann–Sassaman key-signing" protocol [1253]. At Network Associates (now McAfee) [1277], he worked together with the early Bitcoin adopter Hal Finney [70, 300, 1058, 1133], whom he also occasionally cited in his papers [1278].

Sassaman was the maintainer of the popular Mixmaster anonymous remailer code [1279, 1280], and operator of the Randseed Remailer [1281]. The remailer technology was suggested by David Chaum [73] together with cryptocurrency; such remailers are specialized servers for sending information anonymously or pseudonymously. They were the engine behind the notorious cypherpunk mailing list [1272]. The author was advised that more cues underpinning for the direct link between the origin of Bitcoin and Sassaman can be spotted in the code of Mixmaster [1279, 1280], where it is clear that the knowledge of C++ [975], plus the understanding of the Win32 [1282] API [1283], were the basis for the first Bitcoin version [1284].

The ancestors of modern cryptocurrencies were already discussed as important constituents of remailers in the 1980s and 1990s [1285, 1286], and decentralized nodes distributed fixed-sized blocks of encrypted info across a P2P [986] network on remailers, and thus showed essential features of Bitcoin. Also the cypherpunk, Bitcoin entrepreneur [1037-1039], and hot Satoshi Nakamoto candidate Adam Back [69] ran a remailer [1287]. In a 2007 paper [1288] at Black Hat [1289, 1290] which showed striking parallels to elements of Bitcoin, Sassaman cited, along with papers from Hal Finney [70], Bram Cohen [288], David Chaum [73], and Tim May [93], a personal communication with Adam Back [69], and acknowledged him, and the cypherpunks mailing list [167]. This gives evidence how closely the cypherpunk community was interconnected at the time.

Sassaman was an avid advocate of information privacy [1167]; at the age of 21, he co-organized the protests against government surveillance and the arrest of Russian programmer Dmitry Sklyarov [1291]. Already in his early 20s, Sassaman set up and presented at technology conferences like DEF CON [1292]. He was the co-founder of CodeCon along with Bram Cohen [288], and the HotPETS symposium [1293] (with Roger Dingledine [1294] of Tor [408] and Thomas Heydt-Benjamin [1295]). Sassaman was a member of the Shmoo Group [1296]. In a 2005 paper [1278], Sassaman cites, amongst many others such as David Chaum [73], Hal Finney [70], Wei Dai [72], and (again), Adam Back [69], who all contributed crucial components that eventually entered Bitcoin.

Open-source software, P2P [986], avoidance of trust in third parties, e.g., administrators of centralized computer networks, privacy and freedom-of-speech (and its legal implications), pseudonymity, and strong encryption, e.g., as implemented in PGP, were major buzz words Satoshi Nakamoto highlighted in the context of his Bitcoin whitepaper.

Never having attended college, Sassaman started his "dream" postgraduate studies at KU Leuven [430] in Belgium [618] (Benelux [401]) in 2004, which he completed with a MSc in 2008 [1297]. One of his academic supervisors was the digital-currency pioneer David Chaum [73, 271, 485], who was based at UC Berkeley [412, 675] and Los Angeles [926]; his other (presumably main) supervisor was Bart Preneel [505, 677, 678], the head of the Computer Security and Industrial Cryptography (COSIC) center [643]. In 2008, Sassaman submitted his Masters' thesis "Toward an Information-Theoretically Secure Anonymous Communication Service" [1297]. Noticeably, the author could not retrieve the PDF on the internet – are there some fingerprints Sassaman wanted to conceal?

Preneel was one of the designers of "RACE Integrity Primitives Evaluation Message Digest" (RIPEMD) [1298], which, in conjunction with "Secure Hash Algorithms" (SHA) [547, 548, 551, 1299], constitute the methods used to make a Bitcoin address from a public key, specifically SHA256 and RIPEMD160 [210]. During Sassaman's postgraduate period in the 2000s, Preneel presented at events in Japan [679], and (co-)supervised a PhD thesis by Kazuo Sakiyama [1300, 1301].

As pointed out previously in the context of the pivotal citation (2) [612] in the Bitcoin whitepaper [175] from the Quisquater [124, 125] group from UC Louvain [637] (i.e., also situated at Leuven [672]), Preneel, and another top-notch researcher Joos Vandewalle [511, 644, 645], co-authored in the same proceedings of the local Benelux symposium [628] on "Anonymity controlled electronic payment systems" [680], and had another contribution [681] that was first-authored by "Jorge Nakahara Jr" [648-650], presumably a researcher from the Japanese-Brazilian [684] minority.

Even though Len Sassaman is very unlikely to have been an attendee of the Benelux meeting, for instance, due to his young age (~19), and his US-based residence in 1999, he might have gained access to the hardcopy [628] containing (2) [612] in his later postgraduate studies through Preneel's personal book collection, or the local library at KU Leuven [430].

Alternatively, Sassaman might have gotten hold of the proceedings [628] featuring (2) [612] as he joined the organizing WIC [629, 630, 686], and also published at the symposium with his wife Meredith Patterson [1302] in 2007 [1303] on "Subliminal Channels in the Private Information Retrieval Protocols", and with his local PhD supervisor Bart Preneel in 2008 [1252] on "The Byzantine Postman Problem", which alludes to the "Byzantine Generals Problem" [1189] in Bitcoin. Sassaman attended [686] CRYPTO 2006 [1304] in Santa Barbara [492], where his supervisor Bart Preneel [505] chaired a session; talks in the program [1305], and conversations with other participants, might have further sparked his journey towards Bitcoin.

Post 2008, Sassaman was a program committee member of the 13th International Conference on Financial Cryptography and Data Security [1306], held in February 2009 at the island country of Barbados [1306], and joined the "International Financial Cryptography Association" (IFCA) [1307] in 2010.

Sassaman was characterized [47] as "always kind of the odd kid because he was smart" and people remembered "intelligent and lighthearted, chasing down a squirrel at a Cypherpunk meeting and speeding around in a sports car with a brazen ‘Get Out of Jail Free’ card in case he was pulled over.” On February 11, 2006, at the 5th CodeCon, Sassaman proposed to returning speaker Meredith L. Patterson [1302, 1303] during the Q&A after her presentation, and they then married [77]. His wife is a linguist by training who was also familiar with the crypto scene and computer science; Patterson was diagnosed with autism [1013] in her adulthood.

Very poignantly, Sassaman decided to end his young life at the age of 31 on 03 July 2011, i.e., just weeks after Satoshi Nakamoto sent his last known email (of widely undisputed origin) on 26 April 2011 [1308, 1309], and his final tweets regarding Bitcoin on 24 June 2011.

Sassaman thus ticks a large number of the boxes, and would have certainly had the vita, intellect, skill set, psychogram and motivation to create Bitcoin. His studies at KU Leuven [430] in Belgium [618] might also explain Satoshi Nakamoto’s occasional British spelling, the headline in “The Times” in the Bitcoin genesis block, the citation of the 1957 book (8) [557] which is / was available in the local library, the DD/MM/YYY date format [913], and reference to the Euro currency [353]. The Japanese last name of his pseudonym might have been inspired by Jorge Nakahara Jr. [648, 649], who was his colleague at the time in the research group around Bart Preneel [505, 677, 678] and Joos Vandewalle [511, 644, 645] at KU Leuven [430], or other (former) COSIC members with Japanese-looking names, and international cryptographers [650, 688, 695-699, 1310].

Certain inconsistencies remain, e.g., his C++ [975] programming in a Win32 [1282] environment while being known as an Apple Macintosh [1311] user, and evidently the Satoshi Nakamoto postings (of debatable genuineness) past his demise [8, 861]. Possible resolutions of these paradoxes have been suggested, for instance, that Bitcoin had a group of creators, e.g., Len Sassaman [77] himself, Hal Finney [70], and Bram Cohen [288], playing different roles [1269].

Nevertheless, Len Sassaman has been frequently nominated as Satoshi Nakamoto [1312, 1313], and a digital obituary has been embedded in the Bitcoin blockchain [46, 47, 1312-1319], and the matching with new evidence presented in this work support this hypothesis. Several open still questions, like the citation (8) of the 1957 edition of a book [557, 712] in the Bitcoin whitepaper [175], the penchant for mathematical puns and historical associations, would be virtually impossible to clarify posthumously.

With this wealth of supporting circumstantial evidence, mostly carved out in an excellent 2021-article [47], it is disillusioning to read Len Sassaman’s tendentially very negative speech on Twitter [1320] (search: from:@lensassaman bitcoin), for instance, referring to Bitcoin as “overhyped”, and a “heist”. His wife even commented to one of them: “The trouble with bitcoin jokes is they're just not worth the time you put into them.” If not a deep fake, hack, bunk, part of his sophisticated disguise, or a sign of his severe depression, these dismissive, very negative sentiments bluntly reveal that Sassaman (and Patterson) did not like Bitcoin (anymore?).

Intriguingly, Sassaman’s tweets on Bitcoin started on 07 December 2010, i.e., just around the time when Satoshi Nakamoto posted last on the Bitcoin forum, and stopped contributing to the code, and ended 24 June 2011, i.e., about a week before he sadly decided to take his own life on Sunday, 03 July 2011 [1321].

It can be speculated that his severe depressive episode might have drastically escalated through his frustration about side effects, like Silk Road [543], and unforeseen shortcomings of his own brainchild; this perception might have utterly flipped his attitude towards Bitcoin, which he had hoped to develop in a different direction for the good of mankind and his ideals as a cypherpunk. Such sea change behavior towards their own masterpieces has been previously observed with geniuses affected by mental disorders [1009].

It would be interesting to know whether Len Sassaman briefed his wife [1302] on his (alleged) involvement in Bitcoin, and handed over private keys to Bitcoin addresses and passwords to his online accounts. In her reply to on Twitter (@maradydd) [1322] to rumors about her late husband being Satoshi Nakamoto [47] in February 2021, Patterson denied: “It's a very well-researched and respectful article, but to the best of my knowledge, Len was not Satoshi.”

His own, and his wife’s more recent, daunting comments substantially taint the hypothesis that Len Sassaman was the creator of Bitcoin. Still, Meredith Patterson tellingly tweeted almost at the same time as Satoshi Nakamoto entirely disappeared from newsgroups on 07 December 2010: “Bitcoin isn't ready for prime time yet, according to its creator. Interested people can help finish it, though!” [1323]. How would she have been able to “know” about the intention of Satoshi Nakamoto in these early days of Bitcoin? Was she referring to one of his online postings, or did she have personal insight?

The author hypothesizes that, if he was Satoshi Nakamoto, and considering that he most probably had a well devised exit strategy at hand from the very outset, Len Sassaman must have undoubtedly anticipated that his profile will, sooner or later, become a prime suspect, so he needed to spread a glaring inconsistency.

#### *Phil Zimmermann*

Born in 1954, the US-American computer celebrity Philip “Phil” Zimmermann [74] obtained a B.S. in computer science at Florida Atlantic University [1324] in 1978. His interest in the political side of cryptography grew out of his background in military policy issues.

He was inducted into the Internet Hall of Fame [1325] in 2012 for having developed “Pretty Good Privacy” (PGP) [689] (supported by Hal Finney [70]), an encryption program that provides cryptographic privacy and authentication for data communication, e.g., email, in 1991. The PGP case against US government [1326] regarding infringement of US export regulations on cryptography [1173, 1326] was dropped in 1996 [1173].

On the academic side, Zimmermann is an Associate Professor Emeritus of cybersecurity at Delft University of Technology [639], in the Netherlands [400], received an Honorary Doctorate from the Université Libre de Bruxelles [433, 706] (Belgium [618]) in 2016, and is a fellow of the Center for Internet and Society [1327] at Stanford [413]. He is the co-inventor Zimmermann–Sassaman key-signing protocol [1253] and was / is involved in “Voice over IP” (VoIP) [1328] and SilentCircle [1329].

Consequently, Phil Zimmermann ticks quasi all essential requirements for being Satoshi Nakamoto, e.g., his qualifications in cryptography, privacy technology and reticence, computer science (programming, networking), his resilience, political attitude, pioneering spirit, extensive experience with the legal system [1326, 1330], his connection with the Benelux ecosystem (2) [612], and his track record in- and outside academia.

His age (about 54 in 2008) might explain the rather ancient list of citations (1-8), culminating in the 1957 book (8) [557], the double-spacing after the period, and the international, American-European projects and involvements might explain time zones, timestamps, and access to “The Times” as an English speaker outside abroad, e.g., in Brussels [706]. No wonder his name has, even though astonishingly rarely, been proposed as Satoshi Nakamoto [1331].

Intriguingly, Zimmermann opined “Anyone who claims to be the inventor of Bitcoin is lying” [1254]. This - only superficially - simple statement intrinsically necessitates the knowledge of a single, exclusive holder of the secret around the origin of Bitcoin, with the distinct exception of himself. Hence, his plain testimony might thus – at least at second thought, and with a grain of salt – be logically construed as either “I am Satoshi Nakamoto, but I will never come out”, or “I know the real identity behind the pseudonym Satoshi Nakamoto, but I am absolutely certain he / she / they are never going to tell”. Otherwise, the man, who is most famous for the encryption program PGP [689], which provides cryptographic privacy and authentication for data communication, would act blatantly untruthful.

The author would personally question that anybody, other than Satoshi Nakamoto himself, whether an individual or team, would know about his / her / their real identity. Deliberately sharing such knowledge would require life-long trust, which is risky, and those who learnt about the secret without consent of the creator of Bitcoin would have probably been already too much tempted to leak.

Anecdotally, Phil Zimmermann externally communicates a rather skeptical attitude towards Bitcoin, for instance, criticizing its energy consumption to sustain PoW.

*Bart Preneel*

Factoring in the deluge of evidence, it seems the elephant in the room is the head of COSIC [643] at KU Leuven [430], Bart Preneel [505, 677-679]. His name popped up in copious independent contexts of this article. Born on 15 October 1963 in Leuven [672], he received an electrical engineering degree in applied science in 1987, and completed his PhD entitled "Analysis and design of cryptographic hash function" in 1993, both at KU Leuven [430]; Preneel's advisor was the internationally highly accomplished Joos Vandewalle [511, 644, 645] (with René Govaerts [1332]) with whom he co-authored numerous publications in the following years. He is a very active member, and acted as president of the International Association for Cryptologic Research (IACR) [494] organizing the CRYPTO conference series [429, 456, 589] in 2008-2013.

There is a long list of Preneel's significant contributions to cryptography. Most relevant to for this article, Preneel is one of the authors of the RIPEMD-160 [1298] hash function, which is a key component of Bitcoin. Along with Shoji Miyaguchi [1333], he independently invented the Miyaguchi–Preneel scheme [1334]. He was also a co-inventor of the stream cipher MUGI [1335] which was among the cryptographic techniques recommended for Japanese government use by CRYPTREC in 2003.

Through his role in KU Leuven [430], Preneel notably led the European FP7 [1336] projects ECRYPT [1337] (2004-2008) I [1338], and ECRYPT II (2008-2013) [1339]. Several players that have appeared in course of this article are amongst the member and associated institutions: UC Louvain [637], TU Eindhoven [431], EPFL [1340], Ruhr-Universität Bochum [1225], Aarhus University [447, 1341], Royal Holloway (University of London) [1342], Universität Duisburg-Essen [651], IBM Research [1343]. Tanja Lange [701, 702], Arjen Lenstra [442], and Nigel Smart [436] show up in the list of researchers. The ECRYPT [1337] projects referred to foundational work on GQ-scheme [1344], which was co-invented by J.J. Quisquater [124, 125] at UC Louvain [672].

Bart Preneel's comprehensive list of publications [679] features several publications on electronic cash / payment systems pre-dating Bitcoin [1345, 1346]. For instance, his postgraduate student Joris Claessens [647] (co-promoter: Joos Vandewalle) submitted a PhD thesis in 2002 entitled "Analysis and design of an advanced infrastructure for secure and anonymous electronic payment systems on the Internet" [1347]; J.J. Quisquater [124, 125] on the jury.

Claessens thesis cites work from researchers that recurrently appeared in this study, especially in the context of e-cash / cryptocurrencies: Dan Boneh [80], , Stefan Brands [403, 1228], David Chaum [73], Jan Camenisch [514, 1348], Ivan Damgård [446], Ian Goldberg [428], Eiichiro Fujisaki [1349], Tatsuaki Okamoto [488, 489, 504], Kazuo Ohta [502], Ralph Merkle [419, 714] (7) [616] Silvio Micali [81], Ronald Rivest [509], Tomas Sander [1350], David Wagner [415], Bruce Schneier [721], Moti Yung [513], and his own group [1351], but not Quisquater (2) [612] pointing to Stuart and Haber (3-5) [613-615] on timestamping, or any e-cash systems from the cypherpunk scene, such as (1) [270] and (6) [269]. Along with Joos Vandewalle [511, 644, 645], Claessens attended the 1999 symposium [628] in the Benelux where (2) [612] was published, and presented on "Anonymity controlled electronic payment systems" [680].

In a 1996-chapter Preneel senior authored [1352], even the "antiquated" book on probability theory (8) [557] cited in the Bitcoin whitepaper [175] is referenced (even though in its slightly "younger" edition from 1968 [1353]). He also co-authored with David Chaum since the early 1990s [1354], collaborated various times with J.J. Quisquater [124], in particular on the TIMESEC project [626] that led to (2) [612], and communicated with [1355], and cited by Haber and Stornetta (3-5) [613-615]. As mentioned above, Preneel published many times with his graduate student Len Sassaman [77, 679, 1259, 1297], e.g., in 2008 [1252] on “The Byzantine Postman Problem” [1189].

As a cryptographer, he was naturally well of Ralph Merkle’s [419, 714] foundational work (7) around 1980 [616], e.g., by manifold citations in his own papers [1356, 1357]. Maybe Preneel collated the missing items that were available to him in the years to come prior to the Bitcoin whitepaper. Starting in 2016/2017, Preneel published also on Bitcoin [1358, 1359].

As a world-leading pioneer, and internationally highly networked researcher in the field of cryptography, Preneel would certainly have (had) the skill set to conceive Bitcoin. He worked near Brussels [706] where he might have been able to grab a copy of UK-print version of “The Times” [303, 304] on 03 January 2009, and his connections into Japan might have inspired his pseudonym. Preneel most likely had a busy travel schedule, which might explain different time zones of his postings. Even J.-J. Quisquater [124] has only quite recently pointed to this option [1360]. From the first two letters of his given name, and with a sense of humor, it can be recognized $(\mathrm{B} \mapsto 2, \mathrm{A} \mapsto 1) \Rightarrow 21 = 42/2$, which plays a huge, and repeated role in the design of the Bitcoin blockchain. Hence, with some humoristic imagination, Bart Preneel might be half of the “Answer to everything” [1101].

However, other “Satoshiness” attributes do not seem to match that well from an external perspective, e.g., his mediocre computer hardware and usage of OpenOffice while presumably well-versed in LATEX. As the head of a large research institute, Preneel might also simple not have had the bandwidth to lead research and represent during daytime, and sneak into the backroom for pulling off Bitcoin afterhours. It would be up to Bart Preneel to comment on his nomination as “Satoshi Nakamoto”; the author’s hunch is that if he is not (part of) Satoshi Nakamoto, he would know the person, possibly not as the architect of Bitcoin, or at least have the most qualified punt on his real identity.

#### *Leuven-Eindhoven Ecosystem*

In any case, it “smells” like the Leuven / Eindhoven ecosystem, that Preneel is / was so evidently an integral part of, and its international collaborators (e.g., US, Canada, Japan), has something to do with the creation of Bitcoin. Maybe Bitcoin was created by all, or a subset of colluding researchers like Bart Preneel, his postgraduate student Len Sassaman [77] with his co-supervisor David Chaum [73], Joos Vandewalle [645], Nigel Smart [436], and J.-J. Quisquater [124] from Leuven [672], plus Daniel Bernstein [122], Tanja Lange [702], and Henk van Tilborg [402] affiliated with TU Eindhoven [431], or Philips [664-667], and Phil Zimmermann [74]. These experts may have played different roles, such as active drivers, or nescient facilitators.

In addition to the already referenced collaboration and co-publishing, there are further links, for instance, that the Grand Opening of Ei/Psi [1361] at TU Eindhoven on 21 April 2008 was hosted / attended by Henk van Tilborg [402], Tanja Lange [702], Daniel Bernstein [122], and Bart Preneel [505], plus cypherpunk Bruce Schneier [721] from Harvard [421] in the Bay Area [675]. David Chaum’s [73] name popped up in connection with Bart Preneel [505], e.g., co-authored publications and conference organization, and the co-supervision of Len Sassaman [77] at KU Leuven. The author deems the odds that the Benelux cryptography ecosystem and its immediate international periphery, especially in the UK [69, 599] (cf. next subsection) and USA [165, 166, 168, 412, 413, 421, 493, 617, 675, 720, 750, 926, 1142, 1144, 1362], did not play a key enabling role in the creation of Bitcoin as very slim.

### Great Britain

Regarding the native-level British English in the Bitcoin whitepaper [175], and the headline in the print-only UK version “The Times” [304] headquartered in London which was engraved in the genesis block [176, 177], activities in the top-notch cryptography arena in the UK [599, 1363] might be of relevance.

Other than the high-ranking Satoshi Nakamoto suspect Adam Back [69] discuss further above, these researchers have not been in the spotlight, and are primarily brought up here as possible links to solve the conundrum; there is no compelling evidence supporting their direct involvement in the creation of Bitcoin.

#### *Malcolm J. Williamson*

Historically, the accomplished cryptographer Malcolm J. Williamson [423, 424], worked at GCHQ [438] in early 1970s to 1982, then moved to the IDA's [1364] Center for Communications Research [424] in La Jolla [1365] near San Diego [1366] (Southern California [493], USA), presumably on classified projects. He was born in 1950, and passed away in 2015, i.e., around the time when Satoshi Nakamoto eventually went silent (so far). Therefore, Williamson's profile offers a few explanations for the mix of British and American English, his intelligence background, his anonymity, and possibly the time zone data.

As Baby Boomer [1183], his advanced age of about 58 years in 2008 can well explain a double-spacing habit, the rather dated reference list in the Bitcoin whitepaper [175], and he might even have naturally come across the 1957-book (8) [557] during his advanced studies. Whether Williamson was a proficient C++ coder, used OpenOffice [579], had access to citation (2) [612] (Figure 1) and the UK print version of "The Times" (probably if he resided in his country of birth on 03/01/2009), and met other "Satoshiness" filters could not be verified.

#### *Ross J. Anderson*

The well-known British cryptographer Ross J. Anderson [437] is a professor in security engineering at the University Cambridge [1363]; He boasts a track record as an industry consultant, and is an outspoken defender of academic freedoms, intellectual property and other matters of university politics. He has published on the bigger picture of Bitcoin, e.g., in a 2018 paper titled "Making Bitcoin legal" [1367].

#### *Sean Murphy*

Having received a PhD from University of Bath in 1989, the British cryptographer Sean Murphy [439] worked on international projects with project with Bart Preneel [505, 677] from COSIC [643] in the early 2000s. He is now faculty at Royal Holloway, University of London [1342]. His publication activity shows a slight dip around the years 2007 and 2008.

#### *Aggelos Kiayias*

The Greek [1368] cryptographer and computer scientist Aggelos Kiayias [1369] is a professor at the University of Edinburgh [1370] in Scotland [1112], and the Chief Science Officer (CSO) at the company behind the cryptocurrency Cardano [23]. His PhD in 2002 was supervised by Moti Yung [513] at the City University of New York [1362]. Note that Moti Yung took on the same role, also in 2002, for Jonathan Katz [417, 1371] at Columbia University [1372] in New York City [750].

### Further Examples

A large number of other Satoshi Nakamoto candidates have been tabled in various media. The following subsection highlights several profiles which meet a good portion of the criteria attributed to the still unidentified father of Bitcoin, while they also feature crucial contradictions. There are also some dark horses, e.g., formerly or actively affiliated with three-letter intelligence organizations [460, 461, 544, 554, 842-849, 851, 852] at the time, or the huge talent pools, e.g., pertaining (at some career stage) to the computer science programs [73, 80, 122, 412, 413, 415, 428, 1373, 1374], and the quite interwoven cypherpunk [165] initiative that spawned from the San Francisco Bay Area [675].

#### *Elon Musk*

The South-African born business magnate Elon Musk [15] (born on 28 June 1971) has been occasionally affiliated with the invention of Bitcoin [1375]. He publicly denied [1376], even saying that he would tell in public if he had been the person behind Satoshi Nakamoto [59, 918, 1376]. While his skill set, e.g., in software, internet, and artificial intelligence (Zip2 [1377], OpenAI [1378]), payment systems (PayPal [1379]), and advanced technology [1380-1382], as well as his dazzling intelligence of a true genius, his autodidactic skills, highly multi-disciplinary "thinking outside the box" attitude, his unparalleled entrepreneurial resolve, his rare openness of taking huge risks against all odds, and even his Asperger condition, would have likely enabled him to invent Bitcoin. Musk is assumed to be proficient in C++ [975], and, being from South Africa [1113], some British / Commonwealth [565] influence in his writing may be explained.

However, it needs to be considered that just his involvement in the pioneering electric car company Tesla [1383] since 2004 might have been, even for him, far too demanding to silently develop Bitcoin as a side activity in the years before 2008, when the company was in the final, highly stressful stages of launching its first automotive product, and Musk eventually assumed the role of CEO.

Moreover, Musk's common, long-term investment approach to business would contrast Satoshi Nakamoto's sudden departure from Bitcoin in 2010/2011, and his distinct psychogram as a very outspoken, constantly tweeting, "rock-star" entrepreneur starkly clashes with the more introvert, technically minded Satoshi Nakamoto who clearly avoided any public exposure. It is also hard to construe how Elon Musk would have gotten hold of the hardcopy of citation (2) [612], and his academic-research style of writing in the Bitcoin whitepaper [175].

*Paul Le Roux*

In a 2019 article in Wired and some follow-ups [42, 58, 76, 1384, 1385], the programmer, former criminal cartel boss, and informant to the US Drug Enforcement Administration (DEA) [1386] Paul Le Roux [1387] was posited as Satoshi Nakamoto. Born on 24 December 1972 in South Africa [1113], most information about him is only known from secondary sources. Le Roux is reported to possess a diplomatic passport of the Democratic Republic of Congo [1388], issued on 05/08/2008 to the name Paul Salotshi (↦ Satoshi?) Calder Le Roux.

After dropping out of school at the age of 16 and embarking on a programmer career, Le Roux developed the open-source disk encryption program for Microsoft Windows E4M [1389], and, arguably, open-source TrueCrypt [1390].

Le Roux was arrested on 26 September 2012 for conspiracy to import narcotics into the United States. He subsequently conceded to planning or partaking in several murders to set up a wide-ranging illegal business organization; Le Roux was sentenced to 25 years of imprisonment in June 2020 [1387].

His astonishing catalogue of presumably self-taught competences, intelligence, and fit with a number of aspects that are attributed to Satoshi Nakamoto, it is hard to envision how the two psychograms would match. Besides, the invention and implementation Bitcoin could hardly have been a side project while building up an illicit empire involving the assassination of people.

Furthermore, the pseudo-academic style of the very technically, and excellently written Bitcoin whitepaper [175], particularly the inclusion of a paper from a local symposium that was mostly available, at the time, as a hardcopy in the Benelux (2) [612], and the lack of overlap with many other points outlined in this article most likely exclude Le Roux a Satoshi Nakamoto candidate.

*Craig Wright*

Upon the release of the preprint of my document, the author was communicating with Dr. Craig Wright [37, 57, 99, 812, 1391-1400] who is, according to the author's best knowledge at the time of writing, the most prominent person who (at least seriously and repeatedly) claims to be the inventor of Bitcoin (in public). Wright's statement is fiercely contested by an apparently solid majority of the Bitcoin and crypto communities [8, 57, 1395, 1401-1413].

However, since his highly publicized, apparently not entirely voluntary coming out in 2015/2016, Wright has underpinned his position in courts, whether as defendant or plaintiff of libel cases [103-108, 1414] against major stakeholders in the crypto industry and journalism. Nowadays, Wright complains that his original idea of Bitcoin has been substantially altered, especially through the SegWit [1415] and Taproot [1416] updates, to an extent that he became an activist against Bitcoin's present implementation, while promoting another electronic payment system [1032].

From the author's neutral position, Wright makes a lot of statements about his unique, amazingly broad portfolio of competences, formal qualifications, and diverse entrepreneurial and employment record that, if confirmed, would indeed make him a strong candidate for Satoshi Nakamoto.

Yet, for instance, given the psychogram transpiring from the Bitcoin whitepaper, it can be assumed that Satoshi Nakamoto would have certainly anticipated the public expectation of a clear proof of identity, primarily through using the private keys for moving coins residing at Bitcoin addresses that are attributed to him [374, 375].

He could also send a message from email accounts he used during his online life as Satoshi Nakamoto. (However, such evidence would still need to factor in that a hoaxer could have gotten control of such digital fingerprints through alternative means.) Wright stubbornly continues to contend that such evidence was unsuitable for proving the real-world identity of Satoshi Nakamoto, a point which is ferociously challenged by major fractions of the community.

Instead, Wright tends to repeatedly produce rather soft evidence, like his exceptionally broad set of Bitcoin-relevant skills, his staggering number of patent filings, and his relevant professional track record and private life. His explanations for the citations of the 1999 Benelux symposium (2) [612], and the 1957-edition of the textbook (8) [557] remained vague. Would he have had access to the UK (print-only) version of "The Times" [747], run Windows XP on a rather low-end hardware around 2008, and used OpenOffice for authoring the Bitcoin whitepaper?

Wright also ran into ostensible contradictions, e.g., by asserting "I had to know what I was coding before a coded.", while Satoshi Nakamoto posted "I had to write all the code before I could convince myself that I could solve every problem, then I wrote the paper." Surprisingly, and in addition to Adam Back's [69] "hashcash" [269] as cited as (6) in the Bitcoin whitepaper [175], Wright attributed inspiration for Bitcoin to a Finnish group around Tuomas Aura [1417, 1418] at Helsinki University of Technology [1419] (since 2011: Aalto University [1420]) who published on "how the robustness of authentication protocols against denial of service attacks can be improved by asking the client to commit its computational resources to the protocol run before the server allocates its memory and processing time" [1421].

Playing devil's advocate, the very scarce information Satoshi Nakamoto revealed about himself left the world with an open canvas of his biography; it can be contended that essentially anybody from a credible age group who understands the Bitcoin technology platform through the whitepaper and related books, and had studied Satoshi Nakamoto's public postings, could have wrapped a rather plausible, but fictional story around this skeleton. This makes it also difficult to formally disprove a claim on being Satoshi Nakamoto.

In a final attempt, the author just accepted that Wright would, for some reason, have lost possession of the widely demanded, Bitcoin-blockchain based proof of identity. Wright could have then still made his case at least more credible if he was to provide evidence dating prior to August 2008 when he first used his pseudonym, or the release of the whitepaper on 31 October 2008; the author suggested that Wright ought to provide authentic documentation, witness, or a detailed description of key "Eureka" moments while conceiving Bitcoin, e.g., when he came up with the unprecedented solution for the notorious double-spending or single-point-of-failure issues through (a clever combination of) concepts like a distributed ledger, proof-of-work ("mining"), and a hash-linked chain of timestamped blocks.

Overall, Wright became increasingly fluffy when approaching hard proof, which, he argued, will be produced at some stage in courtrooms by “people”. His exit strategy was that he is actually not interested in confirming his statements on being Satoshi Nakamoto, and rather quickly transformed into an activist against the present form of Bitcoin, while promoting his new, competing system [1032], and other project(s).

In the extended interaction with the author of this article, Wright produced an unnecessarily emotionalized atmosphere, with a hot-tempered rhetoric spiced with many hot-headed swearwords and rather despicable language, and his serial filing of libel and defamation cases including personal intimidation and threats (against the author) are in stark contrast to the characteristics and mindset of Satoshi Nakamoto that the author sensed when reading his Bitcoin whitepaper and 2009/2010 online postings.

His multi-day chat with Wright left the author quite perplexed: In general, tabling a claim for the identity of an(other) individual, as per definition of this very word, can only have a binary outcome: Either Wright is indeed the creator of Bitcoin, meaning the genius who already changed the age of information, or he is an imposter with a confusing strategy and objectives. What is his benefit of filing libel cases if he is not willing to prove his identity according to widely acceptable standards?

Why would Wright impose the stress upon himself of coming out or assuming another identity for the rest of his life, knowing that he will never be able to substantiate his claim to the satisfaction of the community? If he was a pretender, why is Wright not afraid that the real “Satoshi Nakamoto” would appear – does he know about his fate of never being willing or able to ever appear on stage, or does he even collude with him? In fairness, if Wright is not able or willing to produce more compelling evidence, the overwhelming fraction of the crypto community will most likely remain in serious doubt.

A more recently launched, rather adventurous hypothesis [67] interweaves the assumed involvement of Paul Le Roux in the creation of Bitcoin with the double-team of Craig Wright [96] and Dave Kleiman [117], and their, presumably jointly owned, company “W&K Information Defense Research LLC”. A notorious court case of in Florida [1422, 1423] dealt with the rights to more than 5 billion USD worth of Bitcoin, claiming that Wright defrauded Kleiman of bitcoins and intellectual property rights. Leaked official documents spurred rumors that Wright might have played a role in the arrest of Le Roux. It is surmised that Craig Wright (and possibly Kleiman) had somehow been able to take possession of Paul Le Roux’s encrypted hard drives that store the private keys for controlling the alleged, about 1 million BTC believed to reside in Bitcoin addresses controlled by Satoshi Nakamoto.

Due to Le Roux’s long-term imprisonment, and after Kleiman’s death, Craig Wright reckoned that, with the huge calculation power provided by a sufficiently large computer farm, he could crack the encrypted media within a long-haul, multi-year, or even decade-lasting effort. In case of success, Wright would still have to move the BTC, which would immediately and broadly be noticed by the community, and federal agencies. For cashing out in fiat money, especially in view of Satoshi’s mindboggling assets, would require KYC / AML registered bank accounts, and thus unavoidably reveal his identity. If this bold, and completely unproven story was true, Craig Wright must thus have pondered that putting a claim on being Satoshi Nakamoto gave him a chance to liquidate “his” (~1 million) BTC. A similar plot would apply to the Kleiman case; it has been said that Wright may somehow have acquired Kleiman’s encrypted drive holding the private keys to (half of) the BTC mined by Satoshi Nakamoto. Evidently this latter narrative would require that Wright and / or Kleiman (or their company W&K) were the team behind, or involved in the creation of Bitcoin, or at least early-day miners. The author of this article is neither able to confirm nor dismiss the validity of this purported plot.

The author leaves it up to the esteemed reader to map the narratives dispensed by Craig Wright [812, 1391, 1392] on the profile and (hard) exclusion filters produced in this article.

Why is the Craig Wright example relevant to cryptocurrency community and industry, in addition to the curiosity-driven search for the origins of Bitcoin? Apart from referring to himself as Satoshi Nakamoto, Wright sues major crypto-stakeholders in court for calling him a fraud [103-108, 1414]. Maybe of significantly more practical relevance than the increasingly academic debate on his inventorship of Bitcoin after the blockchain community has transitioned into a life of its own, Wright has filed, and been issued, numerous intellectual property rights for potentially important components of blockchain technology [109-111], which might hamper progress or jeopardize current and future projects in the field.

## Summary & Outlook

This paper has discovered so far mostly overlooked, circumstantial "material" evidence around the origins of Bitcoin. The first revolves about the citation of a 1999 article published in the book of proceedings of a very local (networking) symposium in the Benelux, that was only available, at least before 2008, as a hardcopy to attendees and a small number of libraries. Furthermore, the headline cited by Satoshi Nakamoto in the Bitcoin genesis block was only available in the UK-print edition of "The Times", but was phrased differently in its online, print-only international / US versions. Moreover, the citation of a first, 1957-edition of a standard textbook on probability theory (8) [711] might carry further cues about Satoshi Nakamoto, e.g., about a family tradition in mathematics, or a passion for historic associations.

Just by the sheer capability of inventing and implementing Bitcoin in a code-first approach, Satoshi Nakamoto must have been an expert-level (C++) programmer versed in cryptography, electronic cash, decentralized peer-to-peer networks, cybersecurity, money, banking, law, a skill set possibly acquired in a prodigiously self-taught mode. Even preceding the very first appearance of his persona, Satoshi Nakamoto cunningly scoped and impressively delivered on a well-equipped, long-haul strategy, resorting to privacy-focused internet providers and technologies, combined with strict, long-term sustained reticence, for constructing a (so far) impenetrable citadel for shielding his real identity.

In addition to the suspicious citations (2) and (8) of the Bitcoin whitepaper, several somewhat free parameters that concern: the design of the Bitcoin blockchain, his stage name, nationality, his choice of English orthography and idioms, and certain dates that were (broadly) left to his discretion. Even more, his computing setup looks like he was (initially) running Microsoft Windows (XP) on rather low-end hardware with the rarely used word processor OpenOffice (instead of LATEX), a somewhat odd package for a computer wizard, unless this was owing to a lack of financial means at the time, or a decoy. Satoshi Nakamoto was also not passionate, or capable, of generating eye-catching graphics.

Deeper analysis, also considering time zones, linguistics, and IP addresses, remains somewhat inconclusive, indicating a person having British / Commonwealth provenance, substantial academic links to the Benelux, but also residence or stints across in the entire continental USA. Even beyond the pick of his online name, a noticeable personal connection to, or passion for the Japanese culture gleams through, possibly inspired by peers from Japan. Also traces of a penchant for history, strategic games, like Chess or Poker, and fun with geometries may be inferred from a coherent collection of evidence.

Even though Satoshi Nakamoto never seems to have deposited directions to search for puzzles, such intention may actually be inferred from his enigmatic character, and the baffling twists and turns of his online life. Maybe Satoshi Nakamoto was an undercover entertainer? Hence, such compounding of cues indicates that at least some hidden messages were deliberately arranged by Satoshi Nakamoto.

While each piece of the puzzle and its "exegesis" [794] might justifiably be challenged on an individual level, the striking accumulation of evidence around his pseudonym, the numbers 21 and 42, the binary system, and symbolism match well with a likely personality, as well as locations, dates and happenings that unfolded in politics, the economy, history, geography, culture, society, media, and sports related to Satoshi Nakamoto and Bitcoin; this large stack of diverse findings unveils a very privacy conscious, reclusive, quite omniscient, multi-talented, autodidactic, charismatic, creative, idealistic, sometimes flippant, “nerdy”, slightly unsociable, and rather factually oriented, than ideologically evangelizing tech genius who might suffer from a (possibly faint) mental condition.

Dovetailing these discoveries with the slew of preceding research on Satoshi Nakamoto amalgamates to a still hypothetical, but overwhelmingly consistent partial biography and psychogram of an intellectually abundantly endowed, altruistic, sensitive and, at least in the eyes of the general public, erratic person with exceptional work ethic towards his libertarian-, possibly cypherpunk- minded ideals who is, by embodying this highly unique spectrum of traits, unprecedented in the history of science and technology.

Even though not pursued in this work, the opulent diversity and (partial) inconsistency of certain information might also be attributed to a confined group, e.g., with a shared track record in a federal or corporate cryptography department, or even an academic, entrepreneurial or family combo that conspired for the development of Bitcoin, possibly coordinated by a designated, *de-facto* leader. This assumption of an expert squad might also explain the unusual blend of primarily US-American, British, European and Far-Eastern artefacts attributed to Satoshi Nakamoto. However, a reference to a group, such as an accidental “we” or “us”, has never (accidentally) slipped through in his communication related to the creation of Bitcoin.

From its outset, this work was not driven by pitching new or supporting existing names, or from infringing with the privacy of the real person(s) that has (so far) successfully perched behind the pseudonym Satoshi Nakamoto. Unsurprisingly, as the evidence discovered in the fabric of Bitcoin were deployed by himself, Satoshi Nakamoto would hardly have deliberately included anything that unwraps his shrewdly refined disguise of identity. His encrypted messages were mostly providing a peek into important happenings, his witty humor, and possibly exposing certain sentiments. In the final section, the compiled evidence was probed against a list of past nominations, and the cohort of candidates could be narrowed down significantly; however, as expected, contradictions remained, and a final call could thus not be made.

This may be owing to the fact that the real Satoshi Nakamoto has not been screened, yet, as a candidate, and further, dark-horse candidates may surface; the real Satoshi Nakamoto is very likely to match the profile that unfolded in this article, resolve the apparent contradictions, and dismiss certain overinterpretations. Even in a dearth of such a “perfect fit”, the developed vita and psychogram can be employed to debunk false nominations, and also self-appointed “Faketoshis”; while the story drafted here will certainly not hold in all its detail, possibly also because even simple explanations have fatally been overlooked, good arguments would have to be offered for deviations, and convincing alternatives would have to be presented in order to support any “Satoshi” hypothesis.

The motivation and psychology of an identity-stealing Satoshi Nakamoto imposter, if not, for instance, a prank, cover-up alibi, or part of a con scheme, forms a complex topic of its own, presenting interesting facets, like a gradually increasing believe in self-fabricated story, and growing identification through adoption of an alter ego [1424], resembling known cases [1425], e.g., in art [1426-1428] and history [1429, 1430], most prominently around the Grand Duchess Anastasia Nikolaevna of Russia (“Anastasia”) [1431]. A telediagnosis of a candidate is deemed impossible, but in psychology, there are dispositions [1432] referred to as fantasy prone personality (FPP) [1433], or dissociative identity disorder (DID), previously known as multiple personality disorder (MPD) [1434].

For now, the choice to apply the catalogue of characteristics excavated in this article to “Satoshi Nakamoto” candidates is up to vexed readers. Rather than being driven by selling names, matches to the “1-in-a-billion” profile proposed here, with rather moderate and reclusive characters, should be con-sidered. When coming under scrutiny, it is to be anticipated that the real Satoshi Nakamoto would bluntly refute his involvement in the creation of Bitcoin, even deny their interest in electronic cash / cryptocurrencies, or just not even respond at all, and likely rebuke researchers for having fielded their name.

Moreover, given his highly rational, unflustered, technical approach, it is unlikely that, if at all, Satoshi Nakamoto would step into the public without being able of producing compelling evidence for convincingly identifying him as the inventor of Bitcoin. It is also improbable that he would move his ("Patoshi" [216, 374, 375]) BTC assets for conversion into fiat money [411] to a centralized exchange, and then transfer them to a regular, and thus know-your-customer (KYC) [1435] and anti-money laundering (AML) [1436] monitored, and official proof-of-identity requiring bank account.

After all, there seems to be no panacea or silver bullet for reaching finality on Satoshi Nakamoto's real identity and aliveness, given that his private keys may have been, whether intentionally, accidentally, or forcefully, burned, physically destroyed, shared, or transferred within a cohort of creators, their families, entourage, or to externals. If Satoshi Nakamoto was not to step out, or in case he was incapacitated, or even deceased, the paradox he left may never be resolved; in this, not so unlikely, case, the foundational mystery of Bitcoin may become an eternal conundrum akin to its historical precedents, but in the digital age of information [874].

Unless it delivers inputs directly obtained from the true Satoshi Nakamoto, the author would personally stay highly skeptical whether brief articles can uncover Satoshi Nakamoto's identity coup; they might still be very captivating, providing captivating entertainment, or a boiler room for candidates motivated by a wide range of possible objectives.

The author believes that the ingenious mastermind of Bitcoin would nowadays be more than happy to chime in with the motto "We are all Satoshi!" For those who are desperately curious to uncover this very fascinating foundational mystery: "There's more work to do …" [260].

## Acknowledgements

Many thanks to some of the grands of cryptography and blockchain the author got the opportunity to have quite revealing and inspiring communication with. He also enjoyed the help of various the readers of drafts, whether experts of the topic, and nocoiners without a background in blockchain.

# Appendix

## Historical Excerpts

This section lists events that possibly impacted Satoshi Nakamoto for conceiving Bitcoin in 2007 [1437], leading to the publication of the Bitcoin whitepaper on 31 October 2008 [522], containing an unusual, historic reference to 1957-book (8) [724], the launch of the Bitcoin blockchain on 03 January 2009 [1438], his postings during his active years of Bitcoin development with the early adopters in 2009 [1438] and 2010 [1439], his final email messages in 2011 [1440], his rather persistent silence after his mysterious exit. Also, the Satoshi Nakamoto self-set day, month and year might of his birth may contain some information.

### Year 1957 [724]

- March 04 – Standard & Poor's first publishes the S&P 500 Index in the United States.
- April – IBM sells the first compiler for the Fortran scientific programming language.
- May 03 – Brooklyn Dodgers owner Walter O'Malley agrees to move the team from Brooklyn, New York, to Los Angeles.
- September 09 – The Civil Rights Act of 1957 is enacted, establishing the United States Commission on Civil Rights.
- October 04 - Space Age – Sputnik program: The Soviet Union launches Sputnik 1, the first artificial satellite to orbit the earth.
- October 10 - U.S. President Dwight D. Eisenhower apologizes to the finance minister of Ghana, Komla Agbeli Gbedemah, after he was refused service in a Dover, Delaware, restaurant.
- October 31 – Toyota begins exporting vehicles to the United States, beginning with the Toyota Crown and the Toyota Land Cruiser.
- November 03 – Sputnik program: The Soviet Union launches Sputnik 2, with the first animal to orbit the Earth (a dog named Laika) on board; there is no technology available to return it to Earth.

### Year 1975 [876]

- January 01 – Watergate scandal [877] (United States): John N. Mitchell, H. R. Haldeman and John Ehrlichman are found guilty of the Watergate cover-up.
- January 08 – U.S. President Gerald Ford [897] appoints Vice President Nelson Rockefeller to head a special commission looking into alleged domestic abuses by the CIA [554].
- February 21 – Watergate scandal: Former United States Attorney General John N. Mitchell, and former White House aides H. R. Haldeman and John Ehrlichman, are sentenced to between 30 months and 8 years in prison.
- April 03 – Bobby Fischer [555] refuses to play in a chess match against Anatoly Karpov, giving Karpov the title.
- April 04 – Bill Gates [13] and Paul Allen [870] found Microsoft [871] in Albuquerque, New Mexico [1441].
- April 05 – Satoshi Nakamoto is "born" [861, 862].
- June 10 – In Washington, D.C. [749], the Rockefeller Commission issues its report on CIA [554] abuses, recommending a joint congressional oversight committee on intelligence.
- October 01 – Thrilla in Manila: Muhammad Ali defeats Joe Frazier in a boxing match in Manila, Philippines.
- October 31 – The Queen single "Bohemian Rhapsody" is released. It later becomes one of their most popular songs.

- November 29 – The name "Micro-soft" (for microcomputer software) is used by Bill Gates [13] in a letter to Paul Allen [870] for the first time (Microsoft [871] becomes a registered trademark on November 26, 1976).
- December 08 – New York City is approved for bailout of 2.3 billion each year through 1978 – 6.9 billion total.

### Year 2007 [1437]

- May – Forerunners of financial crisis [520, 738]
- June 29 – The iPhone [1442], the first modern smartphone, is released in the United States. It was later released in the United Kingdom, France, Germany, Portugal, the Republic of Ireland, and Austria in November 2007.
- November 07 – Whistle-blower website WikiLeaks [359, 362, 538] publishes the standard US army protocol at Guantanamo Bay [537].

### Year 2008 [522]

- January 01 – Cyprus and Malta adopt the Euro [1443] currency.
- January 17 – the American chess grandmaster Bobby Fischer [555] dies in Iceland
- January 21 – Stock markets around the world plunge amid growing fears of a U.S. Great Recession, fueled by the 2007 subprime mortgage crisis.
- January 21 – Online activist group Anonymous [1444] initiates Project Chanology, after a leaked interview of Tom Cruise by the Church of Scientology is published on YouTube, and the Church of Scientology issued a "copyright infringement" claim. In response, Anonymous sympathizers took to the streets to protest outside the church (after February 10), while the church's websites and centers were getting DoS attacks, phone line nukes, and black faxes.
- February 18 – WikiLeaks [359] releases allegations of illegal activities carried out by the Cayman Islands branch of Swiss banking corporation Julius Baer; a subsequent lawsuit against WikiLeaks prompts a temporary suspension of the website, but uproar about violations of freedom of speech causes WikiLeaks to be brought back online.
- September 15 – Stocks fall sharply Monday on a triptych of Wall Street woe: Lehman Brothers' [525] bankruptcy filing, Merrill Lynch's acquisition by Bank of America, and AIG's unprecedented request for short-term financing from the Federal Reserve.
- September 29 – Following the bankruptcies of Lehman Brothers and Washington Mutual, The Dow Jones Industrial Average falls 777.68 points, hitherto the largest single-day point loss in its history.
- October 03 – Global financial crisis : U.S. President George W. Bush signs the revised Emergency Economic Stabilization Act into law, creating a 700-billion-dollar Treasury fund to purchase failing bank assets.
- October 31 – Satoshi Nakamoto releases Bitcoin whitepaper
- November 04 – Democratic U.S. Senator Barack Obama is elected the 44th President of the United States, making him the first African-American president.
- Beijing Olympics (08-24 August [1445])
  - Olympics were big mainstream news
  - Registering domain bitcoin.org [257, 258] on 18/08/2008 [1446]
  - Communication with Adam Back and Wei Dai [267]
  - Final stages of authoring whitepaper and coding Bitcoin

### Year 2009 [1438]

- January 28 – WikiLeaks [359] releases 86 intercepted telephone recordings of politicians and businessmen involved in the 2008 Peru oil scandal

- June 25 – The death of American pop star Michael Jackson [1447] triggers an outpouring of worldwide grief. Online, reactions to the event cripple several major websites and services, as the abundance of people accessing the web addresses pushes internet traffic to unprecedented and historic levels.
- July 22 – The longest total solar eclipse [1448] of the 21st century, lasting up to 6 minutes and 38.86 seconds (0.14 seconds shorter than 6 minutes and 39 seconds), occurs over parts of Asia and the Pacific Ocean.
- September 22 – WikiLeaks [359] exposes the contents of Kaupthing Bank's internal documents prior to the Icelandic Financial Crisis. These documents showed suspicious amounts of money were loaned to bank owners, and debts being written off.
- WikiLeaks [359] posted the membership listing of a radical political group known as the British National Party.

### Year 2010 [1439]

- April 03 – The first iPad [1449] was released.
- April 05 - Julian Assange [360] leaks footage of a 2007 airstrike in Iraq titled "Collateral Murder" on the website WikiLeaks.
- May 02 – The Eurozone [1185] and the International Monetary Fund agree to a €110 billion bailout package for Greece [1368]. The package involves sharp Greek austerity measures.
- May 06 – The 2010 Flash Crash, a trillion-dollar stock market crash, occurs over 36 minutes, initiated by a series of automated trading programs in a feedback loop.
- July 13 – Microsoft [871] ends extended support for Windows 2000.
- October 06 – Instagram was launched.
- October 23 – In preparation for the Seoul summit, finance ministers of the G-20 agree to reform the International Monetary Fund and shift 6% of the voting shares to developing nations and countries with emerging markets.
- November 17 – Researchers at CERN trap 38 antihydrogen atoms for a sixth of a second, marking the first time in history that humans have trapped antimatter.
- November 21 – Eurozone countries agree to a rescue package for the Republic of Ireland from the European Financial Stability Facility in response to the country's financial crisis.
- November 28 – WikiLeaks [359] releases a collection of more than 250,000 American diplomatic cables, including 100,000 marked "secret" or "confidential".
- November 29 – The European Union agree to an €85 billion rescue deal for Ireland [1115] from the European Financial Stability Facility, the International Monetary Fund and bilateral loans from the United Kingdom, Denmark and Sweden.

### Year 2011 [1440]

- January 01 – Estonia officially adopts the Euro [1443] currency and becomes the 17$^{th}$ Eurozone [1185] country.
- April 24 – The 2011 Guantanamo Bay files leak occurs, WikiLeaks and other organisations publishing 779 classified documents about Guantanamo Bay [537] detainees, and it had been exposed 150 innocent citizens from Afghanistan and Pakistan were held in the camp without trial and detainees being as young as 14 years old.
- May 01 – U.S. President Barack Obama announces that Osama bin Laden, the founder and leader of the militant group Al-Qaeda, was killed on May 2, 2011 (PKT, UTC+05) during an American military operation in Pakistan.

- May 16 – The European Union agrees to a €78 billion rescue deal for Portugal [1450]. The bailout loan will be equally split between the European Financial Stabilization Mechanism, the European Financial Stability Facility, and the International Monetary Fund (IMF) [1451].
- August – Stock exchanges worldwide suffer heavy losses due to the fears of contagion of the European sovereign debt crisis and the credit rating downgraded as a result of the debt-ceiling crisis of the United States.
- September 17 – Occupy Wall Street [1452] protests begin in the United States. This develops into the Occupy movement which spreads to 82 countries by October.
- October 27 – After an emergency meeting in Brussels [706], the European Union [597] announces an agreement to tackle the European sovereign debt crisis which includes a writedown of 50% of Greek bonds, a recapitalization of European banks and an increase of the bailout fund of the European Financial Stability Facility totaling to €1 trillion.

## Sum Fun Facts

### 21 [1150, 1453]

- 21 [1004] is the name of an ancient card game (s. Movie "21" [731]), with its most modern variant played in casinos all over the world. This more commonly known version is the American game Blackjack [1002].
- 21 is the atomic number of the chemical element scandium [1454]. Discovered in 1879, it is classified as a rare-earth element.
- The 21st amendment of the US constitution [529] actually negated the effects of the 18th amendment [1455], which was originally put in place to prohibit alcohol on a national scale.
- **According to the Bible there were 21 acts of rebellion committed by the Israelites to break free from Egyptian control.**
- **In Tarot [1456], the number 21 is related to the card The World. One of the many meanings of this card is to have the world at your feet!**
- **There were 21 shillings in a guinea, a form of currency used by Great Britain from 1663 all the way until 1816, when they were replaced by the pound.**
- During the First World War the Japanese Empire sent a list of 21 demands [1151] to the Chinese government. These demands were related to the control of the region of Manchuria, and were opposed by the UK [599] and the US [1151].
- 21 guns [1457] are fired to honor heads of state, such as royalty in the UK, and the President of the USA. Known as the 21-gun salute, this tradition goes all the way back to the times when the navies of the world still shot cannonballs.
- The summer and winter solstices [1458] take place (usually) on June 21 and December 21 respectively.
- There is a lake in the US state of Minnesota [1459] called Twentyone Lake [1460]. A retreat for Satoshi Nakamoto?
- **21 is a triangle number [1060] – it is the sum of the first six whole numbers ().21 is a triangle number [1060] – it is the sum of the first six whole numbers ($1 + 2 + 3 + 4 + 5 + 6 = 21$).**
- A game of badminton [1461] (and, until 2~~00~~1, also table tennis [1462]) ends when a player reaches 21 points.
- Three-on-three basketball [1463], aka 3x3, is a single period of 10 minutes, with the winner the first team to score 21 points (or be closest to 21 at the end of 10 minutes).
- 21 in Roman numerals is XXI.
- Forever 21 [954] in Tokyo [760], Japan [454] (closed 2019 [953])

### 4 and 44

- 4 sounds like "death" in Chinese, making it an unlucky number [1464].
- Now, at least two new spires in Boston are skipping the 44th floor (and one the fourth floor) in deference to the superstitions of expected buyers from East Asia, particularly China [453].

### Numerology

Possibly just for the fun of it, Satoshi Nakamoto might also resorted to numerology [1465], which is the pseudoscientific belief in a divine or mystical relationship between a number, and one or more attributes. It is also the study of the numerical value of the letters in words, names, and ideas. Numerology is often associated with the paranormal, alongside astrology and similar to divinatory arts.

While this connection is highly speculative, it is still interesting, or at least entertaining, to look at a rather abstruse article in advance of 11th anniversary of the registration of bitcoin.org on 18/08/2008 [1466]. This publication establishes a relationship between Bitcoin and the (undocumented) numerology [1467] of the Chaldean or Neo-Babylonian Empire [1468]: "The compound number of 21/3 means that one wants to crack the code into gaining ultimate success that will be ever ongoing throughout one's entire life. However, if it is unbalanced, one will have the feeling of often being close to the total success, but the success seems further away than initially anticipated. Therefore, if it's unbalanced, one's life might feel like walking through a tunnel of a constant learning process."